\documentclass[a4paper,10pt]{article}
\usepackage{graphicx,xcolor,geometry,url,multicol}
\usepackage[round,authoryear]{natbib}
\bibpunct{(}{)}{;}{a}{}{,}
\usepackage{amssymb}
\usepackage[fleqn]{amsmath}
\usepackage[T1]{fontenc}
\usepackage[font=small,labelfont=bf]{caption}

\usepackage{floatrow}

\usepackage{array,multirow}
\usepackage{txfonts}
\usepackage{lmodern}
\usepackage{relsize}
\usepackage{bm}
\usepackage{listings}
\usepackage{bold-extra}
\usepackage{upgreek}
\usepackage{rotating}
\usepackage{outlines}

\usepackage{titlesec}
\titlespacing*\section{0pt}{12pt plus 4pt minus 2pt}{2pt plus 2pt minus 2pt}
\titlespacing*\subsection{0pt}{12pt plus 4pt minus 2pt}{2pt plus 2pt minus 2pt}
\titlespacing*\subsubsection{0pt}{4pt plus 2pt minus 2pt}{2pt plus 2pt minus 2pt}

\titleformat{\section}[hang]{\uppercase}{\large\thesection}{6pt}{\large}[]
\titleformat{\subsection}[hang]{\bf}{\thesubsection}{6pt}{}[]
\titleformat{\subsubsection}[hang]{\it}{\thesubsubsection}{6pt}{}[]

\usepackage[bookmarks=true,breaklinks=true,hypertexnames=false,bookmarksnumbered=true]{hyperref}
 \hypersetup{
pdfauthor = { Adam R.~H.~Stevens },
pdftitle = { Dark Sage TNG },
pdfsubject = { Astronomy },
pdfkeywords = { -- },
pdfcreator = {LaTeX with hyperref package},
pdfproducer = {pdflatex},
colorlinks,
    linkcolor={red!50!black},
    citecolor={blue!50!black},
    urlcolor={blue!50!black}}

\newcommand{\m}{\bar{\mathcal{M}}_{\star}}
\newcommand{\mrem}{\bar{\mathcal{M}}_{\rm rem}}
\newcommand{\zo}{$z\!=\!0$}
\newcommand{\HI}{H\,{\sc i}} 
\newcommand{\Htwo}{H$_{2}$}

\newcommand{\ds}{{\sc Dark Sage}}

\newcommand{\ib}{\operatorname{i-bulge}}
\newcommand{\mb}{\operatorname{m-bulge}}
\newcommand{\iSF}{\operatorname{i-SF}}

\newcommand{\pSF}{\operatorname{p-SF}}
\newcommand{\pSFR}{\operatorname{p-SFR}}
\newcommand{\iSN}{\operatorname{i-SN}}
\newcommand{\pSN}{\operatorname{p-SN}}

\begin{document}

\begin{center}
{\LARGE \bf {D{\Large ARK}\,S{\Large AGE}:~Next-generation semi-analytic galaxy evolution with multidimensional structure and minimal free parameters}}\\[0.4cm]
\today\\[0.1cm]
\end{center}
{\Large Adam R.~H.~Stevens,$^{1,2\dagger}$ Manodeep Sinha,$^{2,3\ddagger}$ Alexander Rohl,$^{4,5}$ Mawson W.~Sammons,$^6$\\
Boryana Hadzhiyska,$^{7,8}$ C\'{e}sar Hern\'{a}ndez-Aguayo,$^9$ and Lars Hernquist$^{10}$}\\[0.2cm]
{\small $^1$International Centre for Radio Astronomy Research, The University of Western Australia, Crawley, WA 6009, Australia\\
$^2$Australian Research Council Centre of Excellence for All Sky Astrophysics in 3 Dimensions (ASTRO 3D)\\
$^3$Centre for Astrophysics and Supercomputing, Swinburne University of Technology, Hawthorn, VIC 3122, Australia\\
$^4$School of Computer and Mathematical Sciences, The University of Adelaide, Adelaide, SA 5000, Australia\\
$^5$Department of Mathematics and Statistics, The University of Western Australia, Crawley, WA 6009, Australia\\
$^6$International Centre for Radio Astronomy Research, Curtin University, Bentley, WA 6102, Australia\\
$^7$Miller Institute for Basic Research in Science, University of California, Berkeley, CA 94720, USA\\
$^8$Physics Division, Lawrence Berkeley National Laboratory, Berkeley, CA 94720, USA\\
$^9$Max-Planck-Institut f\"{u}r Astrophysik, D-85741 Garching, Bayern, Germany\\
$^{10}$Institute for Theory and Computation, Harvard--Smithsonian Center for Astrophysics, Cambridge, MA 02138, USA\\[0.1cm]
$^\dagger$adam@a4e.org, \url{https://orcid.org/0000-0003-1908-2168} \\
$^\ddagger$msinha@swin.edu.au\\[0.3cm]
}
{{\bf keywords:} galaxies:~evolution, galaxies:~haloes, galaxies:~interactions, galaxies:~ISM}\\[0.1cm]

\hrule

\section*{Abstract}
After more than five years of development, we present a new version of {\sc Dark Sage}, a semi-analytic model (SAM) of galaxy formation that breaks the mould for models of its kind.
Included among the major changes is an overhauled treatment of stellar feedback that is derived from energy conservation, operates on local scales, affects gas gradually over time rather than instantaneously, and predicts a mass-loading factor for every galaxy.
Building on the model's resolved angular-momentum structure of galaxies, we now consider the heating of stellar discs, delivering predictions for disc structure both radially and vertically.
We add a further dimension to stellar discs by tracking the distribution of stellar ages in each annulus.
Each annulus--age bin has its own velocity dispersion and metallicity evolved in the model. 
This allows \ds~to make structural predictions for galaxies that previously only hydrodynamic simulations could.
We present the model as run on the merger trees of the highest-resolution gravity-only simulation of the MillenniumTNG suite.
Despite its additional complexity relative to other SAMs, \ds~only has three free parameters, the least of any SAM, which we calibrate exclusively against the cosmic star formation history and the \zo~stellar and \HI~mass functions using a particle-swarm optimisation method.
The \ds~codebase, written in C and {\sc python}, is publicly available at \url{https://github.com/arhstevens/DarkSage}.


\begin{multicols}{2}

\section{Context}
\label{sec:intro}

Research into galaxy formation remains among the top priorities in the field of astronomy \citep{astro2020}.
With the likes of the James Webb Space Telescope in full operation and construction of the Square Kilometre Array underway, this is a promising decade for the field.
The endgame of these multi-billion-dollar enterprises is to develop a comprehensive working theory for how the Universe works, of which galaxy formation is a major, crucial part.
But the flood of incoming high-quality observations is only as powerful as the models we use to interpret those data.
Galaxy formation is a highly complex problem to model, which requires us to `solve' it with numerical simulations.
While every model is flawed, simulations are the closest that extragalactic astrophysicists will ever come to conducting a controlled experiment.
Having a variety of galaxy formation models, frameworks, and simulations\,---\,alongside observations\,---\,is paramount to test and develop our theoretical understanding of not just galaxies themselves, but cosmology too.

Two methods for simulating the formation of galaxies in a cosmological context have stood out in recent decades:~semi-analytic models and hydrodynamic simulations. 
Hydrodynamic simulations simultaneously and self-consistently account for gravity, fluid dynamics, and all astrophysical processes deemed relevant to the formation of galaxies.
Semi-analytic models separate the formation of the universe's large-scale structure from that of galaxies by first constructing a universe with only gravity, and subsequently evolving galaxies in the gravitationally bound structures in that universe\,---\,referred to as `haloes'\,---\,with relatively macroscopic descriptions of baryonic astrophysics.
In truth, hydrodynamic simulations are still `semi-analytic' at the level of a particle, which is generally comparable in mass to a giant molecular cloud or globular cluster of stars.
Or, to frame this in reverse, all baryonic processes are `subgrid' at the level of the (sub)halo in a semi-analytic model.
A holistic summary of these methods of galaxy formation modelling can be found in a number of review articles \citep[e.g.][]{somerville15,vogelsberger20}.

In this paper, we present a new version of the semi-analytic model known as `\ds.'
\ds~has been in development since 2015, with two main versions previously released \citep{stevens16,stevens18}.
The model started as a modified version of SAGE \citep[Semi-Analytic Galaxy Evolution;][]{croton16}.
While the code architecture is still based on SAGE, the physical models implemented in \ds~are almost entirely unique.
SAGE was itself built from the \citet{croton06} version of the Munich semi-analytic model\,---\,from which the likes of {\sc Gaea} \citep[][and references therein]{xie20} and {\sc L-galaxies} \citep{henriques20} also stem\,---\,with many versions preceding that too \citep*{kauffmann99,springel01,delucia04}.
While the new \ds~that we present here brings together five years of development since its last release, it builds on over 25 years of research in the community.

Outside of the main model papers \citep{stevens16,stevens18}, \ds~has been used to investigate the effects of environment and feedback on the gas content of galaxies \citep{sb17}, research the angular momentum of galaxies at both high redshifts \citep*{okamura18} and high stellar masses \citep{porras21}, explain the origin of \HI-excess galaxies \citep{lutz18}, confirm the origin of the \HI~size--mass relation \citep{stevens19b}, predict the results of \HI~intensity-mapping experiments \citep{wolz22}, develop an \HI-dependent halo occupation distribution model for cosmology applications \citep{qin22}, and more.%
\footnote{For a manually updated list of papers that have used results from any of the SAGE family of models, see \url{https://ui.adsabs.harvard.edu/public-libraries/JslFrCAlSE2wzNqL5trCwg}.}
The significant updates that we present in this paper will not only expand the range of use cases for \ds, but also improve the predictive power for the use cases it was originally designed for.

We have endeavoured to adopt a new philosophy with how semi-analytic models should be designed with this new version of \ds.
Our main goal is to re-derive the core aspects of galaxy-evolution physics from as near to first principles as possible;
there are fewer \emph{ad hoc} descriptions left, and the ones we do use are updated to modern measurements where possible, often over broader mass and redshift ranges.
This has meant increasing the coupling between physical prescriptions, which results in the elimination of a large number of free parameters.
Simultaneously, we have \emph{increased} the detail and complexity of the galaxies we evolve, building their structure numerically in more than one dimension.
Feedback from stars and active galactic nuclei\,---\,arguably the most important processes in galaxy evolution\,---\,are derived from the law of energy conservation.
While the model is certainly not free of phenomenology, our philosophy has meant we can capture the uncertainty in our model almost completely with only three free parameters (a factor 3--6 lower than comparable models tabulated in Table \ref{tab:pros}), each with well defined boundaries, which we successfully calibrate to observational statistics of galaxies with an objective, automated method.

\ds~is a publicly available, version-controlled code \citep{dsascl}, open for the community to use and modify as they desire.
The aim of this paper is to describe how the model works, and should serve especially well for those using the code itself.
While we present some predictions of the model here, we deliberately do not linger on them.
Science questions and specific results will be better served as the subject of future, focussed papers.


\section{`The Next Generation' \newline N-body simulations}

The results of \ds~in this paper are produced by running the model on the merger trees of the main gravity-only MillenniumTNG $N$-body simulation.
The MillenniumTNG simulations\footnote{\url{https://www.mtng-project.org/}} \citep{barrera23,bose23,contreras23,delgado23,ferlito23,hadzhiyska23a,hadzhiyska23b,hernandez23,kannan23,pakmor23} 
are the latest addition to `The Next Generation' (TNG) simulation suite, which began with IllustrisTNG\footnote{\url{https://www.tng-project.org/}} \citep{weinberger17,marinacci18,naiman18,nelson18,nelson19b,pillepich18a,pillepich18b,springel18}.
Each simulation in the TNG suite has paired $N$-body and magnetohydrodynamic versions with identical initial conditions,
run with the same code [{\sc arepo} \citep{springel10}, although $N$-body runs have also been completed with {\sc gadget-4} \citep{springel21}],
and all carry the same $\Lambda$CDM cosmology:~$\Omega_{\rm m} \! = \! 0.3089$, $\Omega_\Lambda \! = \! 0.6911$, $\Omega_{\rm b} \! = \! 0.0486$, $h \! = \! 0.6774$, $\sigma_8 \! = \! 0.8159$, $n_{\rm s} \! = \! 0.9667$ \citep{planck15}.
We assume this cosmology for all results in this paper (including observational data that we compare to, where we have adjusted published data accordingly).
The MillenniumTNG run used in this paper (hereafter `MTNG,' referred to as `MTNG740-DM-1' in other papers) has a comoving box length of $500\,h^{-1}$\,cMpc (nearly 740\,cMpc) with $4320^3$ particles of mass $1.329 \! \times \! 10^8\, h^{-1}\, {\rm M}_\odot$ ($ = \! 1.962 \! \times \! 10^8\, {\rm M}_\odot$), and was run with {\sc arepo}.

There are many advantages to pairing a semi-analytic model like \ds~with MTNG:
\begin{itemize}
\item Both the volume and resolution are high enough to simultaneously capture a large number of galaxy clusters with small enough galaxies to study environmental effects on galaxy evolution.
\item Simulations with different mass resolutions have been run, allowing for future resolution studies.
\item Objects from the semi-analytic model can be directly compared against the galaxies formed in the hydrodynamical run \citep{pakmor23} in future studies.
\item Galaxy catalogues from at least one other semi-analytic model have already been produced \citep[][another descendant of the Munich model]{barrera23}, allowing for additional future model comparison.
\item With an equivalent volume and eight times the number of particles, MTNG is a natural successor to the original Millennium simulation \citep{millennium}, which many semi-analytic model results in the literature are based on, including previous iterations of \ds~\citep{stevens16,stevens18}.
\item If a galaxy of a given stellar mass being `resolved' in a simulation requires it to have a minimum number of particles, then by virtue of the fact that only a fraction of baryon particles in a hydrodynamic-simulation halo will become star particles, semi-analytic galaxies built in a simulation of fixed volume and dark-matter particle number are, in principle, resolved (and therefore complete) to lower stellar masses than a hydrodynamic simulation with those same specifications.
\end{itemize}

Haloes and their substructure (`subhaloes') are identified in each snapshot of MTNG with {\sc subfind-hbt}, described in section 7 of \citet{springel21}.
A minimum of 20 particles must be connected in a friends-of-friends (FoF) group to form a halo.
When using the terms `halo mass' or `virial mass,' we mean the mass enclosed by a sphere whose mean density is 200 times the critical density of the (simulated) universe (denoted by subscript `200c').  This has a one-to-one relationship with both virial radius and virial velocity:
\begin{equation}
\label{eq:M200c}
M_{\rm 200c} = \frac{100\, H^2(z)\, R_{\rm 200c}^3}{G} = \frac{V_{\rm 200c}^3}{10\, G\, H(z)}\,,
\end{equation}
where $H(z)$ is the redshift-dependent fractional expansion rate of the universe or `Hubble parameter.'

The merger trees for MTNG were originally constructed with the new {\sc gadget-4} tree builder \citep[see section 7.4 of][]{springel21}.
This is similar to the format of {\sc lhalotree}, originally presented in \citet{springel01} and used for the Millennium simulation \citep[see the supplementary material of][]{millennium} and is now in HDF5 format.
\ds~requires input merger trees in the original binary format of {\sc lhalotree}.
We thus converted the MTNG trees back to the old {\sc lhalotree} format for this project.

While the \emph{final} results in this paper are presented with the full MTNG box, all preliminary results and model calibration were performed with a smaller simulation for practical purposes.
`Mini-MTNG,' as we refer to it, had identical parameters, including mass resolution, but only included 540$^3$ particles in a $62.5\,h^{-1}$\,cMpc box (MTNG's equivalent of `Mini-Millennium').


\section{Definitions and design \newline of baryonic reservoirs}
\label{sec:reservoirs}

Baryonic matter (gas, stars, and black holes) in \ds~galaxies and haloes is broken into discrete components.
Indeed, the primary function of a semi-analytic model is to calculate how mass and metals move between these components.
For basic context and to familiarise the reader with the nomenclature used throughout this paper, we briefly describe each component in this section.

\subsection{Gas components}

The gas reservoirs in \ds~galaxies and haloes include:
\begin{itemize}
\item The interstellar medium (ISM) or `cold gas,' which refers to gas in a galaxy's disc.
Each disc is broken into $N_{\rm ann}$ annular segments, each fixed \emph{a priori} in terms of specific angular momentum.
Each segment has fields for its mass, metallicity, and radial boundaries.
The ISM is assumed to be axisymmetric and thin, but its radial structure is built numerically over time in the model.
We use the subscript `cold' to refer to ISM properties in equations, even though this medium includes some ionized gas.
\item The `hot gas' reservoir, which represents the bulk of the circumgalactic medium (CGM); i.e.~the gas in the halo that is distinct from that in the galaxy.
This reservoir is always assumed to be homogenous in terms of its temperature and metallicity, and therefore is described entirely by two fields:~its total mass and total metal content. It is assumed to follow an analytic density profile.  `Hot gas' represents that which is available to be accreted onto the galaxy, subject to cooling (i.e.~be it through the `hot mode' or `cold mode').
\item The `fountain' reservoir, which accounts for gas in the CGM that has recently been reheated out of the ISM and is in the process of mixing with the hot gas.
It is assumed to follow the same density profile and have the same specific energy as the hot-gas reservoir.  We label this reservoir `fount' in equations.
\item The `outflowing' reservoir, abbreviated to `outfl' in equations.  Similar to the fountain reservoir, this represents gas in the CGM that has been reheated due to feedback.
But its specific energy is higher, such that it will escape the halo after an appropriate time.
\item The `ejected' reservoir, denoted by subscript `ejec,' which represents gas that has been expelled from the halo's virial radius due to feedback (after transiting the outflow reservoir).
It is reincorporated into the CGM over time.
\item The `local intergalactic medium' (LIGM), which represents the remaining gas in the halo that is outside the virial radius of all its subhaloes.
This is a single field per halo; in practice, it is associated with the central subhalo.
The LIGM primarily provides a means of tracking metal enrichment in cosmological gas that can eventually be accreted by the central galaxy.
\end{itemize}
Whenever we generically refer to the `CGM' in text and equations, we mean the sum of the hot, outflowing, and fountain reservoirs:
\begin{equation}
m_{\rm CGM} \equiv m_{\rm hot} + m_{\rm fount} + m_{\rm outfl}\,.
\end{equation}
The temperature and density profile of all 3 CGM components is assumed to be the same.


\subsection{Stellar components}
\label{ssec:stellar}

The stellar components in \ds~galaxies and haloes include:
\begin{itemize}
\item The stellar disc, which\,---\,like the ISM\,---\,is broken into a series of annuli.
These annuli are fixed in terms of specific angular momentum in the same way as the ISM, but the stellar disc need not always be coplanar with the ISM.
 We denote fields concerning the stellar disc with the subscript `*,disc.'
\item The instability-driven bulge, which\,---\,as its name suggests\,---\,is built from disc instabilities driving stellar mass into a galaxy's centre.
This component has zero angular momentum by definition, and is therefore supported against gravity entirely by random motions.
\item The merger-driven bulge, which is more akin to a classical galaxy bulge or ellipsoid.
This constitutes stars that enter the bulge as a \emph{direct} result of galaxy mergers.
We calculate a nominal specific angular momentum for the merger-driven bulge.
\item Intrahalo stars, otherwise referred to as the stellar halo or `IHS' for short.
This is built up through tidal stripping and the disruption of satellite galaxies before they have the chance to merge.
By definition, this exclusively tracks stars inside the virial radius of a subhalo.
Satellites can have a non-zero IHS.
\item Local intergalactic stars (LIGS), the stellar analogue for the LIGM.
It is built up similarly to IHS, but accounts exclusively for tidally stripped stellar mass in a halo that falls \emph{outside} the virial radius.
This field is always zero for satellites by construction.
\end{itemize}
One major infrastructural change we have made to the stellar components is that each one is now broken into $N_{\rm age}$ age bins.
This means there are $(N_{\rm ann} + 4) \times N_{\rm age}$ stellar-mass and stellar-metallicity fields for each galaxy (the `$+4$' accounts for the spheroidal components and the LIGS).

The introduction of stellar-age bins serves several purposes that expand \ds's capabilities.
For one, it means it is possible to reconstruct the formation history of any disc segment or bulge component of a galaxy (or the galaxy as a whole) from a single-snapshot output.
This means \ds~can now predict age gradients in stellar discs and even three-dimensional age--metallicity--radius maps of \emph{populations} of galaxies, which can be tested against modern observations (e.g.~with survey data from integral-field spectrographs).
In this sense, \ds~galaxies have `multidimensional' structure, \'{a} la the title of this paper.
Component-wise star formation histories of \ds~galaxies are trivially reconstructed from a single-snapshot output, allowing for easy comparison with the outcome of models applied to the spectral energy distributions of survey galaxies by component \citep*[notably, e.g.,][]{profuse}.
The age bins also allow us to add detail to our treatment of stellar evolution and feedback, where we now model the distribution of supernovae as a function of time since birth for stellar populations (Section \ref{sec:feedback}).

$N_{\rm age}$ is a user-defined integer.
In this work, we take $N_{\rm age} \! = \! 30$, which is high enough to ensure that the delayed-feedback scheme is sufficiently converged with $N_{\rm age}$ (not shown here) and to recover the shape of galaxies' star formation histories, but not so high that it prohibitively slows the code.
The age bracket of each age bin is interpolated from the snapshot times of the underlying $N$-body simulation (see Section \ref{ssec:time}); $N_{\rm age}$ is therefore capped by the number of time-steps in the merger trees.
If $N_{\rm age} \! = \! N_{\rm snap} - 1$, then each age bin has a one-to-one correspondence with snapshot intervals in the mergers trees.
Later, in Fig.~\ref{fig:reSN}, we show the age bins used in this work.

Whenever we refer to `the bulge' without further specificity (or use subscript `*,bulge'), we mean the combination of the instability- and merger-driven bulges.
Whenever we refer to the total stellar content of a galaxy (or shorthand `*'), we mean the combination of the stellar disc and both bulges, but \emph{not} intracluster stars (nor the LIGS).


\subsection{Disc annuli}

We maintain the design of \citet{stevens16} in keeping the annular boundaries for \ds~discs as
\begin{subequations}
\begin{equation}
\label{eq:j_i}
\frac{j_{\rm outer}^{(i)}}{\rm kpc\,km\,s^{-1}} = 1.4^{i-1}\, h^{-1}\,,
\end{equation}
\begin{equation}
j_{\rm inner}^{(i)} = 
\left\{
\begin{array}{l r}
0\,, & i=1\\
j_{\rm outer}^{(i-1)}\,, & i \geq 2\\
\end{array}
\right.\,,
\end{equation}
\begin{equation}
i \in [1, 2, ..., N_{\rm ann}-1, N_{\rm ann}]\,,
\end{equation}
\end{subequations}
with $N_{\rm ann} \! = \! 30$.
Throughout this paper, we use the superscript `$(i)$' to refer specifically to the $i$th annulus.
When no such superscript appears for a disc property, we are referring to that property for the \emph{whole} disc.
For convenience, in parts of this paper, we will also refer to the mean properties of an annulus, e.g.
\begin{equation}
\label{eq:jav}
\langle j^{(i)} \rangle \equiv \frac{j_{\rm inner}^{(i)} + j_{\rm outer}^{(i)}}{2} \,.
\end{equation}

The above applies for both stellar and gas discs.
Whenever we refer to `the disc,' we mean the combination of the ISM and stellar disc. 


\subsection{Black-hole components}

We expand the consideration of where massive black holes can exist in \ds.
Each subhalo now is now granted several black-hole reservoirs:
\begin{itemize}
\item The central black hole (BH) of the galaxy maintains the definition as in previous versions of \ds.
It is what powers AGN feedback.
Each galaxy is effectively assumed to hold a single central BH; when two galaxies merge, their BHs are assumed to merge too.
\item The `intrahalo black holes' reservoir (IHBH) tracks the total number and summed mass of black holes that end up inside the halo but \emph{outside} the galaxy itself.
This reservoir grows through the disruption of satellite galaxies inside $R_{\rm 200c}$ of their parent halo.  This reservoir has no effect on galaxy evolution.
\item The reservoir for `local intergalactic black holes' (LIGBH) similarly tracks the number and summed mass of black holes stripped from satellites outside the virial radius.
Satellite galaxies themselves always have zero LIGBHs by construction.
\end{itemize}


\subsection{Time-steps}
\label{ssec:time}

In the vernacular of this paper, a full `time-step' refers to the interval between snapshots in the merger trees of the $N$-body simulation that \ds~is run on.
MTNG has 264 time-steps, starting at $z \! \simeq \! 29.3$ and finishing at \zo~(the initial conditions were produced at $z\!=\!63$).
The separation between each pair of the first 31 snapshots is $\Delta \log_{10}(a) \! = \! 0.014$, where $a$ is the cosmological expansion factor.
The interval in $\Delta \log_{10}(a)$ is halved for the next 33 snapshots, and halved again for the remainder. 

The movement of mass, metals, angular momentum, et cetera, between reservoirs in \ds~is ultimately governed by a series of coupled differential equations.
These equations are solved numerically and sequentially.
Rather than setting the `unit' of time used in the numerical solution to that of the time-step in the merger trees, 
we break each time-step into three `sub-time-steps' in \ds.

The use of sub-time-steps helps mitigate the fact that we must chose an order in which we apply our astrophysical processes; in principle, most processes should happen simultaneously (though, some have a natural order in which they should be done).
This way, we loop through all physical processes multiple times over a time-step, rather than once.
Each sub-time-step within the same full time-step covers the same interval in cosmic time.
We denote the cosmic-time interval of a sub-time-step as $\Delta t$ throughout this paper.

In earlier versions of \ds, there were 10 sub-time-steps between each time-step.
Those versions were based on the original Millennium simulation, which only had 64 snapshots in its merger trees, versus 265 in MTNG.
The \emph{total} number of sub-time-steps has therefore moderately increased from those versions (640 then versus 792 now).

Rather than explicitly writing out the network of coupled differential equations in \ds~altogether, we present equations in this paper in the form that they are solved within a sub-time-step.
This best reflects how the equations are written in the codebase.


\section{Baryon accretion, gas cooling, \newline and disc formation}
\label{sec:cooling}

When a halo grows in mass\,---\,i.e.~when stepping between snapshots in the underlying merger trees\,---\,mass is added to the CGM (and/or occasionally the IHS; see below) of the central to ensure the baryonic mass fraction of the halo remains in line with the cosmic baryon fraction, $f_{\rm bary}^{\rm cosmic}$, modulo a suppression factor, $f_\gamma^+$, associated with photoionization heating \citep{efstathiou92,gnedin00}.
That is, we enforce
\begin{multline}
\label{eq:fbary}
f_\gamma^+\, f_{\rm bary}^{\rm cosmic}\, M_{\rm 200c} = m_{\rm ejec}^{\rm cen} + \sum_{\rm gal}^{R_{\rm gal}<R_{\rm 200c}} \bigg( m_*^{\rm gal}  + m_{\rm cold}^{\rm gal} \\
+ m_{\rm CGM}^{\rm gal} + m_{\rm IHS}^{\rm gal} + m_{\rm BH}^{\rm gal} + m_{\rm IHBH}^{\rm gal} \bigg)\,,
\end{multline}
where the sum over `gal' includes the central galaxy and all satellites within the virial radius.
There is an important subtlety in how the ejected reservoir of the central is handled in equation (\ref{eq:fbary}).
The entire ejected reservoir is defined to lie \emph{outside} $R_{\rm 200c}$.
It is included in the right-hand side of equation (\ref{eq:fbary}) because that gas \emph{was} inside $R_{\rm 200c}$; this gas is ejected by stellar and AGN feedback (described in Sections \ref{sec:feedback} and \ref{sec:agn}), and there is no physical reason why that gas should be automatically replaced by other baryons (which would happen if it were not in the equation).
In principle, this means that, even in the absence of photoionization heating, the cosmic baryon fraction only provides an \emph{upper limit} on the halo baryon fraction in \ds.
We can straightforwardly calculate the baryon fraction inside the virial radius of a halo as
\begin{equation}
f_{\rm bary}^{\rm 200c} = f_\gamma^+\, f_{\rm bary}^{\rm cosmic} - \frac{m_{\rm ejec}^{\rm cen}}{M_{\rm 200c}}\,.
\end{equation}
The $f_\gamma^+$ factor is adopted from \ds's predecessors \citep{croton06,croton16}, which follow the `filtering mass' fitting function of \citet*[][see their appendix B]{kravtsov04}.
This depends on both halo mass and redshift, and is primarily relevant during the epoch of reionization ($z\!\gtrsim\!6$, for all haloes) and for haloes with $M_{\rm 200c} \! < \! 10^{12}\,{\rm M}_\odot$ (at all epochs).
$f_{\rm bary}^{\rm cosmic}$ is set by the cosmology of the underlying $N$-body simulation ($f_{\rm bary}^{\rm cosmic} \! = \! \Omega_{\rm b} / \Omega_{\rm m} \! = \! 0.1573$ in this work).

In previous versions of the model, when a halo grew, pristine cosmological gas (i.e.~with close to zero metallicity) was added to the hot-gas reservoir.
This still happens as the norm, but with our new framework, preference is first given to mass in the local intergalactic reservoirs to be accreted into the halo.
Should a halo have a non-zero LIGM and/or LIGS,  mass is proportionally and respectively transferred from these to the hot-gas reservoir and the IHS.
Should this be insufficient to raise the halo baryon fraction to $f_\gamma^+\, f_{\rm bary}^{\rm cosmic}$, then pristine cosmological gas is added to the hot gas to top it up.
See Sections \ref{sec:env} and \ref{sec:mergers} for how the LIGM and LIGS are built up in the first place.

Gas must first pass through the CGM before reaching the ISM.
The process for calculating how much mass is transferred from the CGM to the ISM in a given sub-time-step is similar in concept to previous incarnations of the model (which harken back to \citealt{white91}) but is not identical.

Given that we assume hot gas to be spherically symmetric and homogenized, the cooling time for gas at a given radius in the CGM is calculated as
\begin{equation}
\label{eq:tcool}
t_{\rm cool}(R) \equiv \frac{e_{\rm hot}^{\rm therm}}{\dot{e}_{\rm hot}^{\rm rad}(R)} = \frac{3 \, T_{\rm hot} \, \bar{\mu} m_{\rm p} \, k_{\rm B} }{2\, \rho_{\rm hot}(R)\, \Lambda(T_{\rm hot}, Z_{\rm hot})}\,,
\end{equation}
where $e_{\rm hot}^{\rm therm}$ is the specific thermal energy of the CGM, $\dot{e}_{\rm hot}^{\rm rad}(R)$ is the rate of specific energy loss through radiation, $T_{\rm hot}$ is the CGM temperature, $m_{\rm p}$ is the mass of a proton, $\bar{\mu} \! = \! 0.59$ under the assumption that hot gas is fully ionized, $k_{\rm B}$ is the Boltzmann constant, $\rho_{\rm hot}(R)$ is the hot-gas density profile, $\Lambda(T_{\rm hot}, Z_{\rm hot})$ is the tabulated cooling function from \citet{sutherland93}, and $Z_{\rm hot}$ is the hot-gas metallicity.
We maintain the assumption that the CGM temperature is the same as the virial temperature:%
\footnote{In practice, the temperature of the CGM should vary with radius. But cosmological hydrodynamic simulations have shown that it remains within a factor of 2 of $T_{\rm vir}$ out to $R_{\rm 200c}$, even with variation in how feedback is modelled \citep{stevens17}.}%
\begin{equation}
\label{eq:Thot}
T_{\rm hot} = T_{\rm CGM} = T_{\rm vir} \equiv \frac{\bar{\mu} m_{\rm p} \, V_{\rm 200c}^2}{2\, k_{\rm B}}\,.
\end{equation}
This formula for $T_{\rm CGM}$ is defined for centrals; after infall, satellites are assumed to have fixed $T_{\rm CGM}$.
We further maintain the definition of the `cooling radius' as being that where the cooling time equals the halo dynamical time; i.e.
\begin{equation}
\label{eq:tdyn}
t_{\rm cool}(R_{\rm cool}) = t_{\rm dyn} \equiv \frac{R_{\rm 200c}}{V_{\rm 200c}} = \frac{1}{10\,H(z)}\,.
\end{equation}
However, what we \emph{have} changed is the analytic profile for hot gas.

While previous iterations of SAGE and \ds~assumed the CGM to follow the density profile of a singular isothermal sphere, we instead now adopt the so-called `beta' profile.
Nominally, a beta profile follows the expression
\begin{equation}
\rho_{\beta}(R) = \rho_0 \left[1 + \left(\frac{R}{R_c}\right)^2 \right]^{-3\beta/2}\,,
\end{equation}
where $\rho_0 \! \equiv \! \rho_{\beta}(0)$ and $R_c$ is the `core' radius.
For the application of this profile to the {\sc galform} semi-analytic model, modulo changes to the profile from feedback, \citet{benson03} assume that $\beta \! = \! 2/3$ and $R_c$ is a fixed fraction of the virial radius of a halo.
We follow this idea, but instead of allowing for the profile shape to be modified by galaxy evolution processes, we set an explicit redshift dependence on the latter; i.e.~$R_c = c_\beta(z)\,R_{\rm 200c}$.
Finally, with recognition that the CGM is defined in this paper to be \emph{within} $R_{\rm 200c}$ [i.e.~a volume integral of $\rho_{\rm CGM}(R)$ out to $R_{\rm 200c}$ must return $m_{\rm CGM}$], and that each component of the CGM is assumed to follow the same density profile, we implement
\begin{subequations}
\label{eq:rhohot}
\begin{equation}
\rho_{\rm hot}(R) = \frac{m_{\rm hot}}{m_{\rm CGM}}\, \rho_{\rm CGM}(R)\,,
\end{equation}
\begin{equation}
\rho_{\rm CGM}(R) = \frac{m_{\rm CGM}\, \mathcal{C}_\beta(z)}{4 \uppi \, c_\beta^2(z) \, R_{\rm 200c}^3} \left[1 + \left(\frac{R}{c_\beta(z)\,R_{\rm 200c}} \right)^2 \right]^{-1}\,,
\end{equation}
\begin{equation}
\mathcal{C}_\beta(z) \equiv \left\{1 - c_\beta(z)\, \tan^{-1}\!\left[c_\beta^{-1}(z) \right] \right\}^{-1}\,.
\end{equation}
\end{subequations}
For the functional form of $c_\beta(z)$, we use the fitting function of \citet{stevens17} to gas haloes from the EAGLE\footnote{Evolution and Assembly of GaLaxies and their Environment \citep{crain15,schaye15}.} simulations:
\begin{equation}
\label{eq:cbeta}
c_\beta(z) = {\rm max}\left[0.05, ~0.2\,\exp(-1.5\,z) - 0.039\, z + 0.28 \right]\,.
\end{equation}
The lower limit of 0.05 is imposed by hand; this represents the 16th percentile of haloes at the highest redshift ($z\!\simeq\!4$) studied in \citet{stevens17}.
As a point of comparison, in their models, \citet{benson03} and \citet{font08} adopt \emph{default} (but not fixed) $c_\beta$ values of 0.07 and 0.1, respectively. 

For the CGM of satellite galaxies, we still assume a truncation radius.
While it does not make complete physical sense to define $R_{\rm 200c}$ for a satellite subhalo, we adopt $M_{\rm 200c}$ as the smaller of either the present total subhalo mass or its pre-infall virial mass, then solve for an equivalent $R_{\rm 200c}$ through equation (\ref{eq:M200c}).
This ensures that as satellites are tidally stripped, their CGM truncation radius decreases too, which is physically expected.
By contrast, as noted above, $T_{\rm vir}$ is deliberately \emph{not} recomputed after infall; if it were, we would be implicitly assuming that tidal stripping causes a decrease in the average CGM temperature, which seems non-physical.

Combining equations (\ref{eq:tcool}--\ref{eq:cbeta}), one can solve for $R_{\rm cool}$.
Because of the update in the hot-gas density profile, this solution is different from earlier versions of \ds.

Two modes of gas accretion onto a galaxy can take place; the `cold mode' of accretion takes place when $R_{\rm cool} \! > \! R_{\rm 200c}$, otherwise the galaxy is in the `hot mode' \citep{white91}.
In concept, these are unchanged from earlier models.
That is, in the cold mode, gas in the CGM that is en route to the ISM is limited more by the time it takes to fall onto the ISM than any time required to cool.
In this instance, it becomes somewhat of a misnomer to call the CGM `hot gas,' but we nevertheless maintain the nomenclature for consistency with the literature.
For a sub-time-step $\Delta t$, the mass deposited onto the ISM in the cold mode is taken as
\begin{equation}
\label{eq:coldmode}
\Delta m_{\rm hot \rightarrow cold}^\textrm{cold-mode} = m_{\rm hot} \frac{\Delta t}{t_{\rm dyn}} - \Delta m_{\rm heat}^{\rm radio}\,.
\end{equation}
The term $\Delta m_{\rm heat}^{\rm radio}$ represents an offset in accretion onto the ISM from radio-mode AGN feedback (defined and discussed in Section \ref{sec:agn}).
While this term applies for both the cold and hot mode, it is normally only important in practice for the hot mode.

For the hot mode, the flux of mass cooling through $R_{\rm cool}$ is taken as equal to the mass deposition rate onto the galaxy \citep[based on the results of][]{bertschinger89}.
To calculate this, we solve
\begin{multline}
\Delta m_{\rm hot \rightarrow cold}^\textrm{hot-mode} = 4\uppi \, \rho_{\rm hot}(R_{\rm cool})\, R_{\rm cool}^2 \\ 
\times \left(\left. \frac{{\rm d} t_{\rm cool}}{{\rm d} R} \right|_{R \rightarrow R_{\rm cool}} \right)^{-1}\, \Delta t - \Delta m_{\rm heat}^{\rm radio}\,.
\end{multline}
Assuming equation (\ref{eq:rhohot}), the solution can be written as
\begin{equation}
\label{eq:hotmode}
\Delta m_{\rm hot \rightarrow cold}^\textrm{hot-mode} = \frac{m_{\rm hot}\, R_{\rm cool}\, \mathcal{C}_\beta(z)}{2\, R_{\rm 200c}\, t_{\rm dyn}}\, \Delta t - \Delta m_{\rm heat}^{\rm radio}\,.
\end{equation}

With $\Delta m_{\rm hot \rightarrow cold}$ determined, the final thing to consider is the fraction of that mass that each annulus of the gas disc receives.
As each annulus is bound by fixed values of specific angular momentum, this means defining a probability distribution function (PDF) of $j$ for the cooling gas.
We base this on the PDF of $j$ for an exponential disc with a constant rotational velocity; i.e.
\begin{equation}
\label{eq:PDFj}
{\rm PDF}_{\rm cool}(j) = \frac{4\, j}{j_{\rm cool}^2} \exp \left(\frac{-2\, j}{j_{\rm cool}} \right)\,,
\end{equation}
where $\int_0^\infty {\rm PDF}_{\rm cool}(j)\, {\rm d} j = 1$ by definition, and $j_{\rm cool}$ is the mean of the PDF.
Equation (\ref{eq:PDFj}) here is simply a more correct and flexible way of writing equation (6) of \citet{stevens16}.
We emphasise that even though equation (\ref{eq:PDFj}) is the same as an exponential disc with a constant rotational velocity, that does \emph{not} mean it is the only type of disc consistent with this PDF.
That is, for any arbitrary surface density profile $\Sigma(r)$, one can find a rotational velocity profile $v_{\rm circ}(r)$ that gives the same PDF of $j$ as in equation (\ref{eq:PDFj}).
We are therefore not making any explicit assumptions about the form of $\Sigma(r)$ or $v_{\rm circ}(r)$ as gas cools/accretes onto a galaxy; in fact, we solve for these profiles (using additional information per Section \ref{sec:j2r}).
Even \emph{if} one were to make assumptions about the surface density profile or velocity profile of freshly cooled gas, one would still end up with a predictive model for the resulting disc structure after several time-steps, as both the value of $j_{\rm cool}$ and the vector along which accretion takes places change between time-steps.

\begin{figure}[H]
\centering
\includegraphics[width=0.925\textwidth]{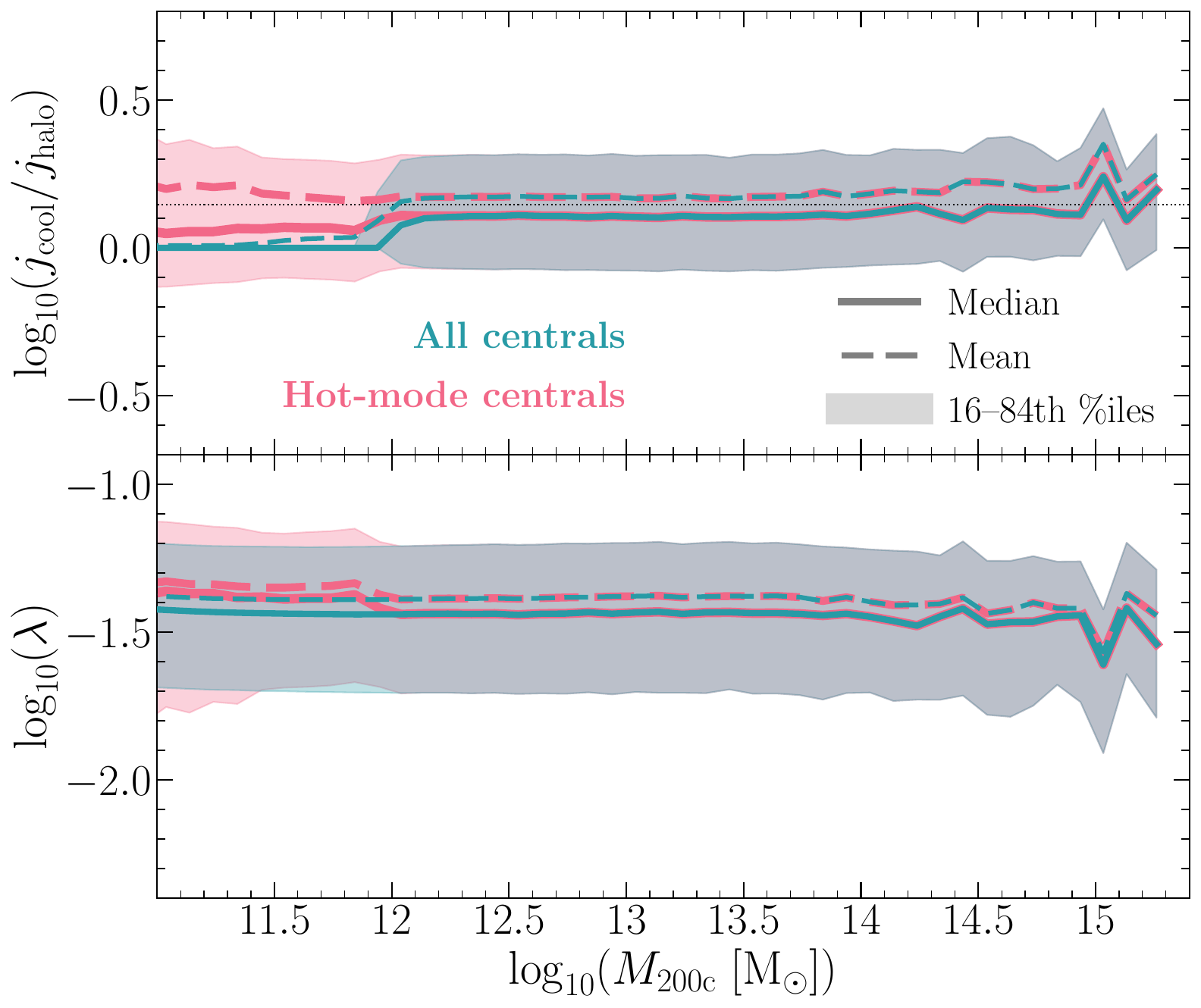}
\vspace{-0.2cm}\caption{{\bf Top panel}: Host halo mass versus the ratio of the specific angular momentum of gas cooling from the CGM onto the ISM of the central galaxy in \ds~to that of the halo from MillenniumTNG at \zo.
For haloes accreting in the `hot mode,' this ratio is determined by equation (\ref{eq:jcool}).
The scatter and mild slope in this relation is driven entirely by the distribution of halo spins, as seen in the {\bf bottom panel}. For cold-mode haloes, it is assumed that $j_{\rm cool} / j_{\rm halo} \! = \! 1$.
Running percentiles and means are shown for the full population (thinner lines in the foreground) and when exclusively selecting those in the hot mode (thicker, behind).
The dotted line in the top panel marks where $j_{\rm cool} / j_{\rm halo} \! = \! 1.4$, the average value predicted for hot-mode haloes by \citet{stevens17}.
Satellite subhaloes are excluded from this plot.}
\label{fig:jcool}
\end{figure}

What then sets $j_{\rm cool}$?  
The answer to this also depends on the accretion mode.
In the cold mode, gas is assumed to be efficiently transported from the halo outskirts to the galaxy via cold streams \citep[as shown to occur in hydrodynamic simulations, e.g.][]{keres05,dekel06,voort11}.
In the absence of evidence to the contrary, we set 
\begin{equation}
j_{\rm cool}^\textrm{cold-mode} \! = \! j_{\rm halo}\,, 
\end{equation}
maintaining the status quo from several semi-analytic models \citep[e.g.][]{shark,henriques20}.  
$j_{\rm halo}$ is the specific angular momentum of the halo inside the virial radius, which we take from the input merger trees.
For the hot mode, we modify the fitting function of \citet{stevens17}\,---\,which was first implemented in \ds~in \citet{stevens18}\,---\,by rewriting it in terms of $j$:
\begin{equation}
\label{eq:jcool}
\log_{10}\!\left( \frac{j_{\rm cool}^\textrm{hot-mode}}{2\, R_{\rm 200c}\, V_{\rm 200c}} \right) = 0.23\, \log_{10}(\lambda) - k_d\,, 
\end{equation}
where we adopt the \citet{bullock01} approximation for halo spin,
\begin{equation}
\lambda = \frac{j_{\rm halo}}{\sqrt{2}\, R_{\rm 200c}\, V_{\rm 200c}}\,.
\end{equation}
While the slope of equation (\ref{eq:jcool}) originates from equation (19) of \citet{stevens17}, the normalisation is set by hand to recover the result of \citet{stevens17} that $\langle j_{\rm cool}^\textrm{hot-mode} / j_{\rm halo} \rangle \! \simeq \! 1.4 \! \simeq \! \sqrt{2}$ when one only considers gas inside the virial radius.
Rearranging the above, one can then solve for $k_d$ to force this outcome.
Doing so gives
\begin{equation}
k_d \simeq \log_{10}\! \left\langle \lambda^{-0.77} \right\rangle\,.
\end{equation}
To match the low-redshift sample used in \citet{stevens17}, we calculate the average of $\lambda^{-0.77}$ for haloes with $10^{11.5} \! \leq \! M_{\rm 200c}/{\rm M}_\odot \! \leq \! 10^{12.5}$ at \zo~in Mini-MTNG.
This returns $k_d \! = \! 1.15$.%
\footnote{In \citet{stevens18}, $k_d$ was set to 1.0.  This was a calculation error.}
Because of the explicit dependence on halo spin in equation (\ref{eq:jcool})\,---\,which is known to have a scatter of $\sim$0.26\,dex and only systematically depend on halo mass to second order in $N$-body simulations \citep{bullock01,knebe08}\,---\,we naturally have a scatter in $j_{\rm cool}^\textrm{hot-mode} / j_{\rm halo}$ in \ds~of $\sim$0.2\,dex.
This is shown as a function of halo mass in Fig.~\ref{fig:jcool}.
As can be seen by comparing the curves and shaded regions in the top panel for all centrals with those exclusively in the hot mode, $M_{\rm 200c} \! \simeq \! 10^{12}\,{\rm M}_\odot$ marks the transition between haloes predominantly in the cold mode and those in the hot mode.
There are, nevertheless, some haloes in the hot mode all the way down to the resolution limit of the simulation.

With the above pieces in place, we can calculate the amount of cooling gas that goes into each annulus of the new-gas disc.
For the $i$th annulus, the mass received is 
\begin{equation}
\label{eq:cool}
\Delta m_{\rm hot \rightarrow cold}^{(i)} = \Delta m_{\rm hot \rightarrow cold} \int_{j_{\rm inner}^{(i)}}^{j_{\rm outer}^{(i)}} {\rm PDF}_{\rm cool}(j)\, {\rm d} j\,.
\end{equation}
If this is the first cooling episode in a halo or $m_{\rm cold}$ was 0 at the start of the sub-time-step, this completes the relevant description of how a gas disc is formed.
If this new gas is adding to a pre-existing disc in the galaxy (as is the case for most galaxies at most snapshots), there is one more step to consider.

While the \emph{magnitudes} of the cooling gas and halo's specific angular momenta are not always assumed to be the same, their \emph{directions} are.
The orientation of a galaxy's disc in \ds~is always tracked by its angular-momentum vector.
When a new gas disc forms, the direction of its angular-momentum vector is set parallel to that of the halo; i.e.~$\hat{J}_{\rm cold} \! = \! \hat{J}_{\rm cool} \! = \! \hat{J}_{\rm halo}$.
However, $\hat{J}_{\rm halo}$ changes over time for any given halo, meaning sequential accretion episodes are unlikely to be parallel.
When a pre-existing gas disc is present, a vector sum of $\vec{J}_{\rm cold}$ and $\vec{J}_{\rm cool}$ is done to define the new orientation of the resulting gas disc.
Both the pre-existing and freshly cooled gas discs are then projected onto this new plane, which uses the same discretised $j$ structure per equation (\ref{eq:j_i}).
This means that the mass in the first annulus of the resulting disc will be the sum of the first annuli of the pre-existing and freshly cooled discs \emph{plus} some fraction of their second (and possibly their third, fourth, etc.) annuli, depending on the angles involved.
This aspect of \ds~remains unchanged from \citet{stevens16}.

It is worth mentioning that $\vec{J}_{\rm halo}$ is typically a noisy quantity in an $N$-body simulation \citep*[e.g.][]{benson17,contreras17}.
While this induces some randomness to the precise orientation and magnitude of the angular momentum of cooling gas in a \ds~at a particular instant, this randomness should be averaged out after several time-steps with our method.

\section{Computing potentials and \newline converting from specific angular momentum to radius}
\label{sec:j2r}

Recall that disc annuli in \ds~galaxies are fixed in terms of specific angular momentum.
While this design choice was deliberate, it nevertheless remains crucial to compute the \emph{radii} of annuli for several of the model's modules.
Annulus radii must be (and are) regularly updated in the model, at least once per sub-time-step per galaxy.  
To convert $j$ to $r$, one must first compute the potential of the galaxy and its (sub)halo in the plane of the disc, $\Phi(r)$.
In \citet{stevens16}, this was achieved by approximating the potential as spherically symmetric and the disc to have precisely centripetal motion at all radii, then iteratively solving for $r_{\rm outer}^{(i)}$ for each annulus through
\begin{equation}
\label{eq:j2r}
j_{\rm outer}^{(i)} = r_{\rm outer}^{(i)}\,v_{\rm circ}( r_{\rm outer}^{(i)}) \,.
\end{equation}
\citet{stevens18} later dropped the assumption of perfect centripetal motion by adding a radially variant factor for the fraction of gravity balanced by circular motion (as opposed to random radial motion), $f_{\rm circ}(r)$:  
\begin{equation}
\label{eq:vcirc}
v_{\rm circ}^2(r) = f_{\rm circ}(r)\,v_{\rm cent}^2(r) = f_{\rm circ}(r)\, r\, \frac{{\rm d}\Phi}{{\rm d}r} \,,
\end{equation}
where subscripts `circ' and `cent' are respective shorthands for `circular' and `centripetal.'
As noted in \citet{stevens18}, the spherical-potential approximation is accurate in the outskirts of a galaxy, but becomes exponentially inaccurate towards the centre.
Moreover, solving equation (\ref{eq:j2r}) iteratively was a bottleneck in the code.
For these reasons we have overhauled the $j$-to-$r$ conversion.

With the new version of the model, we treat the shape of the potential as a combination of spherically symmetric components (the dark-matter halo, stellar bulges, intracluster stars, and hot gas; shorthand `sph') and two axisymmetric, infinitesimally thin discs (one each for stars and cold gas).
This can be written in terms of contributions to the centripetal velocity:
\begin{equation}
\label{eq:vcent}
v_{\rm cent}^2(r) = v_{\rm sph}^2(r) + v_{\rm *,disc}^2(r) + v_{\rm gas,disc}^2(r) \,.
\end{equation}
The challenge comes in the implementation of the two lattermost terms.
While one can numerically solve for a disc's potential as a function of radius from first principles, the only way to self-consistently do this in the framework of \ds~is by maintaining the iterative $j$-to-$r$ process \emph{and} increasing the computational demand of each iteration.
For practical reasons, we therefore employ an approximation of a different kind, but one that remains faithful to the general shape of a disc potential.
Namely, we assume the analytic approximation of \citet[][their equation 37]{ob09}:
\begin{multline}
\label{eq:vdisc}
v_{\rm disc}^2(r) = \frac{G\, m_{\rm disc}}{r_{\rm s}} \biggl[ 1 + 4.8 \exp \left(-0.35\,\frac{r}{r_{\rm s}} - 3.5\,\frac{r_{\rm s}}{r} \right) \biggl]\\
\times \biggl[ \frac{r}{r_{\rm s}} + \left(\frac{r_{\rm s}}{r}\right)^2 + 2\sqrt{\frac{r_{\rm s}}{r}}\biggl]^{-1}\,.
\end{multline}
This approximation works well for exponential discs with scale radius $r_{\rm s}$.
Clearly, it is less accurate for discs with non-parametric disc structure as in \ds.
Nevertheless, after testing, we found this to be the best compromise between accuracy and practicality.

Because there is no requirement for \ds~discs to be precisely exponential, we need to calculate an analogue for $r_{\rm s}$ to use in equation (\ref{eq:vdisc}).
For exponential discs, there is a one-to-one relation between the scale radius and any radius that encloses $x$ per cent of the disc's mass, $r_x$.
For each of $x \! = \! 10, 20, ..., 80, 90$, we calculate the effective exponential scale radius that corresponds to the disc's $r_x$.
We then set $r_{\rm s}$ to be the average of these nine scale radii.
This is a small shift from \citet{stevens18}, who only used $r_{50}$ and $r_{90}$ to inform $r_{\rm s}$, which relied on an otherwise unnecessary and now non-existent free parameter.
The calculation of $r_{\rm s}$ and subsequent application of equation (\ref{eq:vdisc}) are \emph{separately} done for \emph{each} of the stellar and gas discs, fulfilling the right-hand side of equation (\ref{eq:vcent}).

The contribution to the potential from all other baryons and dark matter is then straightforwardly calculated as:
\begin{subequations}
\begin{equation}
v_{\rm sph}^2(r) = \frac{G\,m_{\rm sph}(<\!r)}{r}\,,
\end{equation}
\begin{multline}
m_{\rm sph}(<\!r) = 4 \uppi \int_0^r \biggl(\rho_{\rm DM}(R) + \rho_{\rm CGM}(R) \\ 
+  \sum_{\otimes} \rho_{\otimes}(R) \biggl) R^2\, {\rm d}R\,, 
\end{multline}
\end{subequations}
where $\otimes \! = \! \ib, \mb, {\rm IHS}$.
Each of the profiles that contributes to $M_{\rm sph}(<\!r)$ is analytic. 
Per \citet[][appendix B]{stevens16}, dark matter is assumed to follow an NFW profile \citep*{nfw96} with a baryon-influenced mass--concentration relation \citep{dicintio14,dutton14}:
\begin{subequations}
\begin{multline}
\rho_{\rm DM}(R) =  \frac{M_{\rm DM}^\prime\, c^2_{\rm NFW}(z)}{4 \uppi\, R_{\rm 200c}^2\, R}   \biggl[1 + \frac{c_{\rm NFW}(z)\, R}{R_{\rm 200c}} \biggl]^{-2} \\
\times \biggl[ \ln\bigl(1+c_{\rm NFW}(z)\bigl) - \frac{c_{\rm NFW}(z)}{1 + c_{\rm NFW}(z)} \biggl]^{-1}\,,
\end{multline}
\begin{multline}
c_{\rm NFW}(z) = \Biggl(1 ~ + 3 \times 10^{-5}\, \exp\biggl\{ \\
3.4 \left[ \log_{10}\left(\frac{m_*}{M_{\rm 200c}}\right) + 4.5 \right] \biggl\} \Biggl)\, c_{\rm DMO}(z)\,,
\end{multline}
\begin{equation}
c_{\rm DMO}(z) = 10^{a_{\rm NFW}(z)} + \left(\frac{M_{\rm 200c}\, h}{10^{12}\, \mathrm{M}_{\odot}}\right)^{b_{\rm NFW}(z)}~,
\end{equation}
\begin{equation}
a_{\rm NFW}(z) = 0.520 + 0.385\, \exp\left(-0.617\, z^{1.21}\right)~,
\end{equation}
\begin{equation}
b_{\rm NFW}(z) = -1.01 + 0.026\, z~.
\end{equation}
\end{subequations}
In principle, the dark-matter mass of the halo should be readily calculated as $M_{\rm DM} \! = \! (1 - f_{\rm bary}^{\rm 200c}) M_{\rm 200c}$ (equation \ref{eq:fbary}).
But, again for the sake of computation efficiency, we approximate the matter bound in subhaloes to also be distributed according to the same NFW profile as the dark matter (despite knowing that this is not the most accurate description of how satellites are distributed in \ds\,---\,see \citealt{qin22}).
We therefore effect
\begin{multline}
M_{\rm DM}^\prime \equiv M_{\rm 200c} - \big( m_* + m_{\rm cold}\\
 + m_{\rm CGM} + m_{\rm IHS} + m_{\rm BH} + m_{\rm IHBH} \big)\,.
\end{multline}

Stellar spheroids (the instability-driven bulge, merger-driven bulge, and intrahalo stars) all follow \citet{hernquist90} profiles that are truncated at the virial radius:
\begin{equation}
\label{eq:hernquist}
\rho_{\otimes}(R) = \frac{m_{\otimes}\, a_{\otimes}\, (R_{\rm 200c} + a_{\otimes})^2}{2\uppi\, R\, (R + a_{\otimes})^3\, R_{\rm 200c}^2}\,.
\end{equation}
While we have not changed the functional form of the stellar-spheroid profiles, we now \emph{calculate} each of their scale radii, $a_{\otimes}$, rather than assuming a scaling relation as done previously.
These are discussed in the appropriate sections for each component later in the paper (see equations \ref{eq:a_ib} \& \ref{eq:a_IHS}).

For the final piece of equation (\ref{eq:vcirc}), we assume a function of the form
\begin{equation}
\label{eq:fcirc}
f_{\rm circ}(r) = 1 - \exp \left(-r / r_{\lambda *} \right)\,.
\end{equation}
In \citet{stevens18}, the scale radius $r_{\lambda *}$ took the form $r_{\lambda *} = r_{\rm s,*} / \left\{ 3 \left[1 - m_{\rm *,bulge}/\left(m_* + m_{\rm cold} \right) \right] \right\}$,%
\footnote{Note the correction from equation (A5) of \citet{stevens18}, where $m_{\rm cold}$ was accidentally missed from this expression.}
based loosely on the observational results of \citet{bellstedt17}.
Instead, we now calculate $r_{\lambda *}$ in a more self-consistent way, but maintain the assumption from \citet{stevens18} that $f_{\rm circ}$ should scale with the local stellar spin parameter,
which is defined for an annulus as
\begin{equation}
\lambda_*^{(i)} \equiv \frac{ \langle v_{\rm circ}^{(i)} \rangle }{ \sqrt{\langle v_{\rm circ}^{(i)} \rangle^2 + \sigma_*^{(i)\,2}} }\,.
\end{equation}
First, we initialise $r_{\lambda *}$ as $ r_{\rm s,*} / 3$.
Then, for every subsequent sub-time-step, we update $r_{\lambda *}$ to
\begin{equation}
r_{\lambda *} = \frac{- r\{\lambda_* \! = \! 0.5\}}{\ln(0.5)}\,,
\end{equation}
where $r\{\lambda_* \! = \! 0.5\}$ is smallest radius where $\lambda_*$ rises to 0.5 (interpolated from the annular $\lambda_*^{(i)}$ values).
The factor of $-1/\ln(0.5) \! \simeq \! 1.44$ translates the factor-of-two scale radius into an exponential scale radius appropriate for equation (\ref{eq:fcirc}).

With all of the above, we can analytically calculate $v_{\rm circ}(r)$\,---\,and therefore $j$\,---\,for any $r$.
To go in the opposite direction, i.e.~from $j$ to $r$ for annuli, we calculate $j(r)$ for an array of $r$ values.
Using this array, we interpolate the $j$ boundaries of the disc annuli to find their corresponding $r$.

We approximate the surface density of a given disc component to be constant \emph{within} an annulus; i.e.
\begin{equation}
\Sigma^{(i)}_{\rm disc} \equiv \frac{m^{(i)}_{\rm disc}}{\uppi \left( r_{\rm outer}^{(i)\,2}  - r_{\rm inner}^{(i)\,2} \right)}\,.
\end{equation}
With this, it follows that the mass in an annulus has an \emph{average} radius of
\begin{equation}
\langle r^{(i)} \rangle \equiv \sqrt{ \frac{ r_{\rm inner}^{(i)\,2}  + r_{\rm outer}^{(i)\,2} }{2} } \,.
\end{equation}
To be consistent with equation (\ref{eq:jav}), we define the average circular velocity of an annulus as
\begin{equation}
\langle v_{\rm circ}^{(i)} \rangle \equiv \frac{\langle j^{(i)} \rangle}{\langle r^{(i)} \rangle} \,.
\end{equation}
Finally, for simplicity and convenience, we define an approximate average potential of each annulus as
\begin{equation}
\label{eq:potcold}
\langle \Phi^{(i)} \rangle \equiv \frac{\Phi(r_{\rm inner}^{(i)}) + \Phi(r_{\rm outer}^{(i)})}{2} \,.
\end{equation}

\begin{figure}[H]
\centering
\includegraphics[width=0.9\textwidth]{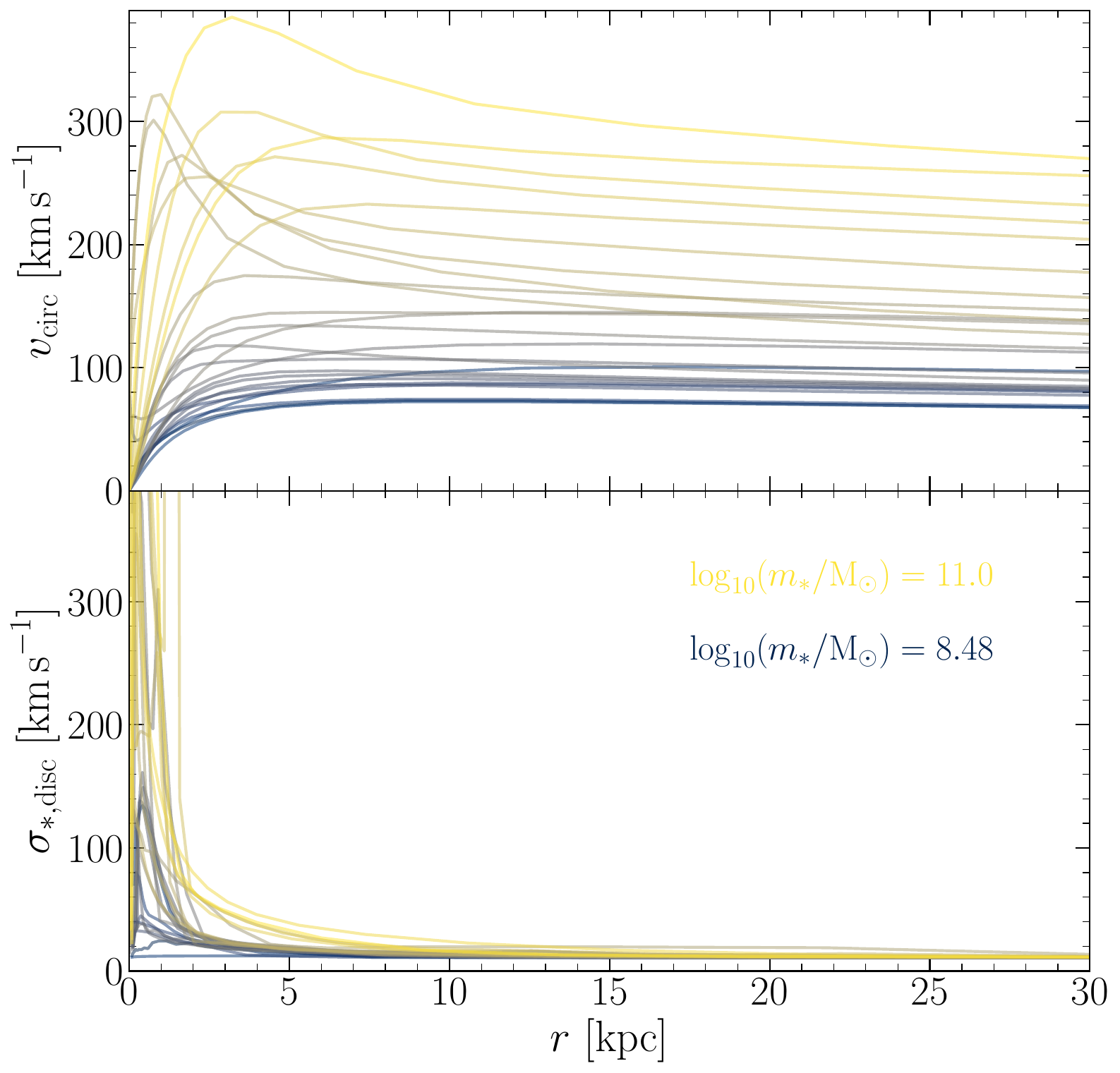}
\vspace{-0.3cm}\caption{
Circular velocity profiles (top) and stellar velocity dispersion profiles (bottom) of \ds~discs at \zo.
We sample one random central galaxy with ${\rm sSFR} \! = \! 10^{-10.5}\,{\rm yr}^{-1}$ and $m_{\rm *,bulge} \! < \! 0.3\,m_*$ for each stellar mass in intervals of 0.1\,dex between $10^{8.5}$ and $10^{11}\,{\rm M}_\odot$ with a 0.04-dex search window.
The colour of the curves transitions from dark blue to bright yellow as we move up in mass.
} 
\label{fig:kinematics}
\end{figure}

To demonstrate the resolved kinematics of galaxy discs in \ds, we plot circular velocity profiles and stellar velocity dispersion profiles for random example galaxies across a range of stellar masses in Fig.~\ref{fig:kinematics}.
A description of how $\sigma_{\rm *,disc}$ builds up over time can be found in Section \ref{ssec:star_instab}.

One relatively fundamental scaling relation of galaxies is the baryonic Tully--Fisher relation \citep{mcgaugh00}, a tighter relation than in its original stellar form of \citet{tully77}.
This relates the peak rotational velocity of a galaxy (a proxy for its potential) with its baryonic mass (in principle, another proxy for potential, if baryonic mass and dark-matter mass scale) for rotation-supported (disc-dominated) systems.
Given that we have built the velocity curves of \ds~galaxies from their potential and the stellar and \HI~masses of galaxies are constrained by their observed mass functions at \zo~(Section \ref{sec:cali}), the baryonic

\begin{figure}[H]
\centering
\includegraphics[width=0.5\textwidth]{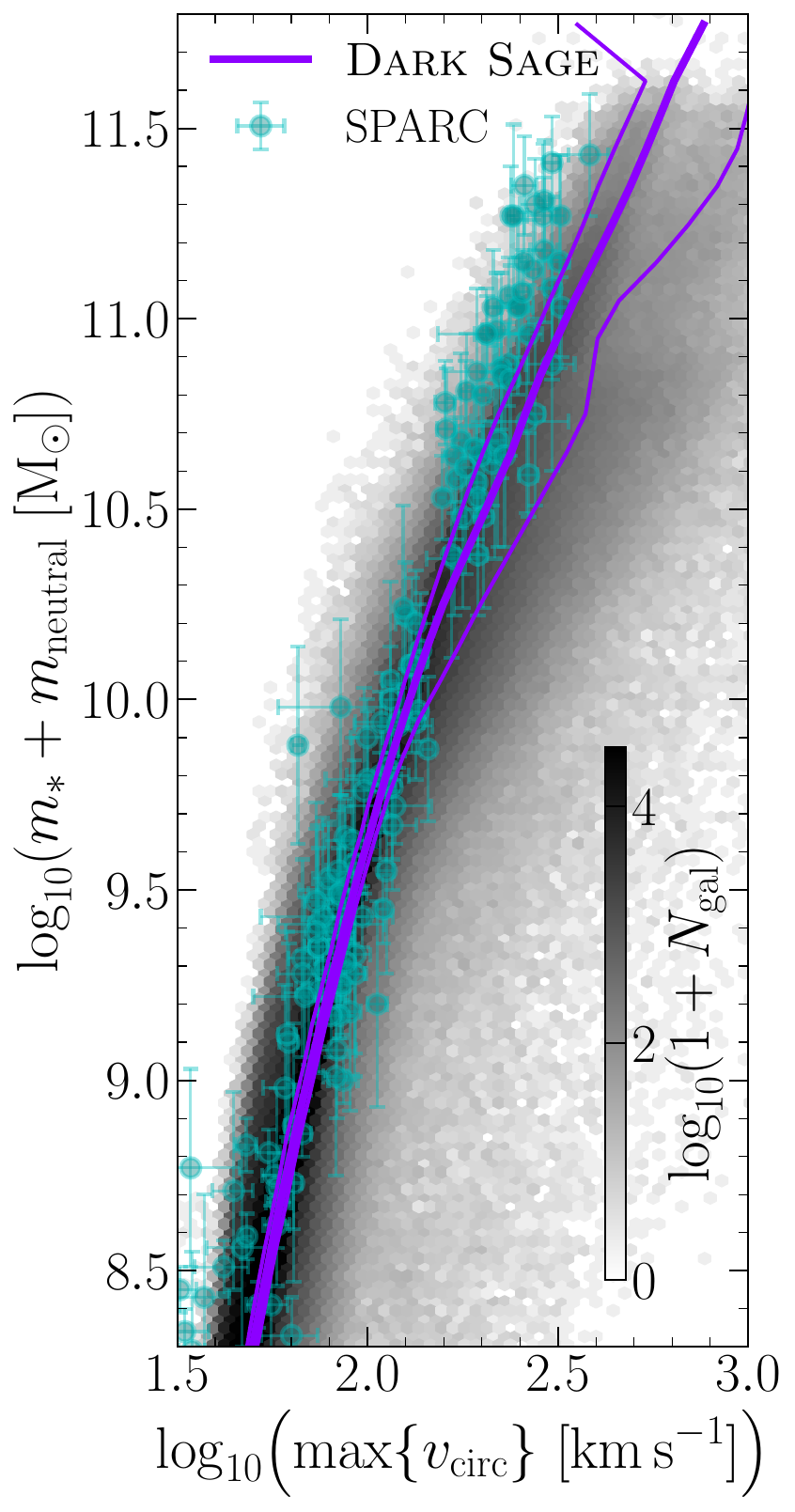}
\vspace{-0.3cm}\caption{
The Baryonic Tully--Fisher relation for \ds~galaxies:~that is, the maximum rotational velocity of each galaxy as a function of its stellar + neutral-gas mass.
Hex bins show the number density of \ds~galaxies with $m_{\rm *,bulge} \! < \! 0.1\,m_*$, 
$m_{\rm H\,{\LARGE{\textsc i}}} \! \geq \! 10^8\,{\rm M}_\odot$, 
and $m_{\rm neutral}/m_* \! \geq \! 10^{-2}$.
The running solid lines are the median (thick) and 16th and 84th percentiles (thin) of the \ds~selection, calculated in bins along the $y$-axis.
The \ds~cuts are a crude attempt at a comparable selection to the SPARC sample of galaxies \citep{lelli19}, compared here.
} 
\label{fig:btf}
\end{figure}

\noindent Tully--Fisher relation should fall out of \ds~as a natural prediction.
As we show in Fig.~\ref{fig:btf} by comparing to data from the SPARC\footnote{\emph{Spitzer} Photometry and Accurate Rotation Curves} sample of galaxies \citep{lelli19}, it does!
Though, we caution that our sample selection to match the SPARC is crude, intended only as a proof of concept here.
The galaxies masses of the SPARC sample also have increasingly larger systematic uncertainties due to the adoption of a morphology-independent mass-to-light ratio \citep[see][]{lelli16,lelli19}.
For these reasons, we should not expect a perfect match.
For example, the scatter at low masses for \ds~of $\sim$0.02\,dex appears low compared to the intrinsic scatter of $0.040 \! \pm \! 0.006$\,dex in SPARC \citep[cf.][]{lelli19}, but we found in testing other cuts that the scatter in \ds~varied up to 0.03\,dex.


\section{Gas chemistry and \newline vertical structure of discs}

\subsection{Velocity dispersion and disc height}

Previously in \ds, it was assumed that the velocity dispersion of gas in galaxy discs, $\sigma_{\rm cold}$, was a universal constant for all cosmic time.
We now introduce a redshift dependence to $\sigma_{\rm cold}$, but maintain the (false but simple) assumption that for a given snapshot this is constant across all annuli of all galaxies.
The function we employ for this is empirical, based on the average redshift evolution for observations of neutral gas shown in fig.~6 of \citet{ubler19}:
\begin{equation}
\label{eq:sigma_cold}
\frac{\sigma_{\rm cold}(z)}{\rm km\,s^{-1}} = 11.0 + 11.3\,z\,.
\end{equation}
Our redshift-zero value of $11\,{\rm km\, s}^{-1}$ is what had been previously assumed at all redshifts.
The redshift coefficient is taken from equation (1) of \citet{ubler19}.
Results from \citet{pillepich19} suggest it might be sensible to add a stellar-mass dependence to equation (\ref{eq:sigma_cold}), but we leave such modifications for future model developments.

We assume that the vertical structure of discs follows a square-hypersecant profile, typical for those that are self-gravitating and isothermal \citep[see][]{benitez18}. 
For a given annulus, we invoke
\begin{subequations}
\begin{equation}
\rho_{\rm cold}^{(i)}(\zeta) = \frac{\Sigma_{\rm cold}^{(i)}}{2\, \zeta_s^{(i)}}\, {\rm sech}^2 \! \left(\frac{\zeta}{\zeta_s^{(i)}} \right) \,,
\end{equation}
\begin{equation}
\zeta_s^{(i)} = \frac{\sigma_{\rm cold}^2(z)}{G\, \Sigma_{\rm cold}^{(i)}}\,.
\end{equation}
\end{subequations}
Here, $\zeta$ represents height from the mid-plane of the disc, where $\zeta_s^{(i)}$ is the local scale height.
By definition, the surface density of an annulus must be $\Sigma_{\rm cold}^{(i)} \! = \! \int_{-\infty}^{\infty} \rho_{\rm cold}^{(i)}(\zeta)\, {\rm d}\zeta$.
The average three-dimensional density of an annulus is therefore
\begin{equation}
\label{eq:rhoav}
\langle \rho_{\rm cold}^{(i)} \rangle = \frac{1}{\Sigma_{\rm cold}^{(i)}} \int_{-\infty}^{\infty}  \rho_{\rm cold}^{(i)\,2} \! (\zeta)\, {\rm d}\zeta = \frac{G}{3} \left( \frac{\Sigma_{\rm cold}^{(i)}}{\sigma_{\rm cold}(z)} \right)^2 \,.
\end{equation}

We do not differentiate between the dispersions (or scale heights) of ionized and neutral gas in \ds~gas discs, despite the theoretical expectation and empirical evidence that suggest these should differ \citep{ubler19}.
Relatively speaking, this is not a great source of uncertainty in the model; the observed systematic difference between the neutral and ionized gas velocity dispersions of galaxies is much less than e.g. the observed scatter in either dispersion at fixed redshift, which we do not model.

When stars are formed (see below), they inherit the velocity dispersion of the gas that formed them.
We record a stellar velocity dispersion for each stellar-age bin within each annulus.
By construction, this means that old stars carry a higher velocity dispersion than young stars in galaxy discs.
As the stellar-age bins are typically wider than the average snapshot interval in the merger trees, each episode of star formation requires updating the velocity dispersion of that age bin.
This is done by taking a mass-weighted average of the squares of the velocity dispersions of the new and pre-existing stars.


\subsection{Hydrogen fraction}

In another update to the model, we adjust the mass fraction of hydrogen, $X$, in a given gas disc annulus.
Naturally, where metallicities are high, one expects to find a lower hydrogen fraction.
We adopt the fitting function of \citet[][their equation B2]{stevens21}, which accurately recovers the hydrogen fraction of gas cells in IllustrisTNG from metallicity alone (i.e.~without needing prior knowledge of the helium mass fraction):
\begin{equation}
X = 
\left\{
\begin{array}{l r}
0.753  - 1.26\, Z \,, & Z \geq 0.025\\
{\rm min}\!\left[0.76,~0.762 - 2.3\, Z + 24.2\, Z^2 \right] \,, & Z < 0.025\\
\end{array}
\right.\,.
\end{equation}
The helium mass fraction is then straightforwardly recovered as 
\begin{equation}
Y = 1 - X - Z \,.
\end{equation}

We impose a metallicity floor of $Z \! = \! 10^{-10}$ on all baryon reservoirs, which accounts for the trace amounts of lithium formed shortly after the Big Bang.
Due to the `primordial lithium problem'\,---\,see \citet{fields11}\,---\,this is lower than what one would expect the initial metallicity of the Universe to be based on Big Bang nucleosynthesis alone. 
Metallicity is then built up from stellar enrichment (see below).
In practice, the precise value of the metallicity floor is irrelevant, provided it is negligible relative to the yield of metals produced by a stellar population, which is of order $10^{-2}$.


\subsection{Neutral fraction}

We introduce a new method to \ds~for calculating the fraction of the ISM that is neutral/ionized.
Previously, this had been assumed to be a constant for all annuli and all galaxies.
We now instead invoke a model of photoionization equilibrium within each annulus, providing a predictive framework for the neutral-fraction radial profiles of galaxies.
This is done in two stages, considering the situations when (i) local sources of ionizing radiation are dominant and (ii) the universal ionizing background is instead dominant.

In what follows, we seek a solution for
\begin{equation}
f_{\rm neutral}^{(i)} \equiv \frac{ m^{(i)}_{\rm neutral} }{ m^{(i)}_{\rm cold} } = \frac{ m_{\rm H\,{\LARGE{\textsc i}}}^{(i)} + m_{\rm H_2}^{(i)} }{ X^{(i)}\, m_{\rm cold}^{(i)} }\,.
\end{equation}
In practice, we calculate this as
\begin{equation}
f_{\rm neutral}^{(i)} = {\rm min} \left[ f_{\rm neutral}^{\star(i)},~f_{\rm neutral}^{{\rm bg}(i)}  \right]\,.
\end{equation}

\subsubsection{Local photoionization}

Let us start with the approximation demonstrated by \citet{rahmati13b} that the local photoionization rate of hydrogen in a galaxy scales linearly with its star formation rate.
Writing this in terms of ionization and star formation rate \emph{densities}, we have:
\begin{equation}
\label{eq:ionization}
\dot{n}_{\rm ion} \simeq q_\gamma\, \rho_{\rm SFR} \,,
\end{equation}
where $q_\gamma \! = \! 2 \! \times \! 10^{53}\, {\rm s^{-1}\, (M_\odot\, yr^{-1})^{-1}} \! = \! 6.31 \! \times \! 10^{60}\, {\rm M}_\odot^{-1}$ \citep{rahmati13b}.
The question then becomes:~how do we implement this on an annulus-by-annulus basis?  

Similar to equation (\ref{eq:rhoav}), we can calculate the average three-dimensional SFR density in terms of an annulus's SFR surface density, given $\rho_{\rm SFR}^{(i)}\! (\zeta) \propto \rho_{\rm cold}^{(i)} \! (\zeta)$:
\begin{equation}
\langle \rho_{\rm SFR}^{(i)} \rangle = \frac{1}{\Sigma_{\rm cold}^{(i)}} \int_{-\infty}^{\infty}  \rho_{\rm cold}^{(i)}\! (\zeta)\, \rho_{\rm SFR}^{(i)} \! (\zeta)\, {\rm d}\zeta\,.
\end{equation}
Approximating $\rho_{\rm SFR}$ in equation (\ref{eq:ionization}) as $\langle \rho_{\rm SFR}^{(i)} \rangle$ above and solving the integral, we take
\begin{equation}
\label{eq:nion}
\dot{n}_{\rm ion}^{\rm (i)} = \frac{q_{\gamma}\, G\, \Sigma_{\rm cold}^{(i)}\, \Sigma_{\rm SFR}^{(i)}}{3\, \sigma_{\rm cold}^2(z)}\,.
\end{equation}

To solve for $f_{\rm neutral}^{(i)}$, we need to balance ionization with the recombination of atoms.
The recombination rate density of hydrogen can be written as
\begin{equation}
\dot{n}_{\rm recom} = n_{\rm e}\, n_{\rm p}\, \alpha\,,
\end{equation}
where $n_{\rm e}$ and $n_{\rm p}$ are the respective number densities of free electrons and free protons, and $\alpha \! = \! 4 \! \times 10^{-13}\,{\rm cm^3\,s^{-1}}$ is the recombination coefficient for hydrogen that we adopt.
For an annulus, we consider \emph{average} three-dimensional number densities, consistent with our earlier assumptions that there is no vertical variation in the neutral fraction (or ionization--recombination balance):
\begin{subequations}
\label{eq:recombination}
\begin{equation}
\dot{n}_{\rm recom}^{(i)} = \langle n_{\rm e}^{(i)} \rangle \, \langle n_{\rm p}^{(i)} \rangle\, \alpha\,,
\end{equation}
\begin{equation}
\langle n_{\rm e}^{(i)} \rangle = \frac{4}{3\,X^{(i)} + 1} \langle n_{\rm p}^{(i)} \rangle\,,
\end{equation}
\begin{align}
\langle n_{\rm p}^{(i)} \rangle &= \left( 1 - f_{\rm neutral}^{(i)} \right)\, X^{(i)}\, \frac{\langle \rho_{\rm cold}^{(i)} \rangle}{m_{\rm p}}\\
 &= \left( 1 - f_{\rm neutral}^{(i)} \right)\, X^{(i)}\, \frac{G}{3\,m_{\rm p}} \left(\frac{\Sigma_{\rm cold}^{(i)}}{\sigma_{\rm cold}(z)} \right)^2 \,.
\end{align}
\end{subequations}
The difference between $\langle n_{\rm p}^{(i)} \rangle$ and $\langle n_{\rm e}^{(i)} \rangle$ is driven by extra free electrons sourced from ionized helium.
For simplicity, we approximate that for every hydrogen atom that is split into a free electron and proton, there is a helium atom split into a free electron and He\,{\sc ii} ion.
In reality, it takes more energy to ionize a helium atom than a hydrogen atom, and the cross-section of the two elements are not identical either,
but these minutiae are unimportant for the detail modelled here.
We also reasonably neglect any contributed free elections from He\,{\sc iii} or ions of other elements, given their low cosmic abundances.

Under photoionization equilibrium,
\begin{equation}
\label{eq:eqm}
\dot{n}_{\rm ion}^{(i)} = \dot{n}_{\rm recom}^{(i)}\,.
\end{equation}
Equation (\ref{eq:recombination}) serves as the right-hand side of equation (\ref{eq:eqm}).
If we take equation (\ref{eq:nion}) as the left-hand side, we are implicitly treating $f_{\rm neutral}^{(i)}$ as $f_{\rm neutral}^{\star (i)}$.
After combining these equations, we can rearrange and solve for the neutral fraction (when local photoionization dominates) in terms of knowns:
\begin{equation}
\label{eq:fnl}
f_{\rm neutral}^{\star (i)} = 1 - \frac{1}{X^{(i)}} \sqrt{ \frac{ 3 \left(3\, X^{(i)} + 1\right) q_\gamma\, m_{\rm p}^2\, \sigma_{\rm cold}^2(z)\, \Sigma_{\rm SFR}^{(i)} }{ 4\, \alpha\, G\, \Sigma_{\rm cold}^{(i)~3} } }\,.
\end{equation}
For convenience, we note that the constants in this equation can be reduced to
\begin{equation}
\frac{3\, q_\gamma\, m_{\rm p}^2}{4\, \alpha\, G} = 0.2495\,{\rm cm^{-6}\,g^2\,s^3}\,.
\end{equation}

One final decision remains:~how do we quantify $\Sigma_{\rm SFR}^{(i)}$ for equation (\ref{eq:fnl})?
While we could tie this to the sub-time-step's instantaneous SFR in the annulus, this would require a simultaneous solution for $\Sigma_{\rm SFR}^{(i)}$ and $f_{\rm neutral}^{(i)}$, because, as we will show in Section \ref{ssec:sf}, local star formation explicitly depends on local molecular fraction, which requires knowing the neutral fraction first.
But a better option exists.
What equation (\ref{eq:ionization}) really symbolises is that \emph{young} stars produce the majority of local photoionizing radiation in galaxies.
The age bins we have in each \ds~disc annulus effectively tell us how many young stars are present.
That is, to be self-consistent with our treatment of stellar evolution (where any stars formed in the same age bin are treated as if they have an identical formation time\,---\,see the next section), we use the annulus's average SFR \emph{for the age bin that the present sub-time-step falls in} to calculate $\Sigma_{\rm SFR}^{(i)}$ for equation (\ref{eq:fnl}).
This is both physically motivated and circumvents a circular solution for $f_{\rm neutral}^{\star (i)}$.

We note that AGN should, in principle, add an additional source of local photoionization.
This is not something we have considered for this version of \ds.
We leave it for future developments of the model.


\subsubsection{Photoionization background}

We adopt a redshift-dependent, universal photoionizing radiation background for neutral hydrogen to place a ceiling on the neutral fraction of the ISM of \ds~galaxies.
When this source of photoionization dominates over local sources, we impose
\begin{equation}
\dot{n}_{\rm ion} = \Gamma_{\rm H\,{\LARGE{\textsc i}}}^{\rm eff}(z)\, n_{\rm H, neutral}\,.
\end{equation}
Here, the effective rate of hydrogen photoionization, $\Gamma_{\rm H\,{\LARGE{\textsc i}}}^{\rm eff}(z)$, is taken from table D1 of \citet{fg20}.
In practice, for \ds~disc annuli, we effect
\begin{equation}
\dot{n}_{\rm ion}^{(i)} = \Gamma_{\rm H\,{\LARGE{\textsc i}}}^{\rm eff}(z)\, f_{\rm neutral}^{{\rm bg}(i)}\, X^{(i)}\, m_{\rm p}^{-1}\, \langle \rho_{\rm cold}^{(i)} \rangle \,.
\end{equation}
Combining this with equations (\ref{eq:recombination} \& \ref{eq:eqm}), after some omitted algebra, we end up with a quadratic equation for $f_{\rm neutral}^{{\rm bg}(i)}$ in terms of knowns.
The requirement that $f_{\rm neutral}^{{\rm bg}(i)} \! \leq \! 1$ by definition eliminates one of the quadratic solutions, leaving
\begin{subequations}
\begin{equation}
f_{\rm neutral}^{{\rm bg}(i)} = 1 + \upgamma - \sqrt{\upgamma^2 + 2\, \upgamma}\,,
\end{equation}
\begin{equation}
\upgamma \equiv \frac{3 \left(3\, X^{(i)} + 1 \right) m_{\rm p}\, \Gamma_{\rm H\,{\LARGE{\textsc i}}}^{\rm eff}(z)}{8\, \alpha\, G\, X^{(i)}} \left( \frac{\sigma_{\rm cold}(z)}{\Sigma_{\rm cold}^{(i)}} \right)^2 \,.
\end{equation}
\end{subequations}


\subsection{Molecular fraction}
\label{ssec:fmol}

We return to using a formula based on the mid-plane pressure of each disc annulus, $P_{\rm mid}^{(i)}$, as the default for calculating the molecular fraction of gas.
Similar to the implementation in \citet{stevens16}, we use the equations from \citet{blitz04} and \citet{elmegreen89}:
\begin{subequations}
\begin{equation}
\label{eq:fH2}
f_{\rm H_2}^{(i)} = \left[ \left(\frac{P_{\rm mid}^{(i)}}{5.93 \times 10^{-13}\, {\rm Pa}}\right)^{-0.92} + 1 \right]^{-1}\,,
\end{equation}
\begin{equation}
P^{(i)}_{\rm mid} = 
\left\{
\begin{array}{l r}
\dfrac{\uppi\, G\, \Sigma_{\rm cold}^{(i)}}{2} \left(\Sigma_{\rm cold}^{(i)} + \dfrac{\sigma_{\rm cold}(z)\, \Sigma_*^{(i)}}{\sigma_*^{(i)}} \right) \! , & \theta_{\rm disc}\! \leq \! 10^\circ \\
\dfrac{\uppi\, G}{2} \Sigma_{\rm cold}^{(i)\,2} \, , & \theta_{\rm disc}\! > \! 10^\circ
\end{array}
\right.\! .
\end{equation}
\end{subequations}
The only practical change%
\footnote{In truth, there is one other, minor change.
Namely, we have removed the rogue factor of $h^2$ that appeared in the normalising pressure in \citet{stevens16}; i.e.~the equivalent of the denominator in the innermost brackets in equation (\ref{eq:fH2}) here.
This was an old error in the code, which was discovered during the proofing stage of that paper.
The text for that paper was modified at the last second before publishing to be consistent with the code.
This was largely inconsequential.
The code has since be fixed.}
from \citet{stevens16} is how $\sigma_{\rm cold}$ and $\sigma_*^{(i)}$ are calculated, which is now consistent with that described above in this paper.
$\theta_{\rm disc}$ is the angular misalignment between the stellar and gas discs.

Other prescriptions for $f_{\rm H_2}^{(i)}$ are available in the code, such as those based on the works of \citet{mckee10} and \citet{gd14}.
We do not adopt them in this paper, though, as we found in testing that extra free parameters were likely required to calibrate the model to observations with these prescriptions.

\begin{figure}[H]
\centering
\includegraphics[width=0.85\textwidth]{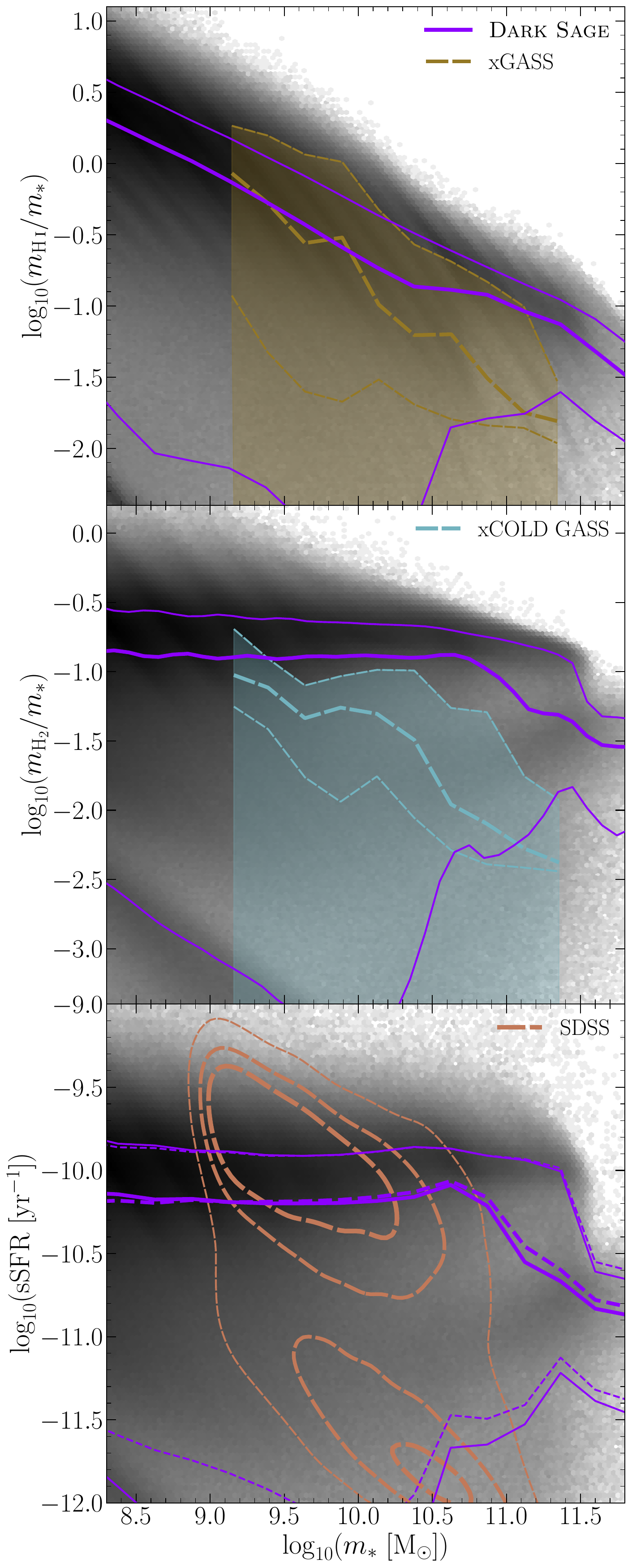}
\vspace{-0.3cm}\caption{
The planes of stellar mass versus \HI~fraction (top panel), \Htwo~fraction (middle panel), and specific star formation rate (bottom panel) in \ds~galaxies at \zo.
Representative observational samples with $m_* \! \geq \! 10^9\,{\rm M}_\odot$ are compared from xGASS for \HI~\citep{catinella18}, xCOLD GASS for \Htwo~\citep{saintonge17}, and SDSS for sSFR (the volume-limited sample used in \citealt{brown17}).
Thick lines represent medians.
Thin lines are 16th and 84th percentiles.
For xGASS and xCOLD GASS, the lines assume non-detections carry their upper-limit value, while the shaded region covers the 16th--84th interpercentile range assuming non-detections have zero mass of that type.
SDSS contours approximately encapsulate 38, 68, and 95 per cent of the sample (thicker to thinner).
\ds~hexbins and solid lines in the bottom panel use the time-step-averaged SFR (almost instantaneous), while the dashed lines represent the average SFR in the last age bin ($\sim$1\,Gyr).
}
\label{fig:ssfr}
\end{figure}

In the top two panels of Fig.~\ref{fig:ssfr}, we show the \HI~and \Htwo~fractions of \ds~galaxies as a function of stellar mass.
These relations are predictions of the model, and are compared against the representative xGASS and xCOLD GASS samples of galaxies \citep{saintonge17,catinella18}.
We note that this raw comparison is simply for a rough reference; to make a fair, accurate comparison between and any simulations and this dataset requires deeper considerations, as has been explored in detail in \citet{stevens19a,stevens21}.


\section{Disc instabilities and \newline star formation}
\label{sec:instab}

The one physical process that arguably is the linch-pin of the entire {\sc Dark Sage} model is gravitational disc instabilities.
The handling of said instabilities drives the majority of star formation, radial redistribution of disc mass, disc heating, bulge growth, and black-hole growth.
While the triggering of an instability in the new {\sc Dark Sage} is very similar to that of \citet{stevens16}, some important changes have been made to how instabilities are resolved.

Previously in {\sc Dark Sage}, stars could form through one of three possible channels.
The `passive' channel prescribed each annulus to form stars at a rate proportional to its local \Htwo~content. 
The `instability' channel represented stars formed in bursts driven by local gravitational instabilities. 
The `merger' channel accounted for extra star formation triggered by gas compression during a galaxy merger.
While the former two were physically well defined to operate in a single annulus, the prescription for merger-driven

\noindent starbursts was an \emph{ad hoc}, purely phenomenological formula, designed originally to work on global galaxy scales instead \citep*{somerville01}:~see section 3.9.3 of \citet{stevens16}.
Because this goes against the ethos of the new model, we have removed the merger channel for star formation.
We maintain the passive and instability channels.
Elimination of the merger-burst channel does not prevent elevated star formation from taking place.
With how mergers are resolved (Section \ref{sec:mergers}), the sudden addition of gas to the descendant's ISM lowers its stability, making it likely that additional star formation occurs through instabilities directly after the merger.

For more than the above reason, instabilities play a more central role in \ds~star formation than they used to (despite their already having been the most important channel for many galaxies, as shown in fig.~1 of \citealt{sb17}).
While passive star formation still occurs, we make an important change in calculating this \emph{after} resolving instabilities, rather than before.
This means that passive star formation is now exclusive to the gas that is stable on the scale of an annulus but still has molecular clouds present.
In hindsight, \ds~should have been originally set up this way.


\subsection{Calculating (in)stability}
\label{sec:calc_instab}

Each annulus of each disc of each galaxy is routinely checked for its gravitational stability according to Toomre's $Q$ criterion.
We calculate this local stability parameter first for each of the stellar and gas components separately, following the respective equations
\begin{subequations}
\begin{equation}
Q_*^{(i)} = \frac{ \kappa^{(i)}\, \sigma_{\rm *,disc}^{(i)} }{ 3.36\, G\, \Sigma_*^{(i)} } \,,
\end{equation}
\begin{equation}
Q_{\rm cold}^{(i)} = \frac{ \kappa^{(i)}\, \sigma_{\rm cold}(z) }{ \uppi\, G\, \Sigma_{\rm cold}^{(i)} } \,,
\end{equation}
\begin{equation}
\kappa^{(i)} \equiv \sqrt{ \frac{ 2\, \langle v_{\rm circ}^{(i)} \rangle \left( j_{\rm outer}^{(i)} - j_{\rm inner}^{(i)} \right) }{ \langle r^{(i)} \rangle^2 \left( r_{\rm outer}^{(i)} - r_{\rm inner}^{(i)} \right) }  }
\end{equation}
\end{subequations}
\citep{toomre64,binney87,pringle07}.
Provided $\theta_{\rm disc} \! \leq \! 10^\circ$, $Q_*^{(i)}$ and $Q_{\rm cold}^{(i)}$ are then combined into a total stability, $Q^{(i)}$, through
\begin{subequations}
\begin{equation}
Q^{(i)} = \left(\frac{1}{{\rm min}\! \left[ Q_{\rm cold}^{(i)}, Q_*^{(i)} \right] } + \frac{W^{(i)}}{{\rm max}\! \left[ Q_{\rm cold}^{(i)}, Q_*^{(i)} \right] } \right)^{-1} \,,
\end{equation}
\begin{equation}
W^{(i)} \equiv \frac{2\, \sigma_*^{(i)}\, \sigma_{\rm cold}(z)}{ \sigma_*^{(i)\,2} +  \sigma_{\rm cold}^2(z)}
\end{equation}
\end{subequations}
\citep{romeo11}.
If $Q^{(i)} \! < \! 1$, the annulus is gravitationally unstable.

By definition, it is not physically possible for something to remain in an unstable state for an extended period of time.
This demands that something must happen to bring the annulus back to $Q^{(i)} \! = \! 1$.
We therefore seek to resolve instabilities as soon as they occur.

Physically, one would expect instabilities to produce star formation from the collapse of unstable gas.
Instabilities can also cause a rearrangement of mass in a galaxy.
In principle, that rearrangement of mass can break the symmetry of a galaxy's gravitational potential, e.g.~through creating a bar, which can create a radially variant torque on the disc, thereby redistributing angular momentum \citep[see, e.g.,][]{anthan03}.
This would, in turn, establish radial migration in the galaxy's disc and alter the orbits of stars, which could also affect the stellar velocity dispersion (and disc scale height) of the disc and create a pseudobulge \citep[see][and references therein]{kormendy04}.
In \ds, we do not explicitly model bars nor asymmetric gravitational potentials, while pseudobulges are inherently considered to be part of the disc \citep[for relevant discussion, see][]{stevens16}.
We nevertheless \emph{do} model the end result that instabilities have on discs, namely in the form of star formation, radial migration, and disc heating.
We do this in a way that is agnostic to the details of \emph{why} each aspect of instability resolution happens the way it does, as it ultimately does not matter for how the model is constructed.

There are two ways to return an unstable disc annulus to stability:~either mass must be removed from the annulus or its velocity dispersion must be raised.
We have already prescribed $\sigma_{\rm cold}$ to be a strict function of $z$, meaning the former must be how gaseous instabilities are resolved.
Cold gas can be removed from an annulus in one of two ways:~by transferring it to adjacent annuli, or by induced star formation triggering stellar feedback.
Stellar mass can only be removed by transferring it to adjacent annuli.
However, we allow the velocity dispersion of stars to also rise.
Each phase therefore has two modes of instability resolution.
The fractional importance of each mode is controlled by a single free parameter, dubbed $f_{\rm move}^{\rm gas}$.

The first step to resolving an instability is deciding which baryon phase the onus is on to solve it.
A return to stability ($Q^{(i)} \! \geq \! 1$) can be guaranteed once \emph{both} $Q_{\rm cold}^{(i)} \! \geq \! 1 + W^{(i)}$ and $Q_*^{(i)} \! \geq \! 1 + W^{(i)}$.
If one phase already satisfies this requirement, the onus is entirely on the other phase to resolve the instability.
If neither criterion is met, both phases have to do some work.

If $\theta_{\rm disc} \! > \! 10^\circ$, the stability of the gas and stellar discs are treated entirely independently.
Respective instabilities are then triggered when $Q_{\rm cold}^{(i)} \! < \! 1$ and $Q_*^{(i)} \! < \! 1$.


\subsection{Resolving gas instabilities}
\label{ssec:gas_instab}

Once a gas instability is identified, it is resolved in two steps.
Both steps involve removing gas from the annulus.
The amount of unstable gas that must be removed from that annulus to return it to stability is
\begin{equation}
m_{\rm cold, unstable}^{(i)} = m_{\rm cold}^{(i)} \left( 1 - \frac{Q_{\rm cold}^{(i)}}{Q_{\rm cold, min}^{(i)}} \right)\,,
\end{equation}
where $Q_{\rm cold,min}^{(i)}$ is greater than $Q_{\rm cold}^{(i)}$ (for a stable disc, $Q_{\rm cold}^{(i)} = Q_{\rm cold, min}^{(i)}$ by our definition) and typically equal to 1 or $1+W^{(i)}$, but can vary based on the circumstance of the instability trigger as described in Section \ref{sec:calc_instab}. 
The first step in resolving this instability is to redistribute gas in the disc.
This is done by moving a controlled fraction of the unstable mass to the two adjacent annuli, in proportion such that angular momentum is conserved.
Mathematically, we can write this as
\begin{equation}
\label{eq:gas_instab}
\Delta m_{\rm cold}^{(i) \rightarrow (i \pm 1)} = \frac{ \langle j^{(i)} \rangle - \langle j^{(i \mp 1)} \rangle }{ \pm \langle j^{(i+1)} \rangle \mp \langle j^{(i-1)} \rangle } f_{\rm move}^{\rm gas}\, m_{\rm cold, unstable}^{(i)} \,.
\end{equation}
Whenever mass is transferred to adjacent annuli in this way, it carries with it metals in proportion to its metallicity.
Strictly, equation (\ref{eq:gas_instab}) only applies for $i \in [2,N_{\rm ann}-1]$.
For the outermost annulus, mass transfer can only happen inwards, meaning we abandon conservation of angular momentum, and simply adopt
\begin{equation}
\Delta m_{\rm cold}^{(N_{\rm ann}) \rightarrow (N_{\rm ann}-1)} = f_{\rm move}^{\rm gas}\, m_{\rm cold, unstable}^{(N_{\rm ann})} \,.
\end{equation}
For the innermost annulus, any inward-moving gas is assumed to be accreted by the black hole.
We approximate the gas accreted onto the black hole to have lost all its angular momentum.
We still adhere to angular-momentum conservation for outward- and inward-moving gas in this case, meaning
\begin{subequations}
\begin{equation}
\Delta m_{\rm cold}^{(1) \rightarrow (2)} = \frac{ \langle j^{(1)} \rangle}{ \langle j^{(2)} \rangle } f_{\rm move}^{\rm gas}\, m_{\rm cold, unstable}^{(1)} \,,
\end{equation}
\begin{equation}
\Delta m_{\rm cold \rightarrow BH}^{(1)} = \frac{  \langle j^{(2)} \rangle - \langle j^{(1)} \rangle }{ \langle j^{(2)} \rangle } f_{\rm move}^{\rm gas}\, m_{\rm cold, unstable}^{(1)} \,.
\end{equation}
\end{subequations}
Note that the growth of the black hole through this channel is independent of its present mass.
This is how \ds~is able to seed zero-mass black holes.
We discuss the consequences of this `quasar' form of black-hole accretion in Section \ref{sec:agn}.

The movement of unstable gas through the disc to the innermost annulus and subsequently onto the black hole is the primary means by which black-hole accretion occurs.
The only other way for black holes to gain a substantial amount of mass is when they merge together.
An important test of this framework is whether \ds~can recover the well-known relation between a galaxy's central black-hole mass and bulge mass.
What makes this particularly exciting is that this is now a prediction of the model; as we show in Section \ref{sec:cali}, this relation has no explicit bearing on the calibration of \ds's free parameters.
By design, instabilities and mergers \emph{can} simultaneously grow both the black hole and bulge in \ds~(if both gas and stars are involved), though this does not guarantee quantitative agreement with the empirical relation between these two properties.

\begin{figure}[H]
\centering
\includegraphics[width=0.95\textwidth]{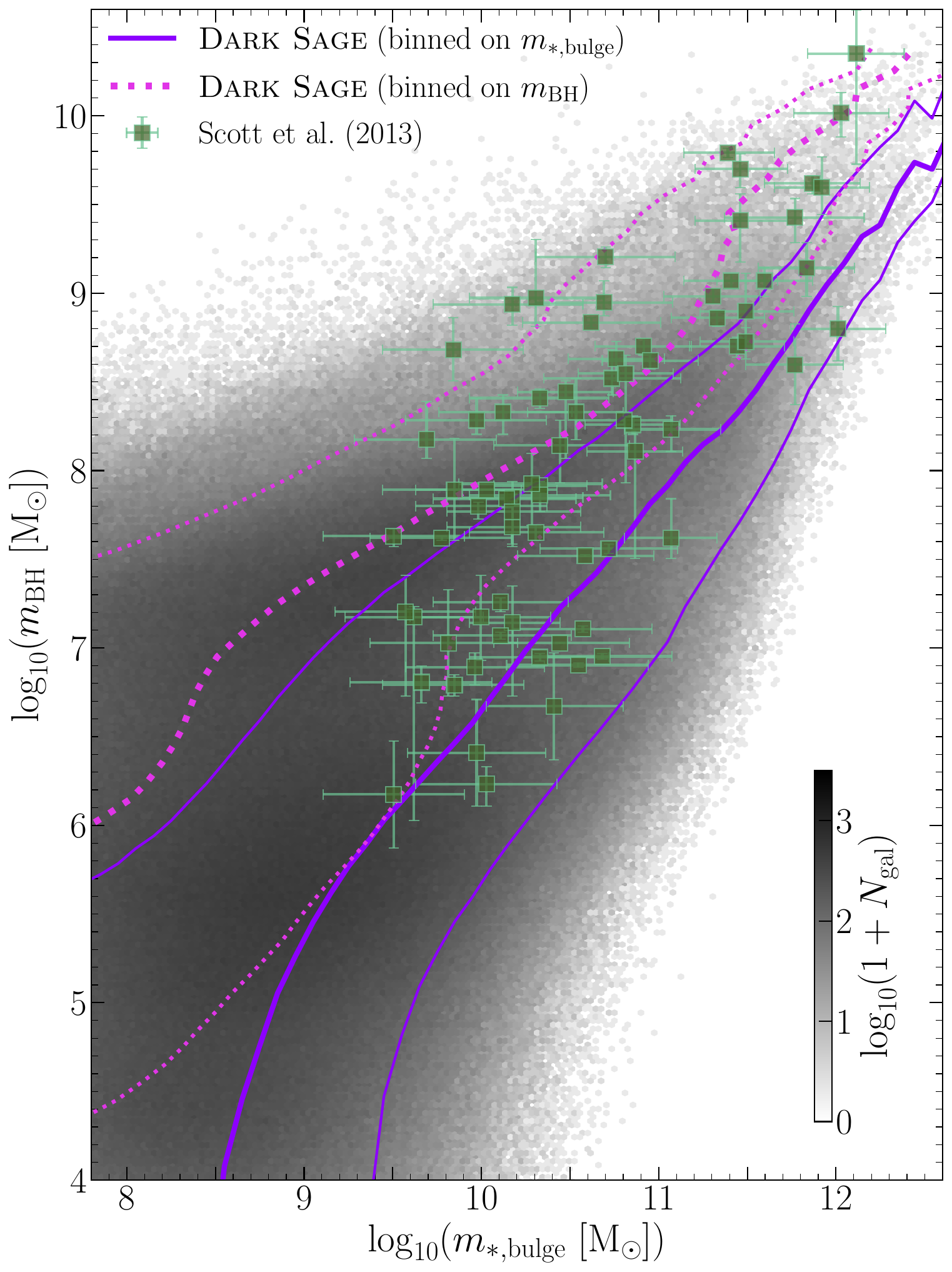}
\vspace{-0.3cm}\caption{Black hole--bulge mass relation for \ds~galaxies at \zo.
The $x$-axis is the sum of the merger- and instability-driven bulge masses.
Because this is not a simple unimodal relation in \ds, we show running percentiles (thick for median, thin for 16th and 84th) when the model's data are binned along the $x$-axis (solid) and $y$-axis (dotted).
Neither median traces the high-number-density ridge of \ds~galaxies.
Compared is a compilation of observational data from \citet{scott13}.
While previous versions of \ds~calibrated to these data, this outcome is now a prediction of the model.
}
\label{fig:bhbulge}
\end{figure}

We show in Fig.~\ref{fig:bhbulge} that \ds~predicts a black hole--bulge mass relation that overlaps favourably with observations \citep*[cf.][]{scott13}.
Interestingly, \ds's relation is not unimodal; this is seen in the intensity of the hexbins and clarified by the misalignment of equivalent running percentiles when the \ds~data are binned along each of the two axes.
No cuts have been applied to \ds~here.
A deeper assessment of this relation relative to observations is left for future work.

After moving gas, part of the remaining unstable gas is then consumed in star formation, with the last of it reheated out of the disc by the instantaneous component of the stellar feedback associated with those new stars:
\begin{multline}
\Delta m_{\rm cold \rightarrow *}^{(i)\iSF} + \Delta m_{\rm cold \rightarrow fount}^{(i)\iSN} + \Delta m_{\rm cold \rightarrow outfl}^{(i)\iSN} \\
 = (1 - f_{\rm move}^{\rm gas})\,  m_{\rm cold, unstable}^{(i)} \,.
\end{multline}
The balance between each of the terms on the left-hand side of this equation is described in Section \ref{sec:feedback}.
Note that $\Delta m_{\rm cold \rightarrow *}^{(i)\iSF}$ represents the \emph{net} production of stellar mass at the end of the present sub-time-step, and is therefore less than the annulus's instability-driven star formation rate multiplied by $\Delta t$ (this is also explained in Section \ref{sec:feedback}). 
The superscripts $\iSF$ and $\iSN$ are intended to differentiate instability-driven star formation and supernovae from the passive channel of SF (Section \ref{ssec:sf}).

After a gas instability is resolved, we recalculate $Q_*^{(i)}$.
This is necessary, as the stellar mass in the annulus will have changed, which inevitably (temporarily) raises not only the probability of a stellar instability triggering but also the mass of unstable stellar mass.


\subsection{Passive star formation}
\label{ssec:sf}

After resolving instabilities (including what follows in Section \ref{ssec:star_instab}), we allow for a second mode of star formation that depends explicitly on the local molecular gas surface density.
This `passive' mode of star formation in {\sc Dark Sage} obeys
\begin{equation}
\label{eq:sf}
\Sigma_{\pSFR}^{(i)} = \epsilon_{\rm H_2}\, \Sigma_{\rm H_2}^{(i)} \,,
\end{equation}
where $\epsilon_{\rm H_2}$ is the star formation efficiency of \Htwo, equivalent to the inverse of a depletion time-scale.
In a change from \citet{stevens16}, $\epsilon_{\rm H_2}$ is no longer treated as a free parameter.
Instead, this is fixed to $\epsilon_{\rm H_2} \! = \! (2.35\,{\rm Gyr})^{-1}$, consistent with the empirical best-fitting relation from \citet{bigiel11}.
Of course, there must be a limit to the amount of star formation in a given annulus set by the total mass of gas present, accounting for the gas that must be removed from the annulus due to feedback from the largest stars that form in that population and effectively go supernova immediately.
In practice, we therefore effect
\begin{multline}
\label{eq:fb_lim}
\Delta m_{\rm cold \rightarrow *}^{(i)\pSF} = {\rm min} \bigg[ \uppi \left( r_{\rm outer}^{(i)~2} - r_{\rm inner}^{(i)~2} \right) \left( 1 - f_\star^{\rm return} \right) \Sigma_{\pSFR}^{(i)}\, \Delta t, \\
m_{\rm cold}^{(i)} - \Delta m_{\rm cold \rightarrow fount}^{(i)\pSN} - \Delta m_{\rm cold \rightarrow outfl}^{(i)\pSN}  \bigg] \,.
\end{multline}
Many of the terms here will be defined in Section \ref{sec:feedback}.

\begin{figure}[H]
\centering
\includegraphics[width=0.95\textwidth]{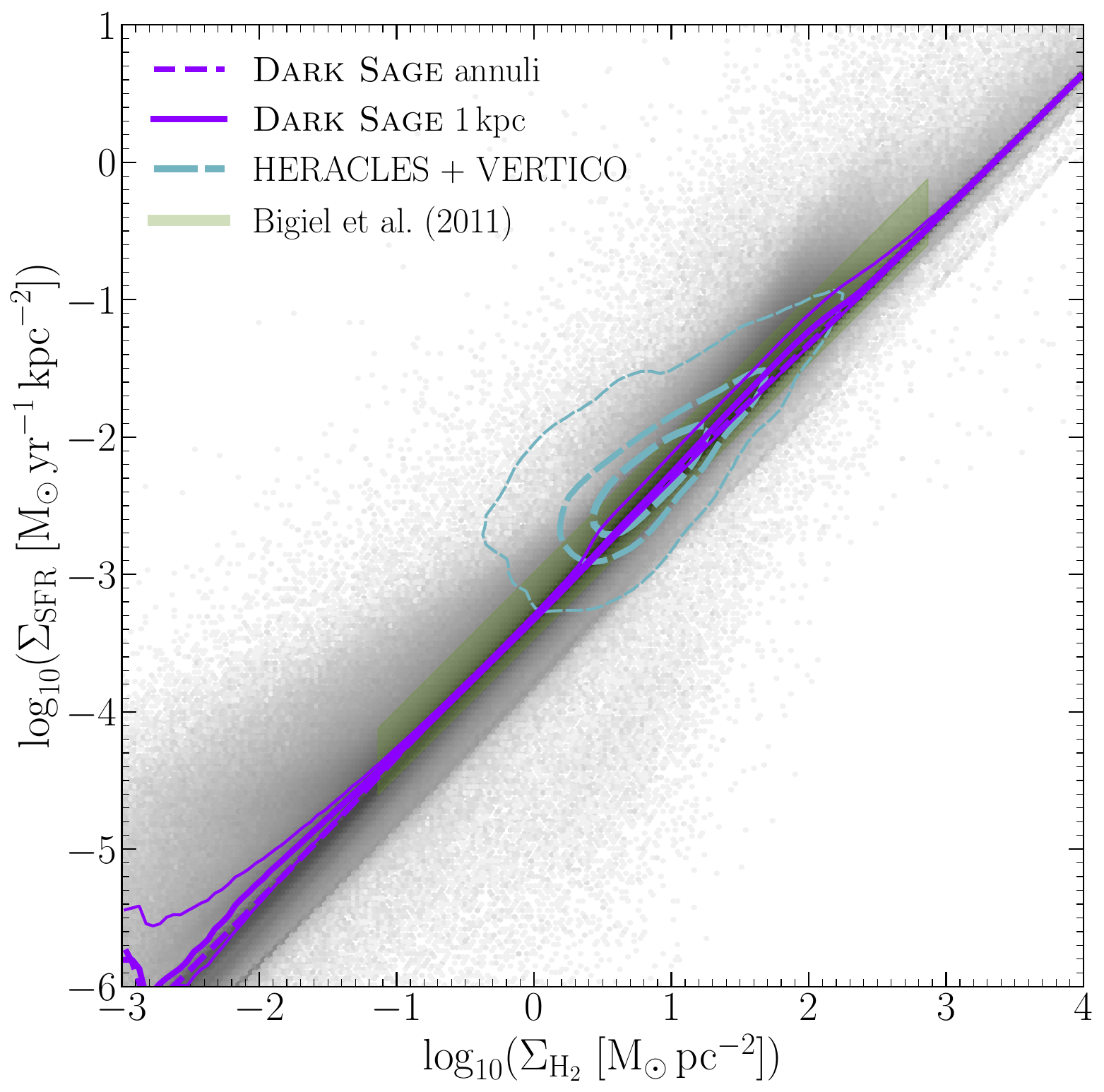}
\vspace{-0.3cm}\caption{The resolved molecular Kennicutt--Schmidt relation for \ds~galaxies at \zo.
The dashed \ds~line is the median relation for all annuli.
The hexbins are grey-scaled according to the logarithmic number of annuli within them.
For a fairer comparison to observations, the hexbins only count annuli with $\Sigma_* \! > \! 10\,{\rm M}_\odot\,{\rm pc}^{-2}$ from galaxies with $10^{8.9} \! < \! m_*/{\rm M}_{\odot} \! < \! 10^{11}$ and $0.02 \! < \! m_{\rm H_2} / m_{\rm H\,{\LARGE{\textsc i}}} \! < \! 1.13$ in (sub)haloes with at least 200 particles at one point in their history.
The solid lines represent the median (thick) and 16th and 84th percentiles (thin) for those binned annuli \emph{after} weighting each annulus by its area.
This weighting further improves the comparison to observations, which typically use pixels of fixed length $\simeq$1\,kpc.
The best-fitting relation from observations \citep{bigiel11} and equivalent scatter is shown by the shaded region.
Contours encapsulate 38, 68, and 95 per cent of pixels across the HERACLES and VERTICO surveys that are detected in both axes.
}
\label{fig:ks}
\end{figure}

Despite equation (\ref{eq:sf}) explicitly tying star formation and H$_2$ surface density together, \ds~predicts a non-trivial resolved molecular Kennicutt--Schmidt (RMKS) relation; i.e. $\Sigma_{\rm H_2}$ versus $\Sigma_{\rm SFR}$.
This is predictive because (i) equation (\ref{eq:sf}) only applies to stars formed in stable annuli, (ii) the majority of star formation happens in unstable annuli, and (iii) there is no explicit tying of the molecular fraction to disc stability.
In Fig.~\ref{fig:ks}, we show the RMKS relation for (a relevant subset of) \ds~galaxies at \zo.
This is compared to the relation fitted to observations by \citet{bigiel11} as well as data from the HERACLES and VERTICO surveys \citep[][respecitvely]{leroy09,brown21}.
Note that, despite HERACLES and VERTICO respectively probing field and cluster environments, these two datasets have been demonstrated to have consistent RMKS relations \citep{jimenez23}.%
\footnote{These data show that environment does, however, affect other local gas scaling relations \citep{watts23}.}


\subsection{Resolving stellar instabilities}
\label{ssec:star_instab}


Similar to gas, there are two steps in resolving a stellar instability.
But these steps are different in detail.
The mass of stars deemed necessary to move out of the annulus to restore stability is set by
\begin{equation}
m_{\rm *, unstable}^{(i)} = m_{\rm *,disc}^{(i)}\, f_{\rm move}^{*(i)} \left( 1 - \frac{Q_{\rm *}^{(i)}}{Q_{\rm *, min}^{(i)}} \right)\,,
\end{equation}
\begin{equation}
\label{eq:fmove*}
f_{\rm move}^{*(i)} = 1 - \frac{\sigma_{\rm cold}(z)}{\sigma_{\rm *,disc}^{(i)}} \left( 1 - f_{\rm move}^{\rm gas} \right) \,.
\end{equation}
%
By construction, it is always true that $\sigma_{\rm cold}(z) \! \leq \! \sigma_{\rm *,disc}^{(i)}$, and therefore $f_{\rm move}^{*(i)} \! \in \! (0,\,1)$.
The unstable mass of stars is moved to adjacent annuli in the same $j$-conserving fashion as done for gas:
\begin{equation}
\Delta m_{\rm *,disc}^{(i) \rightarrow (i \pm 1)} = \frac{ \langle j^{(i)} \rangle - \langle j^{(i \mp 1)} \rangle }{ \pm \langle j^{(i+1)} \rangle \mp \langle j^{(i-1)} \rangle } m_{\rm *, unstable}^{(i)}
\end{equation}

\noindent when $i \in [2, N_{\rm ann}-1]$.
For the outermost annulus, we again neglect angular-momentum conservation by invoking
\begin{equation}
\Delta m_{\rm *,disc}^{(N_{\rm ann}) \rightarrow (N_{\rm ann}-1)} = m_{\rm *, unstable}^{(N_{\rm ann})} \,.
\end{equation}
Inward-moving stars from the innermost annulus are transferred to the instability-driven bulge.
This bulge component has no angular momentum by design.
As such, we effect
\begin{subequations}
\begin{equation}
\Delta m_{\rm *,disc}^{(1) \rightarrow (2)} = \frac{ \langle j^{(1)} \rangle}{ \langle j^{(2)} \rangle }\, m_{\rm *, unstable}^{(1)} \,,
\end{equation}
\begin{equation}
\Delta m_{\rm *,disc \rightarrow \ib}^{(1)} = \frac{  \langle j^{(2)} \rangle - \langle j^{(1)} \rangle }{ \langle j^{(2)} \rangle }\, m_{\rm *, unstable}^{(1)} \,.
\end{equation}
\end{subequations}
As stars transfer from one annulus to the next (or to the instability-driven bulge), they carry with them the same age--metallicity--velocity dispersion distribution.
The metallicity and velocity dispersion squared of each age bin in the receiving stellar component are updated with a mass-weighted sum of the moving stars and that already in the receiving annulus.

By itself, the motion of stellar mass prescribed above does not raise $Q_*^{(i)}$ to its minimum stable value.
To complete this, we invoke `heating' of the stellar disc.
That is, the stellar velocity dispersion of the disc annulus is raised by a factor equal to
\begin{displaymath}
f_{\rm move}^* + \left(1 - f_{\rm move}^* \right)\, \frac{Q_{\rm *, min}^{(i)}}{Q_{\rm *}^{(i)}} \, .
\end{displaymath}
This change applies to each age bin individually.
The functional form of equation (\ref{eq:fmove*}) is something we have put in by hand intentionally so that disc heating does not continue to occur unregulated in a perpetually unstable disc.
The increase to the annular stellar velocity dispersion is applied after moving the unstable mass of stars. 

It is because stars are collisionless and gas collisional that we have different $f_{\rm move}$ values\,---\,i.e.~different radial-migration rates\,---\,for gas and stars.
The functional form of equation (\ref{eq:fmove*}) ensures that stellar velocity dispersions in the disc cannot grow unchecked.
The more stellar discs are heated, the less they will be heated at the next instability.


The size of an instability-driven bulge is set by the velocity dispersion of the stars it acquires.  
This is calculated from the Jeans equation, under the assumptions that the instability-driven bulge is both spherically symmetric and isothermal:
\begin{equation}
\label{eq:jeans}
\frac{{\rm d} \Phi}{{\rm d} R} = -\frac{\sigma_{\ib}^2}{\rho_{\ib}} \frac{{\rm d} \rho_{\ib}}{{\rm d} R}\,.
\end{equation}
Substituting equation (\ref{eq:hernquist})\,---\,and its derivative, where $\otimes \! \rightarrow \! \ib$\,---\,into (\ref{eq:jeans}) and rearranging gives 
\begin{equation}
\label{eq:a_ib}
a_{\ib} = 3 \left( \frac{1}{\sigma^2_{\ib}} \frac{{\rm d} \Phi}{{\rm d} R} - \frac{1}{R} \right)^{-1} - R\,.
\end{equation}
In principle, this expression should be solvable at any $R$.  
However, because of the approximation of isothermality (i.e.~treating $\sigma_{\ib}$ as radially invariant), equation (\ref{eq:a_ib}) does not return identical answers at any two radii.  
For that reason, starting at small radii and working outwards incrementally, we continuously solve for $a_{\ib}$, averaging as we go, until reaching a converged answer.


\section{New stellar-feedback model}
\label{sec:feedback}

We introduce a brand-new model for stellar feedback in \ds.
Gone are the days of explicitly prescribing a mass-loading factor that may or may not be modulated by local surface density.
Instead, we now use a purely energy-based argument to calculate the amount of reheated (and ejected) gas from each episode of star formation, in each annulus.
Gone also are the assumptions of instantaneous recycling, where we have now implemented a delayed, self-consistent scheme for stellar mass loss, metal enrichment, and associated feedback, which we outline below.

\subsection{Energy available from feedback}
\label{ssec:energy}

Let us begin with the assumption that the birth masses of stars, $\mathcal{M}_{\star}$, for any population are distributed according to a \citet{chabrier03} initial mass function (IMF, $\phi$), where we impose
\begin{subequations}
\begin{equation}
\label{eq:IMF}
\phi(\m) \! = \! 
\left\{
\begin{array}{l r}
0\,, & \m \! < \! 0.1 \\
\dfrac{A_\star}{\m}\! \exp\!\left[-\dfrac{1}{2}\! \bigg(\dfrac{\log_{10}\!\big(\frac{\m}{0.08}\big)}{0.69}\bigg)^2 \right], & 0.1 \! \leq \! \m \! < \! 1\\
k_\star\, \m^{-2.3}\,, &1 \! \leq \! \m \! \leq \! 100 \\
0\,, &  \m > 100 
\end{array}
\right.\!,
\end{equation}
\begin{equation}
\m \equiv \mathcal{M}_{\star} / {\rm M}_{\odot}\,,
\end{equation}
\end{subequations}
where $A_\star$ and $k_\star$ are constants.
Recognising that (i) $\phi(\m)$ must be continuous at $\m\!=\!1$ and (ii)
\begin{equation}
\int_{0}^{\infty} \m\, \phi(\m)\, {\rm d}\m \! = \! \int_{0.1}^{100} \m\, \phi(\m)\, {\rm d}\m = 1
\end{equation}
leads to the derivation that $k_\star\!\simeq\!0.238$ and $A_\star\!\simeq\!0.843$.

To calculate the mass fraction of a stellar population returned to the ISM over a specified period of time, one must also know both the lifetimes ($t_{\rm life}$) and end-of-life remnant masses ($\mathcal{M}_{\rm rem}$) of stars as a function of their birth mass.
For the former, we impose the textbook standard
\begin{equation}
\label{eq:tlife}
\frac{t_{\rm life}}{\rm Gyr} = 10\,\m^{-5/2}
\end{equation}
\citep{harwit88}. This is readily derived from the assertions that the luminosity of a star scales as $\mathcal{M}_\star^{3.5}$ and that the Sun has an approximate lifetime of 10\,Gyr.

For the remnant masses of stars, we roughly follow \citet{madau14} in effecting
\begin{subequations}
\begin{equation}
\label{eq:Mrem}
\bar{\mathcal{M}}_{\rm rem}(\m) = 
\left\{
\begin{array}{l r}
\m \,, & \m \! < \! 1\\
0.444 + 0.084\,\m \,, & 1 \! \leq \! \m \! \leq 7\\
0.419 + 0.109\,\m \,, & 7 \! < \! \m \! < \! 8\\
1.4 \,, & 8 \! \leq \! \m \! \leq \! 50\\
\m \,, & \m \! > \! 50
\end{array}
\right.
\end{equation}
\begin{equation}
\bar{\mathcal{M}}_{\rm rem} \equiv \mathcal{M}_{\rm rem} / {\rm M}_{\odot}\,.
\end{equation}
\end{subequations}
The term for $\bar{\mathcal{M}}_{\rm rem}(7 \! < \! \m \! < \! 8)$ is something we have manually introduced as a means of linearly interpolating between the adjacent mass ranges, which more closely follow \citet{madau14}.
The value of $\mathcal{M}_{\rm rem}\!=\!1.4\,{\rm M}_{\odot}$ for $8 \! \leq \! \m \! \leq 50$ assumes that the remnant neutron star or black hole has a mass close to the upper limit of white dwarfs (\'{a} la \citealt{chandra31}).
The basis for $\bar{\mathcal{M}}_{\rm rem} \! = \! \m$ when $\m\!>\!50$ is that the collapse of the star is so rapid, its entire mass is committed to the remnant black hole (which we still class as `stellar mass' in the model).
Given how quickly the IMF diminishes at high stellar masses (equation \ref{eq:IMF}), variations on the assumptions for $\bar{\mathcal{M}}_{\rm rem}(\m\!\gtrsim\!50)$ are inconsequential for our treatment of stellar evolution and feedback.

The mass returned to the ISM%
\footnote{Strictly speaking, the mass is only returned to the ISM for stars in \ds~discs.  Otherwise it directly `returns' to the CGM, as described in Section \ref{sec:fb_CGM}.  But we use the phrase `returned to the ISM' throughout the paper for simplicity.}
for a given star equals $\mathcal{M}_\star - \mathcal{M}_{\rm rem}$.
We approximate that that mass is returned instantaneously at $\Delta t \! = \!t_{\rm life}$ after the star's birth.
We can therefore calculate the fraction of a stellar population's initial mass that is returned to the ISM over a cosmic-time interval $[t_0-t_{\rm form}, t_1-t_{\rm form}]$ as
\begin{subequations}
\label{eq:f_return}
\begin{equation}
f_{\star}^{\rm return} = \int_{\bar{\mathcal{M}}_1}^{\bar{\mathcal{M}}_0} \left[\m - \mrem(\m) \right] \phi(\m)\, {\rm d}\m\,,
\end{equation}
\begin{equation}
\bar{\mathcal{M}}_x = \left( \frac{t_x - t_{\rm form}}{10\,{\rm Gyr}} \right)^{-2/5}\,,
\end{equation}
\end{subequations}
where the latter comes directly from equation (\ref{eq:tlife}), with $x\!=\!0,1$, while $t_{\rm form}$ denotes the time at which the stellar population formed.
Note that because we define $t_1 > t_0$ and higher-mass stars die sooner, it follows that $\bar{\mathcal{M}}_1 < \bar{\mathcal{M}}_0$, which is why $\bar{\mathcal{M}}_1$ is the \emph{lower} bound for the integral.
The fraction of the \emph{remaining} mass of a population (where the returned mass from all stars with $\m\!>\!\bar{\mathcal{M}}_0$ has already been subtracted from the population's total mass) that is returned to the ISM can likewise be calculated as
\begin{subequations}
\begin{equation}
f_{\rm rem}^{\rm return} = f_{\star}^{\rm return} \frac{m_{\rm *,pop}^{\rm form}}{m_{\rm *,pop}^{\rm rem}}\,,
\end{equation}
\begin{multline}
\frac{m_{\rm *,pop}^{\rm form}}{m_{\rm *,pop}^{\rm rem}} = \bigg[ \int_{0.1}^{\bar{\mathcal{M}}_0} \m\, \phi(\m)\, {\rm d}\m \\
+ \int_{\bar{\mathcal{M}}_0}^{100} \mrem(\m) \, \phi(\m)\, {\rm d}\m \bigg]^{-1}\,,
\end{multline}
\end{subequations}
where $m_{\rm *,pop}^{\rm form}$ and $m_{\rm *,pop}^{\rm rem}$ represent the formation mass and remaining mass of the stellar population, respectively. This expression is practically the most useful in the design of \ds, as we only explicitly track the remaining mass of a given population.

When mass is returned from a stellar population to the ISM, metals are returned to the ISM in proportion to the metallicity of that stellar population.
In addition to this, new metals are added to the ISM, borne from fusion in the stars' cores and during supernova explosions.
We maintain that the mass of new metals produced by a stellar population over its lifetime is a fixed fraction of the population's birth mass (ignoring the metals already present).
This fraction is known as the `yield,' which is expected to be $\sim$0.03, based on a \citet{chabrier03} IMF combined with a \citet*{conroy09} simple stellar population model \citep{shark}.
These new metals are deposited into the ISM at a rate proportional to the mass-loss rate of stars in that population.
This occurs \emph{before} any reheating from feedback (Section \ref{sec:fb_ISM}).
Note that we do not track the individual abundances of metal elements, only their summed content.

\begin{figure}[H]
\centering
\includegraphics[width=\textwidth]{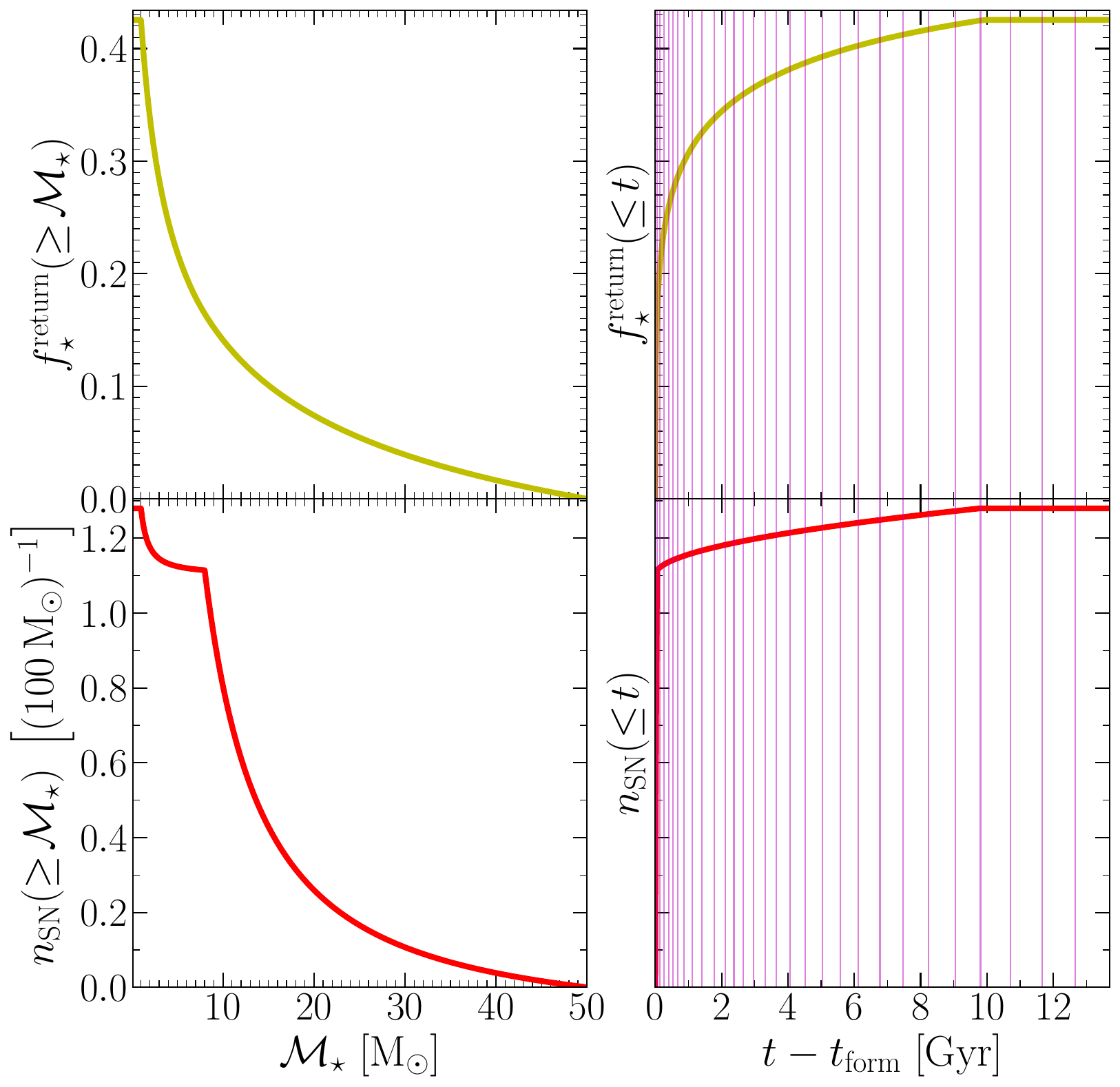}
\vspace{-0.2cm}\caption{Cumulative mass fraction of a stellar population returned to the ISM (top, equation \ref{eq:f_return}) and cumulative number of supernovae per unit initial mass of that population (bottom) as a function of star birth mass (left) and time since the population's formation (right) implemented in our new stellar evolution/feedback model in \ds.
For reference, the vertical lines in the right-hand panels represent the age bins used in \ds~for this work (treating $t_{\rm form}$ as 0; see Section \ref{ssec:stellar}).
For stellar populations born at $t_{\rm form} \! > \! 0$, the vertical lines would effectively be shifted to the left to describe how the evolution of that population is discretised in \ds.
}
\label{fig:reSN}
\end{figure}

Let us hypothesise for a moment that we know the \emph{average} energy released by a supernova \emph{that is used in heating gas out of the ISM and/or ejecting it out of the CGM}, which we label $\mathcal{E}_{\rm SN}$
(note that we make no assumptions about the \emph{total} energy released per supernova).
Then we must ask: how many supernovae will there be for a given stellar population and when will they explode?  
Equation (\ref{eq:tlife}) already provides us the answer to the latter\,---\,any star that produces a supernova does so at the end of its life.
For the former, we assume all stars with $8 \! \leq \! \m \! \leq \! 50$ produce a (core-collapse/Type-II) supernova.
Additionally, we assume some fraction of stars with $1 \! \leq \! \m \! \leq \! 8$ produce a (accretion-driven/Type-Ia) supernova.
This means we effectively assume that each star with $1 \! \leq \! \m \! \leq \! 8$ has a fixed probability, $p_\textrm{SN-Ia}$, of producing a supernova at the end of its life.
The total number of supernovae of each type per unit mass of stars formed in a population, $n_{\rm SN}$, is thus
\begin{subequations}
\label{eq:nSN}
\begin{equation}
\frac{n_\textrm{SN-II}}{{\rm M}_{\odot}^{-1}} = \int_8^{50} \phi(\m)\, {\rm d}\m\,,
\end{equation}
\begin{equation}
\frac{n_\textrm{SN-Ia}}{{\rm M}_{\odot}^{-1}} = p_\textrm{SN-Ia} \int_1^8 \phi(\m)\, {\rm d}\m\,.
\end{equation}
\end{subequations}
To determine $p_\textrm{SN-Ia}$, we look to observations.
Broadly speaking, the rate of Type-II supernovae are observed to exceed that of Type Ia.
For example, \citet{tsujimoto95} note that in the Milky Way, the ratio of Type-Ia to Type-II supernovae\,---\,equivalent to $n_\textrm{SN-Ia} / n_\textrm{SN-II}$\,---\,is $\sim$0.15, but that it is more like 0.2--0.3 in the Magellanic Clouds.
By contrast, \citet{maggi16} find $n_\textrm{SN-Ia} / n_\textrm{SN-II} \! \simeq \! 0.74$ in the Large Magellanic Cloud (LMC), but they expect this to be above the true average ratio due to the short time-scale that their data probe (order $10^4$\,yr) and the shape of the LMC's star formation history.
In the absence of tighter observational constraints, we choose to adopt the Milky Way value of $n_\textrm{SN-Ia} / n_\textrm{SN-II} \! = \! 0.15$.
Through equation (\ref{eq:nSN}), we can therefore solve for $p_\textrm{SN-Ia} \! \simeq \! 0.01$.
In line with equation (\ref{eq:f_return}), we can calculate the total energy imparted on the local ISM by a stellar population over a given time interval, which we frame as a birth-mass interval, as
\begin{subequations}
\begin{equation}
\label{eq:E_SN}
\Delta E_{\rm SN} = \mathcal{E}_{\rm SN}\, m_{\rm *,pop}^{\rm form} \int_{ \bar{\mathcal{M}}_1}^{ \bar{\mathcal{M}}_0} \varphi(\m) \, {\rm d}\m\,,
\end{equation}
\begin{equation}
\varphi(\m) \equiv 
\left\{
\begin{array}{l r}
p_\textrm{SN-Ia}\, \phi(\m)\,, & \m < 8 \\
\phi(\m)\,, & \m \geq 8 \\
\end{array}
\right.\,.
\end{equation}
\end{subequations}
We show how our mass-return and supernova models look as a function of star birth mass and time since population birth in Fig.~\ref{fig:reSN}.

All stars are assumed to form at the middle of an age bin.
Returned mass, the production of new metals, and feedback from stars born within the same age bin occur instantaneously, always assuming half the age bin has elapsed, regardless of where in that bin the current sub-time-step actually is.
This is necessary, as once the model has moved into the time window of the next age bin, all stars formed in the previous age bin are assumed to be part of the same population with a single $t_{\rm form}$.
This is the nature of discretised time; regardless of how many bins one has, one always has to treat the `current' bin in a special way.
This also means that if we were to set the number of age bins to 1, we would return to the classical instantaneous recycling approximation (with the caveat that the recycling fraction is solved for, rather than given as an explicit input).
Whenever star formation occurs, we check if it is necessary to rescale the local SFR to ensure that the mass consumed in star formation plus that associated with the instantaneous portion of stellar feedback does not exceed the mass of the gas annulus (in line with equation \ref{eq:fb_lim}).

Several numbers in our model for stellar feedback could have been treated as free parameters.
For example, the ratio of Type-Ia to Type-II supernovae for a population of stars, the parameters controlling the shape of the IMF, and the normalisation and power index of stellar life-times as a function of birth mass.
Instead, we have opted to fix those numbers, and allow \emph{one} stellar-feedback parameter to vary freely in our calibration:~$\mathcal{E}_{\rm SN}$.
Our rationale is that variation in most of the other parameters would be degenerate with changes to $\mathcal{E}_{\rm SN}$ to first order for calibration purposes.
The exact value of $\mathcal{E}_{\rm SN}$ should, therefore, not be interpreted too literally, as it compensates for uncertainty in numerous assumptions we have made.
Nevertheless, we should not expect $\mathcal{E}_{\rm SN}$ to depart too far from the canonical value of $10^{44}$\,J (for a discussion on the energy output of supernovae, see e.g.~\citealt{smartt09}).
Sure enough, as we show in Section \ref{sec:cali}, we find the best value of $\mathcal{E}_{\rm SN}$ to be $\sim\!0.8 \! \times \! 10^{44}$\,J.

There are further details one may wish to consider in a stellar-feedback model, such as how the metallicity of a stellar population affects the rate of supernovae \citep[e.g.][]{kistler13}, how the energetics of Type-Ia and Type-II supernovae differ, and how the IMF may change with metallicity and formation time \citep[e.g.][]{li23}.
Implicitly, our ignorance of these details is also folded in to the best-fitting value of $\mathcal{E}_{\rm SN}$.


\subsection{Impartation of energy on the ISM}
\label{sec:fb_ISM}

With the energy from stellar feedback determined, the next step is to decide how that energy is used.
The standard practice for doing this is to reheat an appropriate mass of gas out of the ISM and/or eject it.
There are three major differences between how we do this now versus previous versions of \ds:
\begin{enumerate}
\item We do not impose a mass-loading factor.
Instead, we \emph{solve} for the amount of gas reheated out of the ISM using energy-conservation arguments, as we outline below.
\item We split this reheated mass into two categories:~the fraction with enough energy to escape the halo's virial radius, and that which will remain bound to the CGM.
\item We do not instantaneously transfer mass to the hot and/or ejected reservoirs, but instead initially transfer them to the new fountain and outflowing reservoirs, thereby enforcing a delay before that gas is available for cooling.
\end{enumerate}

Let us imagine that a supernova outflow begins carrying the mass prescribed by its return fraction of the source stellar population according to its age and the IMF assumptions laid out in Section \ref{ssec:energy}.
This initial blast then collides with the ISM, picking up mass and driving a wind that removes that mass from the ISM.
Once mixed, the gas must also be heated to the temperature of the CGM, which is assumed to be uniform (equation \ref{eq:Thot}).
Assuming this wind conserves energy, we have
\begin{equation}
\label{eq:SN3}
\Delta E_{\rm SN}^{(i)} = \frac{1}{2}\, \Delta m_{\rm cold \rightarrow CGM}^{(i) \rm SN} \left(V_{\rm 200c}^2 + v_{\rm wind}^{(i)~2}\right)\,.
\end{equation}
Here, we have defined $v_{\rm wind}^{(i)}$ in the annulus's rotating frame as \emph{the equivalent initial wind velocity were the outflow were to hypothetically reach thermal equilibrium with the CGM immediately}.
In reality, we would expect the wind to start colder and faster, gradually heating and slowing (i.e.~exchanging kinetic energy for thermal energy \emph{in addition to} gravitational potential energy) before fully mixing with the CGM.
Defining $v_{\rm wind}$ as we have simply provides a mathematical shortcut to reach our desired outcome (below).
While we \emph{do} model how long it takes an outflow to incorporate into the CGM (or travel beyond; see Section \ref{sec:reinc}), this time-scale is \emph{not} informed by $v_{\rm wind}$.

\ds~is designed such that any feedback-affected gas must have its energy raised to at least that of the hot-gas reservoir.
As such, we can solve for $v_{\rm wind}^{(i)}$ under the condition that 
\begin{displaymath}
\Delta E_{\rm SN}^{(i)}  \rightarrow \left(e_{\rm hot} - e_{\rm cold}^{(i)}\right) \Delta m_{\rm cold \rightarrow CGM}^{(i) \rm SN} \,,
\end{displaymath}
where each $e$ represents the specific energy of the subscripted reservoir.
Combining with equation (\ref{eq:SN3}) and rearranging gives
\begin{equation}
v_{\rm wind}^{(i)} = \sqrt{2 \left(e_{\rm fount} - e_{\rm cold}^{(i)}\right) - V_{\rm 200c}^2}\,.
\end{equation}

The specific energy of the hot-gas reservoir can be broken into three components:~its specific potential energy, specific thermal energy, and specific kinetic energy (i.e.~that which is non-thermal, as noted to be non-negligible by e.g.~\citealt{lochhaas21}); i.e.
%
\begin{equation}
\label{eq:ehot}
e_{\rm hot} = \langle \Phi \rangle_{\rm hot} + e_{\rm hot}^{\rm kin} + e_{\rm hot}^{\rm therm}\,.
\end{equation}
To calculate the average potential of gas in the CGM, we numerically solve
\begin{equation}
\langle \Phi \rangle_{\rm hot} = \frac{4 \uppi}{m_{\rm hot}} \int_0^{R_{\rm 200c}} \rho_{\rm hot}(R)\, \Phi(R)\, R^2\, {\rm d}R\,.
\end{equation}
The potential profile, $\Phi(R)$, is found by numerically integrating equation (\ref{eq:vcirc}).%
\footnote{Because we are only interested in \emph{differences} in specific energy between reservoirs, the integration constant returned when calculating potential energy is irrelevant. But for practical reasons in the codebase, we take $\Phi(r_{\rm max})\!=\!0$, where generally $r_{\rm max} \! \gg \! R_{\rm 200c}$, meaning $\Phi(r \! \leq \! R_{\rm 200c}) \! < \! 0$.}
Note that $\Phi(r)$ is defined \emph{in the plane of the disc} in equation (\ref{eq:vcirc}).
As a necessary approximation for computational efficiency, we assume this functional form works for three-dimensional radii, $R$, as well.
In the limit where the ratio of the disc mass to halo mass goes to zero, this approximation is accurate.
However, in general, it is imperfect.

The specific kinetic energy of hot gas is informed entirely by bulk rotational motion.
We approximate its bulk rotational velocity based on its specific angular momentum (assumed here to be equal to that of the halo) and the average radius of a gas element within it:
\begin{subequations}
\label{eq:ekinhot}
\begin{equation}
e_{\rm hot}^{\rm kin} = \frac{j_{\rm hot}^2}{2\,\langle R \rangle^2_{\rm hot}}  = \frac{j_{\rm halo}^2}{2\,\langle R \rangle^2_{\rm hot}} \,, 
\end{equation}
\begin{align}
\langle R \rangle_{\rm hot} &= \frac{4\uppi}{m_{\rm hot}} \int_0^{R_{\rm 200c}} \rho_{\rm hot}(R)\, R^3\, {\rm d}R \\
 &= \frac{1-\ln(2)}{2}\, R_{\rm 200c}\, c_\beta^2(z)\, \mathcal{C}_\beta(z)\,.
\end{align}
\end{subequations}

As we assume $T_{\rm hot} \! = \! T_{\rm vir}$ throughout the CGM, the thermal specific energy of the hot-gas reservoir is readily found as
\begin{equation}
\label{eq:etherm}
e_{\rm hot}^{\rm therm} = \frac{V_{\rm 200c}^2}{2}\,. 
\end{equation}
For satellites, we use $V_{\rm 200c}$ at infall here, in line with our assumption that $T_{\rm hot}$ remains fixed after infall from Section \ref{sec:cooling}. 

Similar to equation (\ref{eq:ehot}), the specific energy of an annulus in the ISM can be broken down as
\begin{equation}
\label{eq:edisc}
e_{\rm cold}^{(i)} = \langle \Phi^{(i)} \rangle + e_{\rm cold}^{(i)\rm kin}\,.
\end{equation}
No thermal term is added for the ISM; we have assumed that the temperature of the ISM is negligible compared to that of the CGM, such that
\begin{displaymath}
T_{\rm CGM} - T_{\rm cold}  \simeq  T_{\rm vir}\,.
\end{displaymath}
The potential specific energy of an annulus follows equation (\ref{eq:potcold}), while its specific kinetic energy is readily found through contributions from its circular velocity and average vertical motion:
\begin{equation}
e_{\rm cold}^{(i)\rm kin} = \frac{\langle v_{\rm circ}^{(i)} \rangle ^2  + \sigma_{\rm cold}^2(z)}{2}\,.
\end{equation}

While the above provides the necessary equations to calculate $\Delta m_{\rm cold \rightarrow CGM}^{(i) \rm SN}$, we are yet to specify how this feedback-affected gas is handled in detail.
As noted above, we distribute this gas into two transitory reservoirs:~the fountain reservoir and the outflowing reservoir, meaning
\begin{equation}
\label{eq:Delta_mCGM}
\Delta m_{\rm cold \rightarrow CGM}^{(i) \rm SN} = \Delta m_{\rm cold \rightarrow fount}^{(i) \rm SN} + \Delta m_{\rm cold \rightarrow outfl}^{(i) \rm SN}\,.
\end{equation}
The fountain reservoir is defined to have the same specific energy as the hot-gas reservoir:
\begin{equation}
e_{\rm fount} = e_{\rm hot}\,.
\end{equation}
We track a time-scale for which this gas is reincorporated in the hot-gas reservoir, which we describe in Section \ref{ssec:tfount}.
The outflowing reservoir represents gas with energy to escape the (sub)halo's virial radius, and is gradually transferred to the ejected reservoir (see Section \ref{ssec:tejec}).
How, then, do we decide the fraction of $\Delta m_{\rm cold \rightarrow CGM}^{(i) \rm SN}$ that goes to each of the fountain and outflowing reservoirs?

Consider the ideal that any gas kicked from the ISM with a velocity greater than the halo escape velocity,
\begin{equation}
\langle v_{\rm esc}^{(i)}\rangle ^2 = 2 \left[ \Phi(R_{\rm 200c}) - \langle \Phi^{(i)} \rangle \right] \,,
\end{equation}
should be placed in the outflowing reservoir, and that the remainder goes to the fountain reservoir.
For an observer external to the galaxy\,---\,i.e.~in the inertial frame of the halo\,---\,the appropriate velocity to compare against a given $\langle v_{\rm esc}^{(i)}\rangle$ is not simply $v_{\rm wind}^{(i)}$, but rather the vector sum of it and the local rotation velocity (recall that $v_{\rm wind}$ is defined to have already taken care of the fact that the wind slows from heating, which is what allows us to compare it to the escape velocity in the first place).
We assume that the energy from a supernova is emitted with spherical symmetry in the rotating frame of the annulus.
Any gas on the leading side will therefore be kicked with a heightened speed, and conversely that on the trailing side to a lower speed.
One can solve for the angle between these two vectors, where the magnitude of the sum is equal to the escape velocity, through the cosine rule.
The fraction of gas expelled from the ISM with a velocity above the escape velocity of the halo is then equivalent to the fraction of possible angles less than this critical angle.
Recognising that the range of possible angles follows a $\sin(\theta)$ distribution\,---\,where $\theta \! \in \! [0,\uppi]$\,---\,then provides the ratio of ejected gas to reheated gas (after some omitted mathematics): 
\begin{equation}
\label{eq:escfrac}
 \frac{\Delta m_{\rm cold \rightarrow fount}^{(i) \rm SN} }{\Delta m_{\rm cold \rightarrow outfl}^{(i) \rm SN} } 
 = \left( \frac{1}{2} - \frac{ \langle v_{\rm esc}^{(i)}\rangle ^2 - \langle v_{\rm circ}^{(i)} \rangle ^2 - v_{\rm wind}^{(i)~2} }{4\, \langle v_{\rm circ}^{(i)} \rangle \, v_{\rm wind}^{(i)}} \right)^{-1} -1 \,.
\end{equation}
Built into equation (\ref{eq:escfrac}) is the implicit assumption that
\begin{displaymath}
\langle v_{\rm circ}^{(i)} \rangle - v_{\rm wind}^{(i)}  <  \langle v_{\rm esc}^{(i)}\rangle  <  \langle v_{\rm circ}^{(i)} \rangle + v_{\rm wind}^{(i)} \,.
\end{displaymath}
However, there is freedom in the solution for $v_{\rm wind}^{(i)}$ to not obey this restricted range.
If $\langle v_{\rm circ}^{(i)} \rangle + v_{\rm wind}^{(i)}  < \langle v_{\rm esc}^{(i)}\rangle$, then none of the gas kicked out of the ISM is capable of escaping the halo, meaning we set $ \Delta m_{\rm cold \rightarrow outfl}^{(i) \rm SN} \! = \! 0$.
For the inverse reason, if $\langle v_{\rm circ}^{(i)} \rangle - v_{\rm wind}^{(i)}  >  \langle v_{\rm esc}^{(i)}\rangle$, we set $ \Delta m_{\rm cold \rightarrow fount}^{(i) \rm SN} \! = \! 0$.

In our framework, energy conservation dictates that
\begin{multline}
\label{eq:econ}
\Delta E_{\rm SN}^{(i)} = \Delta m_{\rm cold \rightarrow fount}^{(i) \rm SN} \left( e_{\rm fount} - e_{\rm cold}^{(i)} \right) \\
 + \Delta m_{\rm cold \rightarrow outfl}^{(i) \rm SN} \left( e_{\rm outfl}^{(i) \rm SN} - e_{\rm cold}^{(i)} \right) \,.
\end{multline}
The final quantity left to define is $e_{\rm outfl}^{(i) \rm SN}$, which is the specific energy of the material presently being added to the outflow reservoir from a given annulus.
In truth, this mass has a distribution of specific energies, for which we know the minimum and maximum, based on the picture presented above:
\begin{displaymath}
e_{\rm outfl}^{(i) \rm SN} \geq \Phi(R_{\rm 200c}) + e_{\rm hot}^{\rm therm}\,, 
\end{displaymath}
\begin{displaymath}
e_{\rm outfl}^{(i) \rm SN} \leq \langle \Phi^{(i)} \rangle + e_{\rm hot}^{\rm therm} + \frac{1}{2} \left[ \left( \langle v_{\rm circ}^{(i)} \rangle + v_{\rm wind}^{(i)} \right)^2 + \sigma_{\rm gas}^2 \right]\,.
\end{displaymath}
We approximate $e_{\rm outfl}^{(i) \rm SN}$ as the average of these two limits:
\begin{multline}
e_{\rm outfl}^{(i) \rm SN} = e_{\rm hot}^{\rm therm} + \frac{ \Phi(R_{\rm 200c}) + \langle \Phi^{(i)} \rangle }{2} \\
 + \frac{1}{4} \left[ \left( \langle v_{\rm circ}^{(i)} \rangle + v_{\rm wind}^{(i)} \right)^2 + \sigma_{\rm gas}^2 \right]\,.
\end{multline}

Unlike the other reservoirs, the specific energy of the outflow reservoir is updated as mass is added to it:
\begin{multline}
e_{\rm outfl}(t + \Delta t) = {\rm max} \Bigg[ \Phi(R_{\rm 200c}) + e_{\rm hot}^{\rm kin} + e_{\rm hot}^{\rm therm}, \\
\frac{ m_{\rm outfl}(t)\, e_{\rm outfl}(t) + \sum_{i=0}^{N_{\rm ann}} \Delta m_{\rm cold \rightarrow outfl}^{(i) \rm SN}\, e_{\rm outfl}^{(i) \rm SN} }{ m_{\rm outfl}(t) +\sum_{i=0}^{N_{\rm ann}}  \Delta m_{\rm cold \rightarrow outfl}^{(i) \rm SN}} \Bigg]\,.
\end{multline}
The first term inside the `max' function ensures the outflowing energy exceeds the fountaining energy, which it must by definition.
It must at least carry enough energy to reach $R_{\rm 200c}$ and have the equivalent of the thermal and kinetic specific energy of the hot gas.
$e_{\rm outfl}$ is needed in other parts of the model, which we come to later in the paper.

We come back to what happens to the feedback-affected gas next in Section \ref{sec:reinc}.


\subsection{Mass-loading factor}

Our new stellar-feedback model means we now have an entirely predictive framework for determining the mass-loading factors of galaxies experiencing feedback.
The mass-loading factor of a galaxy is defined as the ratio of its gas reheating rate from stellar feedback to its star formation rate:
\begin{equation}
\label{eq:mlf}
\eta \equiv \frac{\dot{m}_{\rm cold \rightarrow fount}^{\rm SN} + \dot{m}_{\rm cold \rightarrow outfl}^{\rm SN}}{{\rm SFR}} \,,
\end{equation}
where this excludes any consideration of gas heating from quasar feedback, but includes all channels of star formation.
Here, $\dot{m}$ and SFR refer to the mass transfer rates and star formation rates averaged over the previous \emph{full} time-step.
This is physically more informative than the most recent sub-time-step, as it is less susceptible to the order in which we resolve galaxy evolution processes (this is why the sub-time-steps are there in the first place).

\ds~is one of few semi-analytic model in the literature to make a true prediction for the mass-loading factor from stellar feedback \citep*[for another example, see][]{lagos13}.
Comparing \ds's mass-loading predictions with observations and other models is not trivial;
definitions of mass-loading factors in the literature are far from uniform.
This is in part because both real and hydrodynamic-simulation galaxies are not so easily broken into discrete reservoirs as SAMs define galaxies and haloes to have.
For the former, it is typically necessary to measure the outflow rate (the numerator in equation \ref{eq:mlf}) at some radius.
Our definition of $\eta$ is effectively measured at a radius of zero.
This means our definition of $\eta$ should be systematically larger than that calculated from observational studies (and hydrodynamic simulations).

\begin{figure*}
\centering
\includegraphics[width=0.85\textwidth]{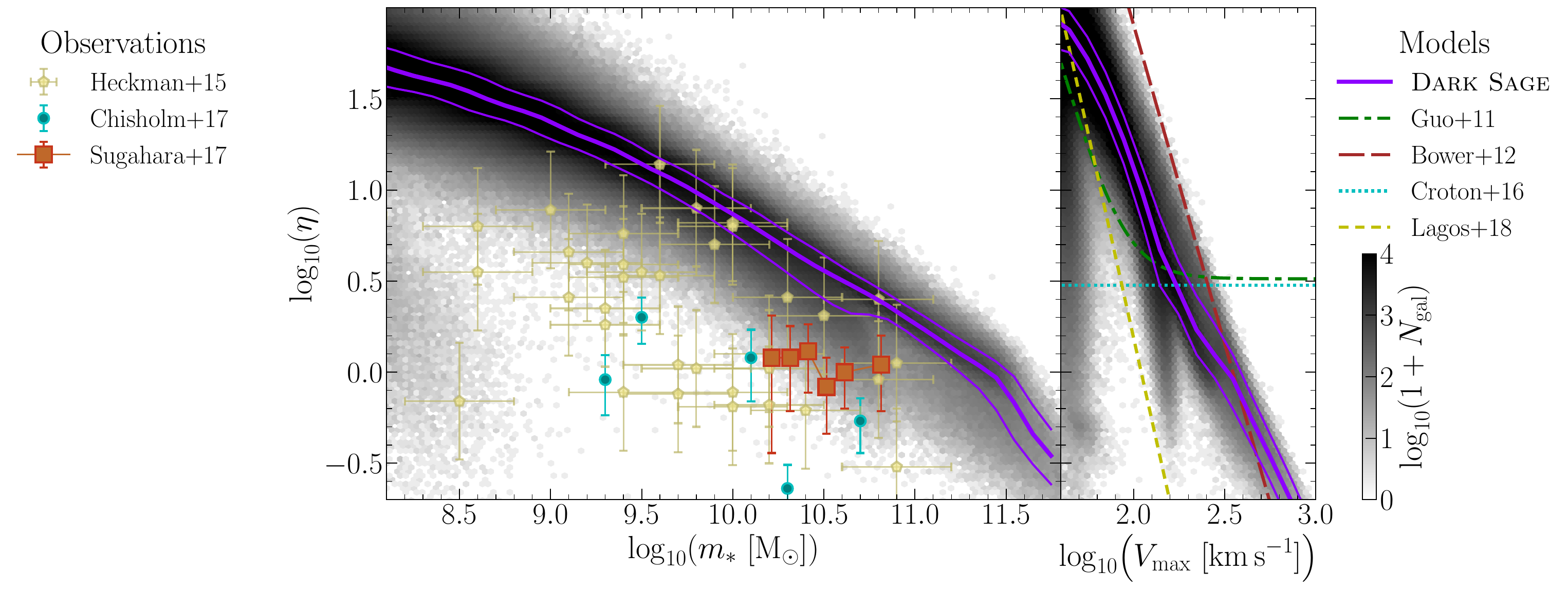}
\vspace{-0.2cm}\caption{Mass-loading factor predicted by \ds~as a function of stellar mass (left) and maximal circular velocity (right-hand panel) for central galaxies with ${\rm sSFR} \! \geq \! 10^{-11}\,{\rm yr}^{-1}$ at \zo.
Greyscale hexbins represent a two-dimensional histogram of \ds~galaxies, where the solid, purple lines follow the running median (thick) and 16th and 84 percentiles (thin).
Compared are observations in the left-hand panel \citep{heckman15,chisholm17,sugahara17}.
Based on the captions of tables 1 and 2 of \citet{heckman15}, we assume an error of 0.3\,dex in stellar mass and $\sqrt{0.2^2 + 0.25^2}$\,dex in $\eta$.
Various other models are compared in the right-hand panel \citep*[including][]{guo11,bower12,croton16,shark}.
Note that $V_{\rm max}$ is from {\sc subfind}, which is different to what we derive from the rotation curves we build; this is the fairest quantity to compare to other semi-analytic models.
}
\label{fig:massloading}
\end{figure*}

While we make a face-value comparison with some observational data from the literature in the left panel of Fig.~\ref{fig:massloading}, for the above reason and others (see discussion in \citealt{nelson19} on this topic), we caution against over-interpreting the apparent offset \ds~has with the data.
Points for individual galaxies at $z\!\simeq\!0$ are shown from \citet{chisholm17}, as they are for $z\!<\!0.2$ galaxies from \citet{heckman15},
while the points from \citet{sugahara17} are bins of galaxies from their $z\!\simeq\!0$ sample.


In the right-hand panel of Fig.~\ref{fig:massloading}, we compare \ds's predicted mass-loading factors as a function of maximum halo circular velocity against the \emph{input} functional mass-loading factors of other semi-analytic models.%
\footnote{Sometimes, semi-analytic models will invoke a mass-loading factor as a function of virial velocity rather than maximum circular velocity.
This might change the preciseness of a comparison with \ds, but it does not change our overall interpretation.}
The assumed functions used in \citet{bower12} and \citet{shark} are similar in slope to what \ds~predicts, but systematically higher and lower, respectively.
Meanwhile, the asymptotic function of \citet{guo11} and the constant used by \citet{croton16} appear both qualitatively and quantitatively inconsistent with our energy-conserving framework.
The comparison with \citet{shark} is particularly interesting, as their fitting function was based on the predictive model of \citet{lagos13}.
There are \emph{many} differences between \ds~and the \citet{lagos13} version of {\sc galform}, placing an elaborative discussion and comparison between the models outside the scope of this paper.

The slope and scatter of $\eta$ as a function of either $m_*$ or $V_{\rm max}$ are the outcome of our complex network of equations in \ds, although they ultimately boil down to the diversity of gravitational potentials at fixed mass and the increase in average potential depth at higher masses.
The normalisation of $\eta$ in \ds~is effectively set by the value of $\mathcal{E}_{\rm SN}$.
As we show in Section \ref{sec:cali}, this is calibrated to other observations in a way that is blind to Fig.~\ref{fig:massloading}.


\subsection{Metal loading}

In a departure from previous versions of \ds, we do \emph{not} assume that gas reheated from the ISM carries the same metallicity as the ISM.
We imagine a scenario where the energy from a stellar-feedback event expands with spherical symmetry (in the ISM's frame) into an ISM where all atoms (metals and non-metals) are distributed randomly and are isothermal.
If the cross section of each atom were the same, we would expect the same amount of energy to be imparted on each atom on average.
While heavier elements may have a greater cross section for interacting with an outflow when stationary, this is offset by the lighter elements having higher thermal motion at fixed temperature.
In the absence of a detailed model for these competing effects, we approximate the energy imparted on each atom to be equal.
Because we have implicitly assumed elsewhere that anything reheated carries a common \emph{specific} energy, this approximation must mean some metals are left behind.
As oxygen is the most abundant metal in the Universe, and carbon the second-most abundant, the average atomic mass of a metal should be a little less than $16\,m_{\rm H}$.
Non-metals (hydrogen and helium) should have an average atomic mass of around $7\,m_{\rm H}/4$.
This implies that for every unit mass of non-metals reheated from an annulus, we should see $\sim$$Z^{(i)}_{\rm cold}/9$ times that mass of metals reheated.
Or, in other words, the metallicity of the locally produced outflow is one tenth that of the local ISM.
This is what we enforce.

One potential issue with our treatment of metals in stellar-feedback outflows is that some observations find evidence that the metallicity of outflows is typically \emph{elevated} relative to the ISM \citep*[e.g.][]{chisholm18}, which is the opposite of what we have imposed.
However, there is once again more nuance than this in comparing our model to observations.
As we explain more in Section \ref{sec:reinc}, the outflow that we follow from a galaxy in \ds~represents that gas that will eventually either redistribute itself within the CGM, with the same energy as the CGM, and/or be ejected from the halo.
We essentially consider all the material in an outflow to carry one of two possible energy states.
This means we lose the information of an energy or velocity \emph{distribution} in the outflow.
In practice, we should expect the low-energy portion of that distribution to never fully reincorporate into the quasi-hydrostatic CGM, but instead to return to the ISM whence it came on a relatively short time-scale.
One might also hypothesise that an outflow has a metallicity distribution, which is anti-correlated with the velocity distribution, by virtue of metals being heavier.
It is therefore conceivable that the metallicity of an outflow observed close to the disc of a galaxy might differ significantly from the metallicity of the portion of that outflow that ends up mixing with the CGM.
Moreover, the statistics on the metallicity of outflows are weak, and the measurement uncertainties are often dominated by systematics. 

\begin{figure}[H]
\centering
\includegraphics[width=\textwidth]{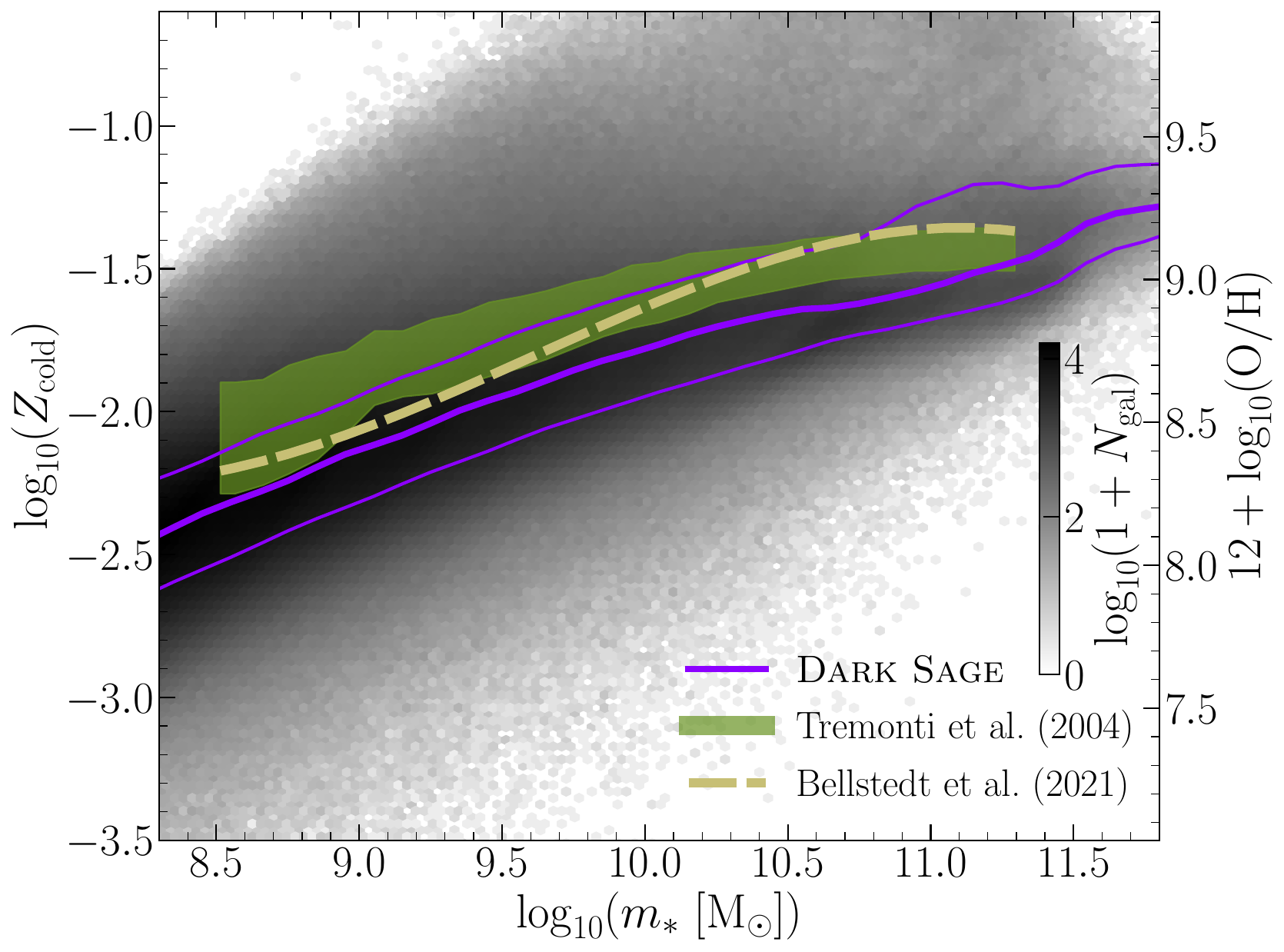}
\vspace{-0.2cm}\caption{
The relation between stellar mass and ISM metallicity for galaxies at \zo.
We only include \ds~galaxies with $m_{\rm neutral} \! \geq \! 10^8\,{\rm M}_{\odot}$ to ensure there is appreciable gas present for a meaningful gas metallicity.
\ds~data are plotted in the same way as Fig.~\ref{fig:massloading}.
Compared is the observed relation from \citet{tremonti04} at $z \! \simeq \! 0.1$, covering their 16--84th interpercentile range, and the running median from \citet{bellstedt21} at $z\!=\!0.07$.
}
\label{fig:MZR}
\end{figure}

One observed relation that is much more enshrined in canon is that between the stellar mass and gas metallicity of galaxies.
In previous versions of the model, this relation informed the calibration of the model's free parameters.
As we discuss in Section \ref{sec:cali}, our calibration method has changed significantly, and the mass--metallicity relation of galaxies is no longer used.
This means it is now a prediction of the model.
We show this relation for \ds~in Fig.~\ref{fig:MZR}.
This is compared to the observation-based relations of \citet{tremonti04} and \citet{bellstedt21}.

We note that there are hidden systematic uncertainties in the compared mass--metallicity relations in Fig.~\ref{fig:MZR}, which should not make the apparent systematic offset between them and \ds~a point of concern.
For \citet{bellstedt21}, assumptions in their method of modelling galaxies' spectral energy distributions are a natural source of systematic uncertainty, such as their enforced proportionality in the evolution of $Z_{\rm gas}$ with stellar-mass growth.
There is also $\sim$1\,Gyr more of cosmic evolution from their $z \! = \! 0.07$ result to our \ds~result at \zo.
For the comparison with \citet{tremonti04}, whose data represent $z \! \simeq \! 0.1$, the largest sources of systematic uncertainty are from the emission-line diagnostics used to measure the oxygen abundance \citep[see][]{kewley08} and the conversion from oxygen abundance to metallicity.
In our comparison, we assume
\begin{equation}
12 + \log_{10}({\rm O} / {\rm H}) = 8.69 + \log_{10}(Z_{\rm gas} / 0.0142)\,, 
\end{equation}
where 0.0142 is the estimated metallicity of the Sun \citep{asplund09}
(we have corrected the \citealt{tremonti04} relation for a \citealt{chabrier03} IMF and our cosmology).
For similar reasons, in their comparison of the mass--metallicity relation of the IllustrisTNG simulation with observations, \citet{torrey19} ignore the normalisation of the \citet{tremonti04} relation, moving it down in their fig.~6.
While we could have taken the same approach in our Fig.~\ref{fig:MZR}, we opted not to for the sake of fidelity.

The change to how we model metal loading in \ds~turned out to be crucial to the outcome shown in Fig.~\ref{fig:MZR}.
That is, had the outflows from the ISM carried higher metallicity, $Z_{\rm cold}$ would have been systematically lower in \ds~galaxies.
While we outlined above why \emph{some} systematic offset between \ds~and observations \emph{as presented in Fig.~\ref{fig:MZR}} is expected\,---\,in this case, an apparent factor of $<$1.4\,---\,larger offsets naturally become increasingly harder to waive away.

\begin{figure}[H]
\centering
\includegraphics[width=\textwidth]{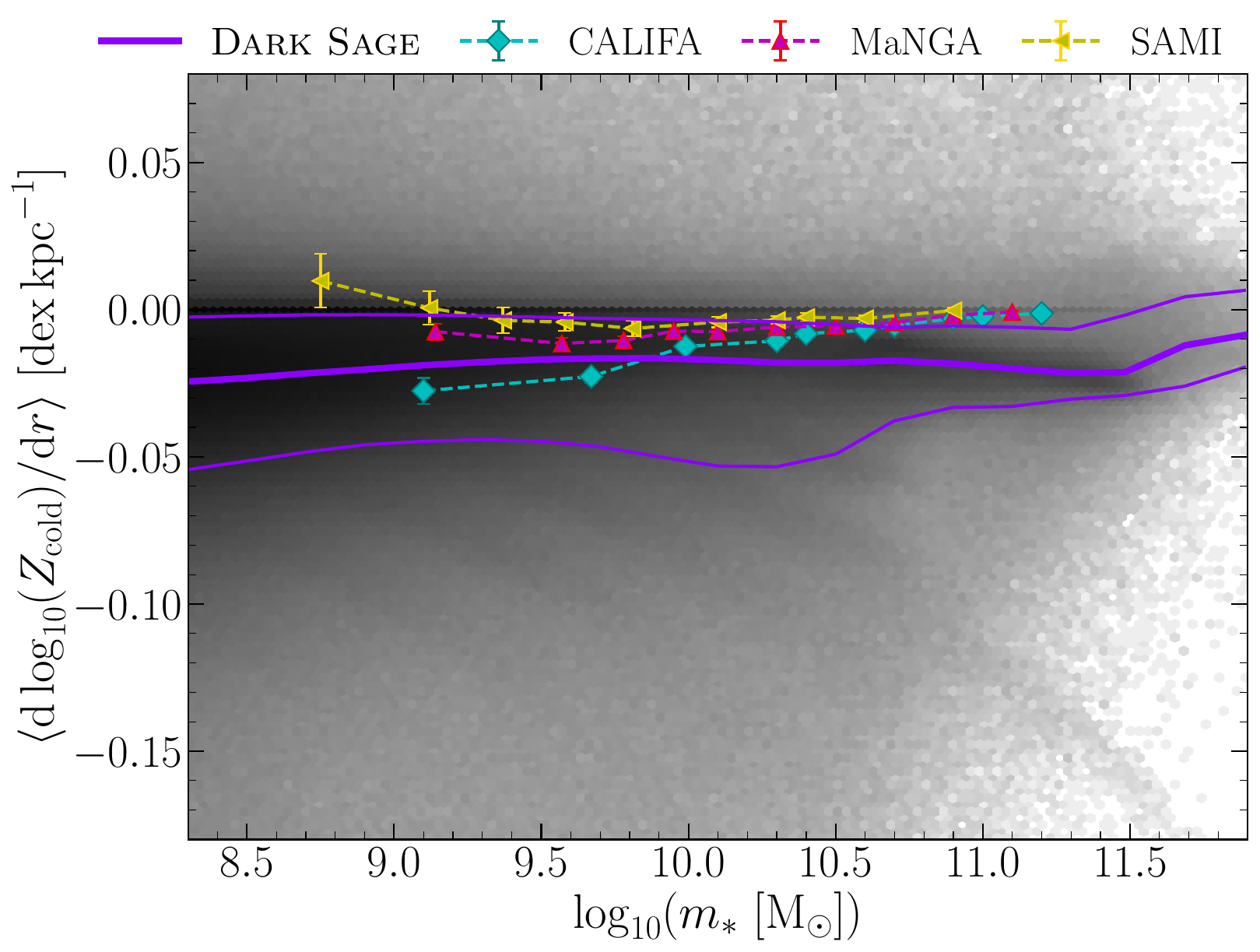}
\vspace{-0.2cm}\caption{
Radius-weighted average of the metallicity gradient in gas discs in \ds~galaxies at \zo.
Data for \ds~are presented in the same way as in previous figures.
Dotted lines with error bars represent the average relations from three IFS surveys:~the `Calar Alto Legacy Integral Field spectroscopy Area' survey (CALIFA; \citealt{sanchez14}), the `Mapping nearby Galaxies at Apache Point Observatory' survey (MaNGA; \citealt{belfiore17}), and the `Sydney-AAO Multi-object Integral-field spectrograph' survey (SAMI; \citealt{poetro21}), which have been homogenized (\citealt{sharda21}; Acharyya et al. in prep.).
}
\label{fig:Zgrad}
\end{figure}

With the resolved detail of \ds~discs, we can also predict the gas metallicity \emph{gradients} of galaxies.
This is a rare quality for a semi-analytic models, which few others have \citep[as another example, see][]{yates21}.
As a first prediction from the new \ds, we show the radius-weighted \emph{average} metallicity gradient of \ds~galaxies as a function of stellar mass in Fig.~\ref{fig:Zgrad}, where

\begin{multline}
\left\langle \frac{ {\rm d} \log_{10} (Z_{\rm cold})}{ {\rm d} r} \right\rangle \equiv \left[ \sum_{i=1,\, m_{\rm cold}^{(i)} \neq 0}^{N_{\rm ann}-1} \log_{10}\!\left(Z_{\rm cold}^{(i)}\right) \right] \times \\
\left[ \sum_{i=1,\, m_{\rm cold}^{(i)} \neq 0}^{N_{\rm ann}-1} \left(\langle r^{(i+1)} \rangle - \langle r^{i} \rangle \right) \right]^{-1}\,.
\end{multline}
We choose to explicitly weight by the radial range of each annulus, as observational gradients are typically \emph{implicitly} weighted in the same way; i.e.~data will be binned by radius and a gradient fitted to those bins.
Such observations from integral-field spectroscopy (IFS) surveys are compared in
Fig.~\ref{fig:Zgrad} (to appear in Acharyya et al.~in prep.),%
\footnote{The same data were meant to be presented in fig.~2 of \citet{sharda21}, but the points shown in the panel with gradients in units dex\,kpc$^{-1}$ were actually from a file in units of dex per effective radius (Acharyya, private communication).}
though we caution that this is still a cursory comparison, and should therefore not be interpreted as a means of falsifying \ds.
While the vast majority of \ds~galaxies have negative metallicity gradients (which is the expected norm), \ds~is also able to produce galaxies with positive metallicity gradients.
The relation between gas metallicity gradient and stellar mass is relatively flat in \ds, as it is in observations.
But this is dependent on how the metallicity gradient is calculated.
Though not shown here, reproducing Fig.~\ref{fig:Zgrad} with a \emph{mass}-weighted average gradient across the annuli in a \ds~disc results in an increasingly steep relation towards low stellar masses.
Metallicity gradients normalised by the stellar-disc half-mass radius look very similar (but for a re-scaled $y$-axis).

\noindent


\subsection{Direct impartation of energy on the CGM}
\label{sec:fb_CGM}

The instability-driven bulge, merger-driven bulge, and intrahalo stars (collectively `spheroid stars') can produce feedback.
While no stars form in spheroids in \ds, the delayed-feedback scheme means that spheroid stars still produce Type-Ia supernovae.
We assume that spheroid supernovae cannot reheat gas from the disc.
Instead, the metals and energy they produce are imparted directly onto the CGM.
Summing over all stellar-age bins, the energy from spheroid supernovae, $\Delta E_{\rm SN}^{\rm sph}$, is balanced by generating an outflow from the CGM. 
In practice, this means transferring mass from the hot and fountain reservoirs to the outflowing reservoir.
Because we have defined $e_{\rm fount} \! = \! e_{\rm hot}$, we transfer mass \emph{proportionally} out of the hot and fountain reservoirs, assuming this gas carries the same specific energy as that pre-computed for the outflowing reservoir:
\begin{subequations}
\begin{equation}
\Delta m_{\rm hot \rightarrow outfl}^{\rm SN} = {\rm min} \! \left[m_{\rm hot}, ~\frac{\Delta E_{\rm SN}^{\rm sph}}{e_{\rm outfl} - e_{\rm hot}} \frac{m_{\rm hot}}{m_{\rm hot} + m_{\rm fount}} \right]\,,
\end{equation}
\begin{equation}
\Delta m_{\rm fount \rightarrow outfl}^{\rm SN} = {\rm min} \! \left[m_{\rm fount}, ~\frac{\Delta E_{\rm SN}^{\rm sph}}{e_{\rm outfl} - e_{\rm hot}} \frac{m_{\rm fount}}{m_{\rm hot} + m_{\rm fount}} \right]\,.
\end{equation}
\end{subequations}


\subsection{Comparative remarks}

Although vastly different in how the feedback is \emph{applied}, a clear parallel can be drawn between our model for delayed stellar feedback and that in the {\sc meraxes} semi-analytic model \citep{mutch16}.
This is not entirely by coincidence; part of the inspiration for our model design came from \citet{mutch16}.
Similar designs of delayed feedback and chemical enrichment are also incorporated into the {\sc galacticus} model \citep{benson12} and in \citet{delucia14}, although again different in implementation to \ds.
We refer the reader to the other papers for more context.

In previous incarnations of SAGE and \ds, the returning of mass from stars to the ISM and the production of new metals was done \emph{after} reheating gas from feedback.
Implicitly, this meant it was guaranteed that some (local, metal-rich) cold gas would forcibly remain after an episode of star formation.
We have deliberately changed this.
Now, mass is returned and metals added to the ISM \emph{before} we apply feedback to the ISM.
Provided there is sufficient energy, this means that returned gas can immediately find its way to the CGM. 

We explored the possibility of a distributive-feedback model in \ds, where the energy from supernovae in one annulus could affect the gas in adjacent annuli.
Our default assumption was that energy would be distributed from an annulus centre isotropically.
For infinitesimally wide annuli, one could imagine that most of the energy from feedback should be directed towards neighbouring annuli instead of out of the disc.
However, with our set-up in \ds, it is almost always true that the width of a disc annulus is much greater than the disc scale height, implying a negligible fraction of feedback energy should transfer through the disc.
As such, in the end, we reasonably approximated the feedback taking place in an annulus to be independent from other annuli in a galaxy.


\section{Active galactic nuclei}
\label{sec:agn}

We maintain a two-mode model for feedback from active galactic nuclei (AGNs) in \ds, but have changed the way both modes operate.


\subsection{Radio mode}

In earlier versions of \ds, the model for radio feedback followed \citet{croton16}.
In this mode, hot gas is passively accreted directly onto the black hole, based on the black hole's \citet{bondi52} accretion rate. 
Combined with the `maximal cooling flow' model of \citet{nulsen00}, the accretion rate can be derived as
\begin{equation}
\label{eq:bondi}
\dot{m}_{\rm BH}^{\rm radio} = \frac{15 \uppi \, G \, \bar{\mu} m_{\rm p}\, k_{\rm B} \, T_{\rm vir}}{16\, \Lambda(T_{\rm vir}, Z_{\rm hot})} m_{\rm BH}
\end{equation}
\citep[cf.][]{croton06}, where $m_{\rm BH}$ is the black hole's mass.  
In \citet{croton16}, an additional free parameter was added to equation (\ref{eq:bondi}).
However, with the changes we have made to the model, we no longer need this.
We note that we explicitly do not allow $\dot{m}_{\rm BH}^{\rm radio}$ to exceed the black hole's \citet{eddington26} accretion rate in the code, but it is typically orders of magnitude below this anyway.

A half-century-old standard, we assume that mass accreting onto a black hole quasi-statically moves through equatorial, circular orbits\,---\,all the while losing energy and angular momentum\,---\,until reaching the orbit of lowest energy, where it is captured \citep*[see][]{lynden69,bardeen70,bardeen72,novikov73,shakura73,stevens15}.
The power emanating from the accretion disc\,---\,which drives an AGN\,---\,is then given directly by the accretion rate:  
\begin{equation}
\label{eq:agn}
\dot{E}_{\rm AGN} = \epsilon_{\rm AGN}\, \dot{m}_{\rm BH}\, c^2\,,
\end{equation}
where $c$ is the speed of light in a vacuum and $\epsilon_{\rm AGN}$ is the ratio of inertial (gravitational) mass lost by that accreted to its rest mass (at infinity), often referred to as the `radiative efficiency.'
Note that $\dot{m}_{\rm BH}$ represents the \emph{rest}-mass accretion rate of the black hole, but $m_{\rm BH}$ represents the \emph{inertial} mass of the black hole.
Over a sub-time-step, the growth of a black hole is therefore
\begin{equation}
\Delta m_{\rm BH} = (1 - \epsilon_{\rm AGN})\, \dot{m}_{\rm BH}\, \Delta t \,.
\end{equation}
The difference between rest mass and inertial mass was not properly accounted for in previous versions of the model.
This means black-hole feedback is now even more self-regulatory.
That is, for a fixed rest mass of accreted material, a black hole can either grow by a lot and produce a weak AGN (low $\epsilon_{\rm AGN}$), or grow less and produce a stronger AGN (high $\epsilon_{\rm AGN}$).
But the less the black hole grows, the weaker any future radio-mode feedback will be. 
In principle, to conserve inertial mass (energy), that lost by the material accreted onto the black hole should be added somewhere.
Without doing so, the baryonic mass of the halo within which that black hole resides will drop.
In practice though, this is inconsequential not only because this comprises a negligible fraction of a halo's baryonic mass, but also because it is automatically accounted for by \ds's infall prescription anyway.
At the next time-step, 
baryonic mass will be added to the CGM of that halo to top it back up to the cosmic baryon fraction (modulo any suppression from photoionization feedback, and including ejected gas in the accounting\,---\,see Section \ref{sec:cooling}).
We implicitly approximate rest mass and inertial mass as equivalent for all other components of galaxies and haloes (as literally all cosmological simulations and semi-analytic models do), which we do not expect to have any adverse effects whatsoever.

Under the above picture, the value of $\epsilon_{\rm AGN}$ depends exclusively on the spin of the black hole (neglecting charge) and the relative orientation of the accretion disc to the spin vector \citep{bardeen72}. While $\epsilon_{\rm AGN} \! = \! 0.0572$ for a non-rotating \citet{schwarzschild16} black hole, $\epsilon_{\rm AGN}$ can theoretically range from 0.0378 to 0.3210, given the anti-parallel and parallel cases for a \citet{kerr63} black hole rotating at the \citet{thorne74} limit, and can reach 0.4226 if the Thorne limit is breached.
However, we do not explicitly consider black-hole spin in \ds~(for an example of a semi-analytic model that does, see \citealt{fanidakis11}).
The addition of such a feature would be an entire project in and of itself.  In the meantime, we opt to treat $\epsilon_{\rm AGN}$ as a free parameter that is applied uniformly for all black holes.

Equation (\ref{eq:agn}) applies to both modes of accretion, but the way that that energy is handled depends on the mode.
For the radio mode, the energy is used as a preventative form of feedback.
This has been standard for semi-analytic models since \citet{croton06} and \citet{bower06}.
After the gross cooling rate of a galaxy is calculated (Section \ref{sec:cooling}), we calculate how much of that cooling would be prevented by the energy released from the black-hole accretion disc:
\begin{subequations}
\begin{equation}
\Delta m_{\rm heat}^{\rm radio} = \frac{\dot{E}_{\rm AGN}^{\rm radio}\, \Delta t}{e_{\rm hot} - \langle e_{\rm cold} \rangle}
 = \frac{\Delta E_{\rm AGN}^{\rm radio}}{e_{\rm hot} - \langle e_{\rm cold} \rangle} \,,
\end{equation}
\begin{equation}
\langle e_{\rm cold} \rangle \equiv \sum_{i=1}^{N_{\rm ann}} e_{\rm cold}^{(i)} \int_{j_{\rm inner}^{(i)}}^{j_{\rm outer}^{(i)}} {\rm PDF}_{\rm cool}(j)\, {\rm d} j\,,
\end{equation}
\end{subequations}
providing the final pieces to equations (\ref{eq:coldmode}--\ref{eq:hotmode}).

The term `radio mode' is derived from the presence of radio jets that are generated from significantly sub-Eddington rates of black-hole accretion.
These pervade the CGM and are what keeps the gas hot, or, rather, prevents it from cooling.
While we have no \emph{explicit} consideration of AGN jets in \ds~\citep[but see][where there is an explicit consideration of this in the {\sc Radio}-SAGE model]{raouf17}, the radio mode still implicitly relies on there being jets.
The influence of these jets persists over extended time-scales.
For this reason, AGN memory in a semi-analytic model is important; without it, a model can be vulnerable to spurious, non-physical increases and decreases in radio-mode heating.

In \citet{croton16}, radio-mode memory was achieved by introducing a `heating radius,' which never decreased with time and effectively prevented gas in the CGM internal to it from cooling.
This radius would increase when periods of radio AGN activity exceeded what the galaxy had had previously.
This model was adopted in previous versions of \ds, and has been adopted in other semi-analytic models too \citep[e.g.][]{shark}.

For the new \ds, we achieve radio-mode memory differently.  
We abandon any consideration of a heating radius and instead move mass from the hot reservoir to the fountain reservoir equal to that offset from cooling:
\begin{equation}
\Delta m_{\rm hot \rightarrow fount}^{\rm radio} = \Delta m_{\rm heat}^{\rm radio}\,.
\end{equation}
This does not affect energy conservation, as the fountain and hot reservoirs are defined to have the same specific energy.
But this \emph{does} create a lasting impact by effectively disabling the AGN-affected gas from cooling for a dynamical time.
This is not only less contrived than the \citet{croton16} heating-radius model, but it also allows the suppression of cooling from radio-mode feedback in a galaxy to wax and wane moderately over time, rather than effectively forcing it to stay the same or increase.

Provided $m_{\rm BH} \! > \! 0$ and  $m_{\rm hot} \! > \! 0$, radio-mode accretion and feedback will occur in a \ds~galaxy at every sub-time-step.


\subsection{Quasar mode}

Whereas the radio mode is coupled to the CGM and is primarily responsible for the quenching of high-mass galaxies, the quasar mode is coupled to the ISM and is the primary growth channel for black holes.
In earlier versions of SAGE and \ds, quasar-mode accretion and feedback were triggered in two instances: from disc instabilities and when galaxy mergers occurred.
Instability-driven accretion in \ds~requires gas to first flow into the innermost annulus; any unstable gas there that is not consumed/removed by a starburst and its associated feedback (Section \ref{ssec:gas_instab}) is accreted onto the central black hole.
The previous treatment of merger-driven black-hole accretion in \ds~was \emph{ad hoc} and not well motivated theoretically; its phenomenological design \citep[see][]{kauffmann00,croton16} was better suited for semi-analytic models that only treat \emph{global} galaxy properties.
We have now abandoned this channel of quasar-mode accretion entirely, leaving only the instability channel.
Though, this is not to say that mergers no longer affect when black holes grow.
The additional gas placed in the relevant annuli after a merger inevitably drives an instability that typically cascades all the way into the galaxy centre.
These merger-induced instabilities play a major role in black-hole growth.
Effectively, all we have done is retire a redundant prescription; this has no adverse consequences on the model's predictions and it means the model has fewer free parameters.

$\dot{m}_{\rm BH}^{\rm quasar}$ is set by the amount of unstable gas in the innermost annulus of a galaxy and one free parameter that controls what fraction of this gas is consumed in a starburst ($f_{\rm move}^{\rm gas}$, see Section \ref{ssec:gas_instab}).
Unlike the radio mode, there is nothing to explicitly prevent quasar-mode accretion from exceeding the \citet{eddington26} rate.
Equation (\ref{eq:agn}) then gives the power that drives quasar-mode feedback.

Quasar-mode feedback is ejective, not preventative.
The energy is first dumped into the innermost annulus of the ISM, working outward until all the energy output over a given sub-time-step is used.
Quasar feedback on the ISM differs from stellar feedback in that all feedback-affected gas is sent to the outflowing reservoir, with none going to the fountain reservoir.
The physical justification for these choices is that the gas nearest the black hole should be maximally affected before energy from it starts coupling to the outer ISM.
Using the pre-computed specific energy of the outflowing reservoir (Section \ref{sec:fb_ISM}), we effect
\begin{subequations}
\begin{equation}
\Delta m_{\rm cold \rightarrow outfl}^{(i) \rm quasar} = {\rm min} \left[m_{\rm cold}^{(i)}, ~\frac{\Delta E_{\rm left}^{(i)}}{e_{\rm outfl} - e_{\rm cold}^{(i)}} \right]\,,
\end{equation}
\begin{equation}
\Delta E_{\rm left}^{(i)} = 
\left\{
\begin{array}{l r}
\Delta E_{\rm AGN}^{\rm quasar}\,, & i=1\\
\Delta E_{\rm AGN}^{\rm quasar} - \sum_{k=1}^{i-1} \Delta E_{\rm used}^{(k)}\,, & i \geq 2 \\
\end{array}
\right.\,,
\end{equation}
\begin{equation}
\Delta E_{\rm used}^{(k)} = \Delta m_{\rm cold \rightarrow outfl}^{(k) \rm quasar} \left(e_{\rm outfl} - e_{\rm cold}^{(k)}\right)  \,.
\end{equation}
\end{subequations}
This equation is solved for each annulus $i$ until either reaching $i \! = \! N_{\rm ann}$ or $\Delta E_{\rm left}^{(i)} \! = \! 0$.
Whenever the former is the trigger to stop, there must still be some non-zero excess energy
\begin{equation}
\Delta E_{\rm excess}^{\rm quasar} = \Delta E_{\rm AGN}^{\rm quasar} - \sum_{i=1}^{N_{\rm ann}} \Delta E_{\rm used}^{(i)}\,.
\end{equation}
Identical to our treatment of excess energy from stellar feedback, we use as much of this energy as possible to eject gas from the hot reservoir:
\begin{subequations}
\begin{equation}
\Delta m_{\rm hot \rightarrow outfl}^{\rm quasar} = {\rm min} \! \left[m_{\rm hot}, ~\frac{\Delta E_{\rm excess}^{\rm quasar}}{e_{\rm outfl} - e_{\rm hot}} \frac{m_{\rm hot}}{m_{\rm hot} + m_{\rm fount}} \right]\,,
\end{equation}
\begin{equation}
\Delta m_{\rm fount \rightarrow outfl}^{\rm quasar} = {\rm min} \! \left[m_{\rm fount}, ~\frac{\Delta E_{\rm excess}^{\rm quasar}}{e_{\rm outfl} - e_{\rm hot}} \frac{m_{\rm fount}}{m_{\rm hot} + m_{\rm fount}} \right]\,.
\end{equation}
\end{subequations}

The main differences from the \citet{stevens16} implementation of quasar-mode AGN feedback is our far more comprehensive consideration of the specific-energy differences between reservoirs, the allowance for an annulus to have its gas partially ejected (as opposed to it being `all or nothing'), and the time-scale on which this gas reincorporates back into the hot reservoir (described in the next section).


\section{The galactic fountain and \newline gas reincorporation}
\label{sec:reinc}

We introduce a new treatment for the galactic fountain, which affects how gas is reincorporated into the hot-gas reservoir to become available for cooling again after being heated by feedback.
Our method is new not only for \ds~but for semi-analytic models in general.

Early in their development, semi-analytic models assumed the reincorporation time-scale to scale linearly with the halo dynamical time (equation \ref{eq:tdyn}) \citep{springel01}, with the proportionality constant often treated as a free parameter \citep[per][]{delucia04}.
Motivated by the difficulty that this simple model posed in recovering observations at multiple epochs \citep*{henriques13,mutch13}, it later became commonplace for reincorporation time-scales to have a direct dependence on virial mass or velocity \citep*[with various functional forms\,---\,e.g.][]{henriques13,white15,croton16}.
However, none of these approaches are in line with our philosophy of having a self-consistent feedback framework that minimises the number of free parameters.
Instead, we introduce a framework where we update time-scales for mass transfer between the CGM and ejected reservoir self-consistently with the energetics of feedback.


\subsection{Fountaining in the CGM}
\label{ssec:tfount}

Gas in the fountain reservoir is intended to reincorporate into the hot-gas reservoir to become available for cooling again.
We must decide the time-scale that that gas remains in the transitory fountain reservoir. 
In the absence of a detailed consideration of how efficiently gas mixes (beyond the scope of this version of \ds), we make a simple assumption that gas takes a halo dynamical time to transit the fountain reservoir.
We track an average `fountain time' whenever mass is added to the reservoir:
\begin{multline}
t_{\rm fount}(t + \Delta t) = {\rm max} \bigg[0, \\
\frac{m_{\rm fount}(t)\, t_{\rm fount}(t) + \Delta m_{\rm \Sigma \rightarrow fount}\, t_{\rm dyn}}{m_{\rm fount}(t) + \Delta m_{\rm 
\Sigma \rightarrow fount}}- \Delta t \bigg]\,,
\end{multline}
where $\Delta m_{\rm \Sigma \rightarrow fount}$ is the sum of contributions to the fountain reservoir for the sub-time-step.

For each sub-time-step, we move a fraction of the fountaining mass to the hot reservoir in proportion to the fraction of $t_{\rm fount}$ that has elapsed:
\begin{equation}
\Delta m_{\rm fount \rightarrow hot} = {\rm min} \! \left[1,~\frac{\Delta t}{t_{\rm fount}}\right]\, m_{\rm fount}\,.
\end{equation}


\subsection{Ejection of outflowing gas}
\label{ssec:tejec}

Gas is in the outflowing reservoir is, by definition, eventually going to escape the halo.
Again, we need a time-scale to decide how long gas added to the outflowing reservoir will remain in it before being transferred to the ejected reservoir.
We estimate the instantaneous outflow time-scale based on the specific energy of the gas added to the reservoir at that time.
If we assume the thermal energy of the outflowing gas remains constant, we can approximate the average velocity of outflowing gas travelling as that which balances energy at the halo centre and at the virial radius:
\begin{subequations}
\begin{equation}
t_{\rm outfl}^{\rm inst} = \frac{2\, R_{\rm 200c}}{v^{\rm inst}_{\rm outfl}(0) + v_{\rm outfl}^{\rm inst}(R_{\rm 200c})}\,,
\end{equation}
\begin{equation}
v_{\rm outfl}^{\rm inst}(R) \equiv \sqrt{2\, \left(e_{\rm outfl}^{\rm inst} - \Phi(R) - e_{\rm hot}^{\rm therm}\right)}\,.
\end{equation}
\end{subequations}
When mass is added to the outflow reservoir from supernova feedback,
\begin{displaymath}
e_{\rm outfl}^{\rm inst} = e_{\rm outfl}^{\rm SN}\,.
\end{displaymath}
Otherwise
\begin{displaymath}
e_{\rm outfl}^{\rm inst} = e_{\rm outfl}\,.
\end{displaymath}

In the same manner as the fountain reservoir, we track an average outflow time as mass is continually added to the outflowing reservoir, ensuring to reduce this as cosmic time elapses:
\begin{multline}
t_{\rm outfl}(t + \Delta t) = {\rm max} \bigg[0, \\
\frac{m_{\rm outfl}(t)\, t_{\rm outfl}(t) + \Delta m_{\rm \Sigma \rightarrow outfl}\, t_{\rm outfl}^{\rm inst}}{m_{\rm outfl}(t) + \Delta m_{\rm \Sigma \rightarrow outfl}} - \Delta t \bigg]\,,
\end{multline}
where $\Delta m_{\rm \Sigma \rightarrow outfl}$ represents the additions to the outflowing reservoir from all sources for the sub-time-step.
For central galaxies and satellites outside their host's virial radius, mass is added to their ejected reservoir according to:
\begin{equation}
\label{eq:out_ejec}
\Delta m_{\rm outfl \rightarrow ejec} = {\rm min} \! \left[1,~\frac{\Delta t}{t_{\rm outfl}}\right]\, m_{\rm outfl}\,.
\end{equation}
Satellites inside the virial radius of their host do not have an ejected reservoir by construction.
For these galaxies, the mass on the left-hand side of equation (\ref{eq:out_ejec}) is instead placed in the fountain reservoir of the central (and its fountain time-scale is updated accordingly). 


\subsection{Effect of halo growth}

As a halo grows in mass, the depth of its potential well and thermal energy both increase.
This can cause outflowing gas that previously had enough energy to escape the halo to no longer have enough.
Whenever a (sub)halo increases in mass, we first check whether \emph{any} of the outflowing gas can still reach the virial radius.
This requires
\begin{displaymath}
\Phi(R_{\rm 200c}, t) + e_{\rm hot}^{\rm therm}(t) > e_{\rm outfl}(t - \Delta t)\,.
\end{displaymath}
If this inequality is not satisfied, all gas in the outflowing reservoir is transferred to the fountain reservoir.
Otherwise, we compute the average specific energy required for the outflowing gas to maintain its outflow time-scale.

While we only explicitly track the \emph{average} outflow time-scale and energy of gas in the CGM, one would physically expect the outflowing gas to have a \emph{distribution} of energy.
The lower-energy gas in this distribution must be that which is moved into the fountain reservoir as the result of halo growth.
This should, therefore, \emph{raise} the specific energy of the material that remains in the outflowing reservoir.
In principle, this should reduce the outflow time.
However, as the halo grows, the assumed temperature of the remaining outflowing gas rises\,---\,which means it gets slower, if we assume its total energy is conserved\,---\,and the distance it must travel to reach $R_{\rm 200c}$ also increases.
Both of these effects should raise the outflow time.
Because there are competing effects that should both raise and lower $t_{\rm outfl}$, we take the agnostic stance of keeping the outflow time constant, and then use energy conservation to calculate the fraction of the outflowing gas that should be moved to the fountain reservoir:
\begin{subequations}
\begin{multline}
\Delta m_{\rm outfl \rightarrow fount}\, e_{\rm fount}(t) + m_{\rm outfl}(t)\, e_{\rm outfl}(t)\\
 = m_{\rm outfl}(t - \Delta t)\, e_{\rm outfl}(t - \Delta t)\,,
\end{multline}
\begin{equation}
e_{\rm outfl}(t) = \Phi(R_{\rm 200c},t) + e_{\rm hot}^{\rm therm}(t) + \frac{1}{2} v_{\rm outfl}^2(R_{\rm 200c},t)\,,
\end{equation}
\begin{multline}
v_{\rm outfl}(R_{\rm 200c},t) = \frac{2\, R_{\rm 200c}(t)}{t_{\rm outfl}}\\
 - \sqrt{2 \left[ e_{\rm outfl}(t - \Delta t) - \Phi(0,t) - e_{\rm hot}^{\rm therm}(t) \right]}\,.
\end{multline}
\end{subequations}
%


\subsection{Reincorporation of ejected gas}

When gas is added to the ejected reservoir from the outflowing reservoir, we calculate an expected time for that gas to reincorporate back into the halo.
From the specific energy of the outflowing reservoir, we can calculate both the speed of the ejected gas at the virial radius, and the radius at which the gas will cease to have kinetic energy and must turn around.
We approximate the average speed the gas moves during this period as half its ejected speed at the virial radius.
With the understanding that the gas must travel the distance from $R_{\rm 200c}$ to $R_{\rm turn}$ twice, we effect
\begin{subequations}
\begin{equation}
t_{\rm reinc}^{\rm inst} = \frac{4 \left( R_{\rm turn} - R_{\rm 200c} \right)}{v_{\rm outfl}(R_{\rm 200c})}\,,
\end{equation}
\begin{equation}
\Phi(R_{\rm turn}) + e_{\rm hot}^{\rm therm} = e_{\rm outfl}\,.
\end{equation}
%
\end{subequations}
To solve these simultaneous equations efficiently in the code, we approximate the shape of the potential outside the virial radius with an analytic NFW profile.

As with the other time-scales, we average the reincorporation time-scale as more mass is added:
\begin{multline}
t_{\rm reinc}(t + \Delta t) = {\rm max} \bigg[0,\\
 \frac{m_{\rm ejec}(t)\, t_{\rm reinc}(t) + \Delta m_{\rm outfl \rightarrow ejec}\, t_{\rm reinc}^{\rm inst}}{m_{\rm ejec}(t) + \Delta m_{\rm outfl \rightarrow ejec}} - \Delta t \bigg]\,.
\end{multline}
Ejected mass is then reincorporated directly into the hot-gas reservoir at a rate set by
\begin{equation}
\Delta m_{\rm eject \rightarrow hot} = {\rm min} \! \left[1,~\frac{\Delta t}{t_{\rm reinc}} \right] m_{\rm ejec}\,.
\end{equation}



\begin{figure}[H]
\centering
\includegraphics[width=\textwidth]{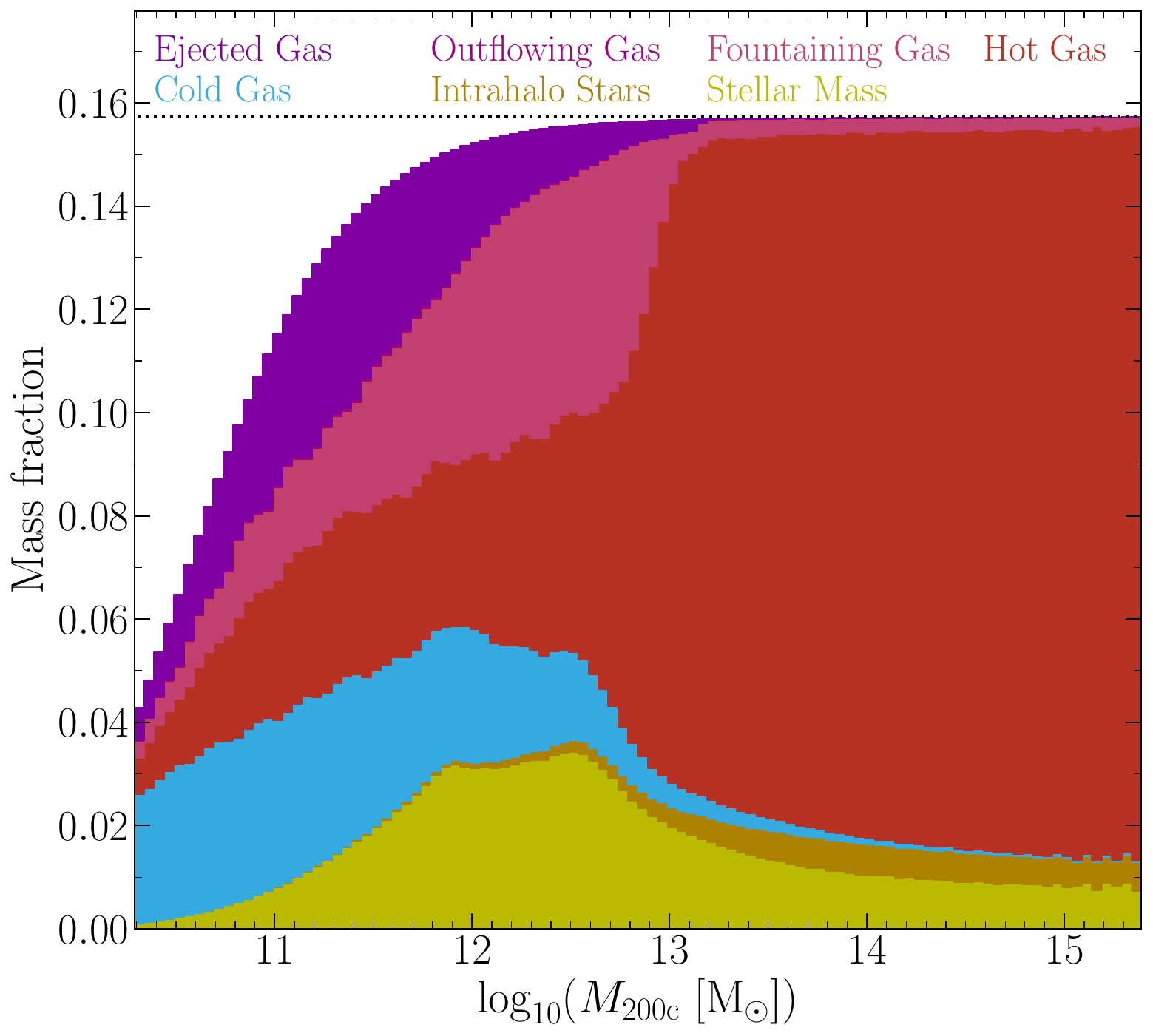}
\vspace{-0.2cm}\caption{
The breakdown of baryons in \ds~haloes at \zo.
We calculate the average mass of each reservoir for all haloes in each halo mass bin of width 0.05\,dex.
Bars are stacked on top of each other, such that the height of all bars in a column, less the ejected gas, gives the total baryon fraction inside the halo.
All matter for satellite galaxies inside the virial radius is included.
Black holes are reasonably neglected from this plot.
The dotted horizontal line represents the cosmic baryon fraction.
Photoionization heating suppresses the total baryon fraction below the cosmic value for $M_{\rm 200c} \! \lesssim \! 10^{13}\,{\rm M}_\odot$.
The ejected gas contributes towards the halo baryon fraction for the purposes of cosmic accretion, even though it is outside the halo (as discussed in Section \ref{sec:cooling}).
The outflowing-gas reservoir is nigh negligible at all halo masses.
}
\label{fig:breakdown}
\end{figure}

\subsection{Baryon breakdown}

With all the baryon reservoirs and the equations for how mass moves between them in \ds~now described, it is informative to see how the baryonic mass inside haloes is broken between them at \zo.
We show precisely this in Fig.~\ref{fig:breakdown}.

The stellar mass fraction as a function of halo mass is as one would expect, peaking at $M_{\rm 200c} \! \simeq \! 10^{12}\,{\rm M}_{\odot}$.
The CGM fraction increases monotonically, dominating the baryon budget at high halo masses.
Of the mass in the CGM, the outflowing reservoir contains little mass at all halo mass scales, showing that the time spent in this transitory reservoir is less than the time spent in the ejected reservoir before reincorporation.
A significant portion of the CGM is in the fountain reservoir for $M_{\rm 200c} \! \lesssim \! 10^{13}\,{\rm M}_{\odot}$.
This highlights that most of the gas reheated from stellar feedback does not have sufficient energy to escape the halo, and that notably less star formation and stellar feedback operates in the group--cluster regime.
At these mass scales, after fountaining gas reincorporates into the hot-gas reservoir, it is offset from cooling from radio-mode AGN feedback.

\begin{figure*}
\centering
\includegraphics[width=\textwidth]{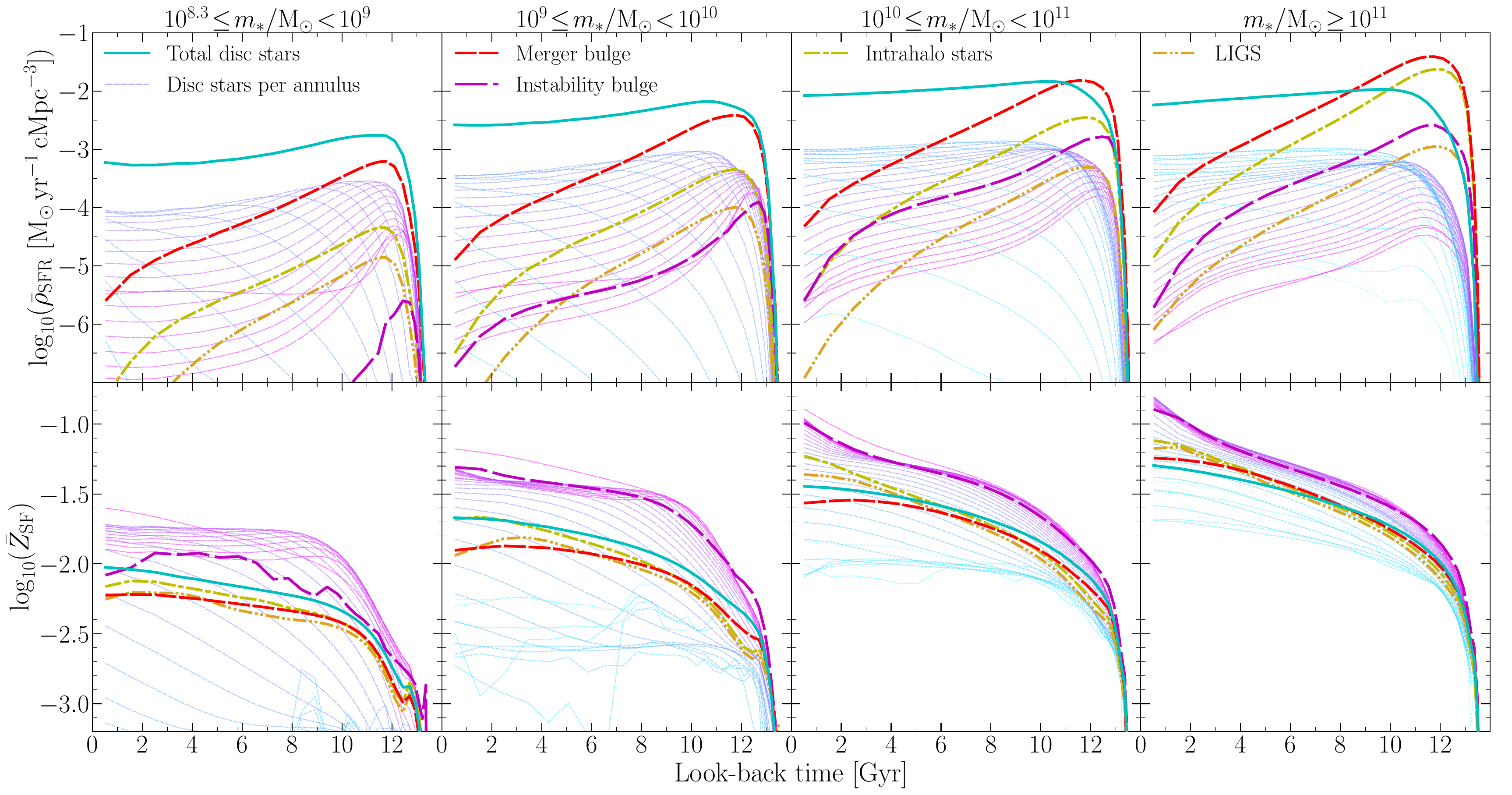}
\vspace{-0.6cm}\caption{
Top panels: Contributions to the cosmic star formation density history based on the stellar components of galaxies in three wide stellar-mass bins.
This is reconstructed from the \zo~output of \ds~using the stellar-age bins in each stellar component after multiplying the mass in those components by $m_{\rm *,pop}^{\rm form} / m_{\rm *,pop}^{\rm rem}$.
Bottom panels:~the average metallicity of those stars, equal to the average metallicity of the star-forming gas at the time of formation.
The dashed, thin lines are the per-annulus contribution to the disc stars, smoothly changing colour from teal for outer annuli ($i \rightarrow N_{\rm ann}$) to blue for intermediate annuli, to magenta for inner annuli ($i \rightarrow 1$; i.e.~towards the instability-driven bulge, which functions like a `zeroth' annulus).
}
\label{fig:sfh}
\end{figure*}

The age bins in the stellar reservoirs allow us also reconstruct the star formation histories of galaxies on a per-component basis.
As an example of \ds's abilities in this area, we use the summed birth mass of stars in common stellar components across galaxies within wide stellar-mass bins to build cosmic star formation density histories in the top panels of Fig.~\ref{fig:sfh}.
Because each stellar component has its own metallicity, we can also reconstruct the average metallicity of the star-forming gas as a function of cosmic time for each component.
The averages of these for the binned galaxies are shown in the bottom panels of Fig.~\ref{fig:sfh}.
This plot shows that high-mass galaxies are dominated in stellar mass by merger-driven bulges, while the stellar mass in low-mass galaxies is predominantly in discs.
It also shows that spheroidal stars (be they in either bulge or the intrahalo stars; and even the LIGS) have a notably higher average age than disc stars.
Except for the highest-mass galaxies, merger-driven bulges tend to be metal-poor relative to discs.
There is also a smooth transition in both the age and metallicity of disc stars, getting younger and more metal-rich from the outskirts to the centre and into the instability-driven bulge.
These predictions from \ds~can be tested with component-based SED-fitting models applied to large-scale galaxy surveys (e.g.~\citealt{profuse}; also see \citealt{bellstedt20}).


\section{Environmental stripping of gas and stars}
\label{sec:env}

When a subhalo becomes the satellite of a host halo, several rules for its evolution change.
Most notably, baryon reservoirs become prone to environmental effects that can deplete their contents.
In \ds, we now account for the effects of starvation, strangulation, tidal stripping, and ram-pressure stripping.

In principle, central haloes can also be affected by the environment of nearby large-scale structure.
For example, a halo falling into a cosmic filament from a void should experience tidal and ram pressure effects from that filament.
In this section, we describe how environmental effects are handled for both central and satellites galaxies in \ds.


\subsection{Central galaxies}
\label{ssec:envcen}

While we have no explicit treatment of large-scale environmental effects on central haloes, this is considered implicitly from how the galaxies are evolved within the merger trees of the $N$-body simulation.
Where an increase in virial mass leads to the addition of new baryons in the halo in order to maintain a target baryon fraction (Section \ref{sec:cooling}), so too are baryons removed when the virial mass decreases to ideally maintain this same fraction.
Only the CGM and IHS of a central are stripped in this circumstance.
This means that the baryon fraction can become higher than the target value in rare circumstances where removing the CGM and IHS outright is not enough to bring it back down.

When the CGM and IHS are stripped this way, they are done so proportionally to their mass.
Those stripped gas and stars are transferred to the central's LIGM and LIGS, respectively.
This means that those baryons will return to the halo preferentially over fresh primordial gas, should the halo later grow (Section \ref{sec:cooling}).
Helpfully, this mitigates against potential halo-finding and/or tree-building issues that can lead to artificial, momentary losses in halo mass \citep[see discussions on this and related issues by][]{behroozi13,srisawat13,elahi19a,elahi19b}; 
should this happen in \ds, baryons will initially be taken out of the halo, but then the \emph{same} baryons (i.e.~gas with the same metallicity and stars with the same metallicity and age distribution) will return a snapshot later.

Note that in a change from previous versions of the model, we do not remove the ejected gas of central haloes when they are tidally stripped, as this mass is already considered to be beyond the virial radius.


\subsection{Satellite galaxies}
Much more detail is given to (and needed for) the environmental effects on satellites.
To start with, immediately when a galaxy becomes a satellite, all gas in its LIGM is transferred to the host's equivalent reservoir.
This is simply in line with how we have defined that reservoir; it does not make sense to track a local intergalactic medium for a satellite.
The same applies for LIGS and LIGBHs.

Satellites are shut off from cosmological gas accretion entirely.
This \emph{starves} the CGM from any replenishment.
However, gas can continue to cool from the CGM of the satellite onto its ISM.
Likewise, while the satellite remains outside its host virial radius, its CGM will continue to reincorporate gas from the satellite's ejected reservoir.
This is physically justifiable from the results from hydrodynamic simulations that have shown that satellites can still continue to accrete new gas, even while they are being affected by environmental effects that remove gas \citep{wright22}.

The CGM of a satellite is gradually stripped, an effect often referred to as `strangulation' in the literature.
In a change from earlier versions of \ds, we now also strip the intrahalo stars of a satellite gradually, alongside the CGM (previously, the IHS was instantly transferred to the central upon infall).
While this change is inconsequential for galaxy evolution, it allows us to track the stellar haloes of groups after they fall into a cluster.
The rate of stripping of the CGM and IHS is based on (as follows) the subhalo mass-loss rate from the $N$-body simulation, which is caused by gravitational tides.
We routinely check the total baryonic mass of the satellite subhalo relative to its expected baryon fraction and total subhalo mass.
Should the former be above expectation, we proportionally remove mass (and metals) from the CGM and IHS to bring the baryonic mass as close to expectation as possible,%
\footnote{We say `as close to expectation as possible' rather than definitively saying it will be equal to expectation because it is possible to remove all the CGM and IHS of a satellite and for that to not be enough to restore the expected baryon fraction.}
transferring that mass to the respective reservoirs of the central.
The decision to strip the CGM and IHS of satellites this way is to be consistent with how those same reservoirs are stripped for centrals when their halo mass drops (Section \ref{ssec:envcen}).
This philosophy extends from \citet{croton16}.

In truth, the above paragraph is only an accurate description of a satellite that has crossed the virial radius of its host.
Until that point, we preference removal of the ejected reservoir of the satellite over the CGM.
Any stripped gas and stars of satellites outside $R_{\rm 200c}^{\rm host}$ actually go to the respective local intergalactic reservoir.
Only if the ejected gas mass of the satellite is zero does \ds~start to strip the CGM.
This is expedited by the choice that
once the satellite crosses $R_{\rm 200c}^{\rm host}$, its ejected reservoir is transferred to the central's CGM immediately and in its entirety.
The assumption is that the gravitational influence of the host halo is sufficiently larger than the satellite, such that any ejected gas can no longer be reincorporated into the satellite's CGM.
Because this gas is lost inside the virial radius of the host, there should arguably be no delay in its incorporation to the host's CGM, hence why we do \emph{not} put it into the central's ejected reservoir.  

Material from satellite \emph{galaxies} themselves (cold gas and stars in the disc) is also stripped, but follows a different procedure.
For this, we calculate the strength of both tides and ram pressure on the satellite galaxy.
Previous versions of \ds~only accounted for ram pressure, which only affects the gas.
However, we only turn these processes on once $m_* + m_{\rm cold} \! > \! m_{\rm hot}$.
The reason for this is that the hot-gas halo should initially shield the cold-gas disc from ram pressure, as should the CGM and IHS experience tidal stripping before the baryons in the galaxy disc that are more tightly gravitationally bound.
Our exact criterion for switching this shielding off is not theoretically derived; it is simply maintained from \citet{sb17}.
Other criteria for the same phenomenon have been used in other semi-analytic models \citep[e.g.][]{cora18}.

We approximate the tidal radius of a satellite galaxy based on the formula of \citet*[][cf.~their equation 6]{tormen98}:
\begin{subequations}
\label{eq:rtidal}
\begin{equation}
r_{\rm tidal} = R_{\rm sat}\, M_{\rm sat}^{1/3}\, \Upsilon^{-1/3}\,,
\end{equation}
\begin{equation}
\Upsilon \equiv
\left\{
\begin{array}{l r}
2\, M_{\rm halo}(<\!R_{\rm sat}) & \\
 - R_{\rm sat} \left. \dfrac{{\rm d} M_{\rm halo}(<\!R)}{{\rm d} R}\right|_{R=R_{\rm sat}}\,, & R_{\rm sat} \! < \! R_{\rm 200c}^{\rm host}\\
  & \\
2\, M_{\rm 200c}^{\rm host}\,, & R_{\rm sat} \! \geq \! R_{\rm 200c}^{\rm host}\\
\end{array}
\right.\,.
\end{equation}
\end{subequations}
Strictly, this is only accurate for satellites with circular orbits where $r_{\rm tidal} \! \ll \! R_{\rm sat}$ and $M_{\rm sat} \! \ll \! M_{\rm halo}(<\!R_{\rm sat})$ (and therefore $M_{\rm sat} \! \ll \! M_{\rm 200c}^{\rm host}$).%
\footnote{\citet[][see their fig.~15]{springel08} show that, on average, equation (\ref{eq:rtidal}) reproduces the edge of subhaloes in simulations as identified by {\sc subfind}.
This supports the use of this equation as an approximate tidal radius, even when orbits are not circular. }
However, we only enforce the lattermost criterion for tidal stripping to take place, requiring that $M_{\rm sat} \! < \! 0.1\,M_{\rm halo}(<\!R_{\rm sat})$ (and by extension that $M_{\rm sat} \! < \! 0.1\,M_{\rm 200c}^{\rm host}$).
With that criterion passed, we remove all gas and stars in each disc annulus where $r_{\rm outer}^{(i)} \! \geq \! r_{\rm tidal}$.
The stripped gas and stars are respectively transported to the CGM and IHS of the central if $R_{\rm sat} \! \leq \! R_{\rm 200c}^{\rm host}$. If the satellite is instead outside the virial radius, the stripped gas and stars go to the respective local intergalactic reservoirs.
In the instance that $r_{\rm tidal} < r_{\rm outer}^{(1)}$, we trigger a tidal disruption event where \emph{all} the satellite's baryons are transferred to the CGM and IHS, and the satellite's existence is no longer considered thereafter.
As we explain in Section \ref{sec:mergers}, this is not the only means by which a disruption event is triggered.
With the exception of tidal disruption events, the stellar bulge of a satellite galaxy is left unaffected by environmental effects in \ds.

With tidal effects resolved, any remaining gas in the ISM of a satellite is then subject to ram-pressure stripping.
Consistent with previous versions of \ds, for each annulus, we calculate the ram pressure exerted on a satellite galaxy as
\begin{equation}
P_{\rm ram} = \rho_{\rm hot}^{\rm cen}(R_{\rm sat})\, V_{\rm sat}^2
\end{equation}
and compare this against the gravitational restoring force per unit area for each annulus, approximated as
\begin{equation}
P_{\rm restore}^{(i)} = 
\left\{
\begin{array}{l r}
2\uppi\, G \, \Sigma_{\rm cold}^{(i)} \left( \Sigma_{\rm cold}^{(i)} + \Sigma_{*}^{(i)} \right)\,, & \theta_{\rm disc} \! \leq \! 10^\circ \\
2\uppi\, G \, \Sigma_{\rm cold}^{(i)\,2} \,, & \theta_{\rm disc} \! > \! 10^\circ \\
\end{array}
\right.
\end{equation}
\citep[cf.][]{gunn72}.
For each annulus where $P_{\rm ram} \! > \! P_{\rm restore}^{(i)}$, its gas is transferred to the central's CGM (or LIGM if it is beyond $R_{\rm 200c}^{\rm host}$).


\section{Galaxy mergers}
\label{sec:mergers}

Similar to \citet{croton16}, a galaxy merger or disruption event occurs in \ds~when the baryon fraction of a subhalo rises above a threshold value, $f^{\rm thresh}_{\rm bary}$.
 In previous versions of SAGE and \ds, the subhalo baryon fraction only considered the cold-gas and stellar mass of the satellite galaxy it housed.
We now further include the CGM, IHS, and black-hole masses of the satellite in this calculation.
Where previously a threshold fraction of 1.0 was used, we now impose a threshold fraction of 0.5.
Recall that the subhalo mass is the total sum of all particles associated with that object from the underlying $N$-body simulation.
Further recall that that mass represents the sum of \emph{both} dark matter and baryons, but that all matter in the $N$-body simulation \emph{behaves} like dark matter.
As a subhalo is tidally stripped, both its dark-matter and hot-gas mass decline.
But the stellar and cold-gas content of a satellite typically take longer to be reduced by environmental effects (Section \ref{sec:env}).
This means the baryon fraction of a subhalo tends to get larger the longer it is a satellite.
In principle, it could continue to rise to arbitrarily large numbers greater than 1\,---\,e.g.~when a subhalo composed of thousands of particles at infall is continually stripped down to the minimum of 20 particles\,---\,unless we impose an extra limit.
Our physical argument for triggering a merger or satellite disruption event once a subhalo's baryon fraction rises to 0.5 is that the implicit assumption in the $N$-body simulation\,---\,that the satellite galaxy behaves like a dark-matter-dominated object\,---\, no longer holds.
It presumably must have reached the point where either the un-modelled influence of baryons on the satellite's orbit should have caused it to merge  with the central galaxy \emph{or} the stellar and cold-gas content of the galaxy should have been tidally stripped.

The above means that satellite galaxies are generally merged or disrupted \emph{before} their subhalo is `lost' in the merger trees of the $N$-body simulation (their being lost in the merger trees triggers a merger or disruption in the same fashion, but these occasions are rare).
This also means there are no `orphan' galaxies in \ds.

Arguably, it would be better to include orphan galaxies (i.e.~detach satellite galaxies from their host subhalo and let them continue to evolve until different criteria are met for merging or disruption), but this requires changes to \ds's framework beyond the scope of this version of the model.
In practice, this means that the intrahalo and local intergalactic stellar and gas components of \ds~centrals include the content that other models might have otherwise associated with orphan galaxies.


\subsection{Merger or disruption event?}

In the above situation, how do we then choose between merging the satellite or disrupting it?  The answer lies in how long the subhalo has been a satellite.
The moment a subhalo becomes a satellite, a predicted merger time-scale is calculated.
If less than this time has elapsed when the above is triggered, the satellite presumably must have been disrupted.
Otherwise, it should have reasonably merged already or be in the process of doing so imminently.
In \citet{croton16}, this predicted time-scale was based on the dynamical-friction model of \citet{binney87}.
However, \citet{poulton20} has since shown that this time-scale\,---\,in addition to several other merger time-scale formulae commonly used in the literature\,---\,only works for a limited range of infall scenarios, and even then carries a factor of $\sim$2 uncertainty.
Building on this, \citet{poulton21} built a more sophisticated model by fitting a function to time-scales derived from the meticulous tracking of subhalo orbits in cosmological simulations with the {\sc OrbWeaver} code.
We now adopt this time-scale in \ds.

The \citet{poulton21} time-scale, hereafter $t_{\rm merge}$, can be calculated as follows (cf.~their equations 4--7, 11--12).
\begin{subequations}
\begin{equation}
\frac{t_{\rm merge}}{5.5\, R_{\rm peri}^{0.2}} = 
\left\{
\begin{array}{l r}
\sqrt{\dfrac{R_{\rm 200c}^{\rm host}\, R_{\rm sat}^{1.6}}{G\, M_{\rm halo}(< \! R_{\rm sat})}}\,, & R_{\rm sat} < R_{\rm 200c}^{\rm host}\\\\
\dfrac{R_{\rm 200c}^{\rm host}\, R_{\rm sat}^{0.3}}{\sqrt{G\, M_{\rm 200c}^{\rm host}}} \,, & R_{\rm sat} \geq R_{\rm 200c}^{\rm host}\\
\end{array}
\right.\,,
\end{equation}
\begin{equation}
R_{\rm peri} = \frac{J_{\rm orb}^2}{(1+\varepsilon)\, G\, M_{\rm host}\, M_{\rm sat}\, M_{\rm redu}} \,,
\end{equation}
\begin{equation}
\varepsilon = \sqrt{ 1 + \frac{2\, E_{\rm orb}\, J_{\rm orb}^2}{(G\, M_{\rm host}\, M_{\rm sat})^2\, M_{\rm redu}} } \,,
\end{equation}
\begin{equation}
J_{\rm orb} = M_{\rm redu}\, | \vec{R}_{\rm sat} \times \vec{V}_{\rm sat} | \,,
\end{equation}
\begin{equation}
E_{\rm orb} = \frac{M_{\rm redu}\, V_{\rm sat}^2}{2} + M_{\rm sat}\, \Phi(R_{\rm sat}) \,,
\end{equation}
\begin{equation}
M_{\rm redu} \equiv \frac{M_{\rm sat}\, M_{\rm host}}{M_{\rm sat} + M_{\rm host}}\,,
\end{equation}
\begin{equation}
M_{\rm host} \equiv
\left\{
\begin{array}{l r}
M_{\rm halo}(<\!R_{\rm sat})\,, & R_{\rm sat} < R_{\rm 200c}^{\rm host}\\
M_{\rm 200c}^{\rm host}\,, & R_{\rm sat} \geq R_{\rm 200c}^{\rm host}\\
\end{array}
\right.\,.
\end{equation}
\end{subequations}
Every term in the above equations is known/calculable for each subhalo in \ds, making this straightforward to implement.
$R_{\rm sat}$ represents the distance between the central galaxy's centre of mass and the satellite galaxy's centre of mass.
$R_{\rm peri}$ represents the expected separation of the galaxies at the pericentre of their orbit with total angular momentum $J_{\rm orb}$, energy $E_{\rm orb}$, and ellipticity $\varepsilon$.
$V_{\rm sat}$ is the relative speed between the two galaxies.
$M_{\rm redu}$ is the `reduced' mass of the two-galaxy system.
The mass terms for the central and satellite\,---\,$M_{\rm host}$ and $M_{\rm sat}$, respectively\,---\,account for all matter, including dark matter.

When a disruption event occurs, if the satellite galaxy is inside the host's virial radius, the remaining entirety of the satellite's CGM, ISM, and ejected gas are transferred to the central's hot-gas reservoir.
Similarly, the entirety of the satellite's stellar mass (including its IHS) is transferred to the central's IHS.
The size of the central's IHS is then updated based on where the satellite was disrupted:
\begin{equation}
\langle R \rangle_{\rm IHS}^{\rm cen} (t + \Delta t) = \frac{m_{\rm IHS}^{\rm cen}(t)\, \langle R \rangle_{\rm IHS}^{\rm cen}(t) + \left( m_*^{\rm sat} + m_{\rm IHS}^{\rm sat} \right) R_{\rm sat}} {m_{\rm IHS}^{\rm cen}(t + \Delta t)}\,,
\end{equation}
where the average size of the IHS relates to its scale radius, $a_{\rm IHS}$, of a truncated \citet{hernquist90} sphere via
\begin{equation}
\frac{\langle R \rangle_{\rm IHS}}{a_{\rm IHS}} =  2 \left(\frac{a_{\rm IHS}}{R_{\rm 200c}} + 1\right)^2 \ln \! \left(\frac{R_{\rm 200c}}{a_{\rm IHS}} + 1 \right) - 2 \frac{a_{\rm IHS}}{R_{\rm 200c}} -3\,.
\label{eq:a_IHS}
\end{equation}
This equation can be solved for $a_{\rm IHS}$ iteratively.
If the disrupted satellite was outside the virial radius, the gas and stars of that satellite are instead transferred to the LIGM and LIGS components of the halo, respectively.

The resolution of a merger is more complicated.
How the remaining baryons are handled depends on whether the merger is `major' or `minor.'

A minor merger represents when one galaxy vastly outweighs the other.
Conversely, a merger is deemed major when the merging systems have comparable mass.
In practice, a major merger is defined in \ds~when
\begin{multline}
{\rm min}\left[m_{\rm *}^{\rm cen} + m_{\rm cold}^{\rm cen}, m_{\rm *}^{\rm sat} + m_{\rm cold}^{\rm sat} \right] \geq \\ f_{\rm major} \times  {\rm max}\left[m_{\rm *}^{\rm cen} + m_{\rm cold}^{\rm cen},  m_{\rm *}^{\rm sat} + m_{\rm cold}^{\rm sat} \right]
\end{multline}
(while the central always has a greater \emph{total} mass than its satellites, it is not strictly guaranteed that it has greater galactic baryonic mass).
The treatment of minor mergers remains similar to \citet{stevens16}, but the handling of the ISM in major mergers has notably changed.
To more accurately reflect the implicit assumption in the handling of minor mergers that $M_{\rm host} \! \gg \! M_{\rm sat}$, we set $f_{\rm major} \! = \! 0.1$.
Previous versions of the model had this set to 0.3.


\subsection{Resolving minor mergers}

The ISM of a satellite is assumed to be dispersed uniformly across a subset of the annuli of the central's disc.  
The orbital specific angular momentum of the satellite is projected onto central disc's rotation axis to find the annulus with equivalent $j$.  
Rather than dumping the entire ISM of the satellite into this one annulus of the central, we deposit it across several annuli.  
This is intended to account for the fact there is actually a \emph{distribution} of orbital $j$ of the satellite's ISM relative to the central, given that the satellite's ISM has its own rotation prior to merging.
We approximate this by selecting the maximum number of annuli ($l-k+1 \! \geq \! 1$) where
\begin{subequations}
\begin{equation}
j_{\rm outer}^{(k) {\rm cen}} > \left( \vec{j}_{\rm orb} \cdot \hat{j}_{\rm cold}^{\rm cen} \right) - j_{\rm spread}\,,
\end{equation}
\begin{equation}
j_{\rm inner}^{(l) {\rm cen}} < \left( \vec{j}_{\rm orb} \cdot \hat{j}_{\rm cold}^{\rm cen} \right) + j_{\rm spread}\,,
\end{equation}
\begin{equation}
\vec{j}_{\rm orb} \equiv \vec{R}_{\rm sat} \times \vec{V}_{\rm sat}\,,
\end{equation}
\begin{equation}
j_{\rm spread} \equiv  \left( \vec{R}_{\rm sat} \cdot \hat{j}_{\rm cold}^{\rm cen} \right)\, \left( \hat{j}_{\rm cold}^{\rm sat} \cdot \hat{j}_{\rm cold}^{\rm cen} \right)\, V_{\rm 200c,infall}^{\rm sat}\,.
\end{equation}
\end{subequations}
For each annulus $i$ of the central's gas disc where $k \! \leq \! i \! \leq \! l$, we deposit $m_{\rm cold}^{\rm sat} / (l-k+1)$ into that annulus.
Metals are deposited in proportion to the satellite's \emph{average} ISM metallicity.
This is identical to the procedure from \citet{stevens16}.

In reality, the manner in which we calculate $j_{\rm spread}$ or how we distribute the satellite's ISM across annuli $k$ through $l$ in the central are of little consequence.  The sudden addition of mass to those annuli typically drives a local instability, which promptly redistributes some of that mass over a larger number of annuli (and also removes some of that gas from those annuli due to feedback from instability-driven star formation\,---\,see Section \ref{ssec:gas_instab}).

The above takes place regardless of whether the orbit of the satellite was prograde or retrograde relative to the central's gas disc.
However, if the orbit was retrograde, this would artificially add angular momentum to the descendent galaxy's disc, when instead there should be a net loss.
To rectify this violation of angular-momentum conservation, after a retrograde minor merger, we effectively shrink the descendant's gas disc by proportionally transferring matter inward, such that the resulting $j_{\rm cold}^{\rm cen}$ is as it ought to be; i.e.
%
\begin{multline}
\vec{j}_{\rm cold}^{\rm cen}(t + \Delta t) = \Bigl[ m_{\rm cold}^{\rm cen}(t) + m_{\rm cold}^{\rm sat} \Bigl]^{-1}
 \Bigl[ m_{\rm cold}^{\rm cen}(t)\, \vec{j}_{\rm cold}^{\rm cen}(t) \\
+ m_{\rm cold}^{\rm sat}\,  \hat{j}_{\rm cold}^{\rm cen}  \left( \vec{j}_{\rm orb} \cdot \hat{j}_{\rm cold}^{\rm cen} \right)\Bigl]\,.
\end{multline}
%
%
This equation is imposed before determining whether any local instabilities are triggered.

We maintain the choice that all stars in the satellite galaxy are transferred to the merger-driven bulge of the central from previous versions of the model.
While we would expect collisional gas to settle into the disc structure of the ISM,
it would be contrived to force the orbits of collisionless stars acquired from a minor merger to match the stellar disc.
The merger-driven bulge exists in the model precisely to account for stars whose orbits are less ordered (there is no analogue for gas).
While this principle has not changed, we have added several details in how the merger-driven bulge is updated after a minor merger.
For example, we now also transfer any remaining IHS of the satellite to the central's merger-driven bulge too.
That is,
\begin{equation}
m_{\mb}^{\rm cen}(t + \Delta t) = m_{\mb}^{\rm cen}(t) + m_*^{\rm sat} + m_{\rm IHS}^{\rm sat}\,.
\end{equation}
In all likelihood, most of the IHS of the satellite would have already been stripped (per Section \ref{sec:env}) and any remaining IHS should be relatively nuclear (and therefore small compared to the size of the central's bulge).

As another new feature in \ds, we now approximately track the angular momentum and velocity dispersion of the merger-driven bulge, which in turn is used to calculate its size.  
After a minor merger, we update the angular momentum of the central by summing the rotational angular momentum of what the bulge originally had with the total rotational angular momentum of the satellite.  
That is, we conserve bulge rotation, but neglect any transfer of orbital angular momentum.  
The assumption is any orbital angular momentum is lost to dynamic friction before stars are added to the bulge.  
The specific angular momentum of the merger-driven bulge after a minor merger is then:
\begin{multline}
\vec{j}_{\mb}^{\rm cen}(t + \Delta t) =  m_{\mb}^{\rm cen~-1}(t + \Delta t) \bigl[ m_{\mb}^{\rm cen}(t)\, \vec{j}_{\mb}^{\rm cen}(t)\\ + m_{\rm *,disc}^{\rm sat}\, \vec{j}_{\rm *,disc}^{\rm sat} + m_{\mb}^{\rm sat}\, \vec{j}_{\mb}^{\rm sat} \bigl]\,.
\end{multline}

For the merger-driven bulge's velocity dispersion, we add the mass-weighted squares of the dispersion of each relevant component and a further term that accounts for the radial motion of the satellite relative to the central.  That is,
\begin{multline}
\label{eq:sigma_bulge}
\sigma_{\mb}^{\rm cen~2}(t + \Delta t) = m_{\mb}^{\rm cen~-1}(t + \Delta t) \Biggl[m_{\mb}^{\rm cen}(t)\, \sigma_{\mb}^{\rm cen~2}(t) \\
+ m_{\mb}^{\rm sat}\, \sigma_{\mb}^{\rm sat~2} + m_{\ib}^{\rm sat}\, \sigma_{\ib}^{\rm sat~2}
+ \left(m_*^{\rm sat} + m_{\rm IHS}^{\rm sat}\right) \\ \times \left(\vec{V}_{\rm sat} \cdot \hat{R}_{\rm sat}  \right)^2
+ \sum_{i=1}^{N_{\rm ann}} m_{\rm *,disc}^{(i) {\rm sat}}\, \sigma_{\rm *,disc}^{(i) {\rm sat}~2} \Biggl] \,.
\end{multline}

The CGM and ejected reservoirs of the satellite are added to the hot-gas reservoir of the central.
In essence, we assume there is no significant delay for that gas to become available for cooling.
There is no impact on whether the satellite was inside or outside the central's virial radius at the time the merger was triggered.


\subsection{Resolving major mergers}

During a major merger, the structure of the remnant galaxy is changed significantly from either of its progenitors.
First, we define the plane for the new gas disc to exist, chosen to be that of the last orbital plane between the two galaxies.
We then calculate the projected specific angular-momentum component of each gas disc annulus in each galaxy \emph{relative to the centre of mass of the two-galaxy system} in the direction of this plane.
The annulus with the corresponding $j$ in the new disc is where the gas from the old annulus is placed.
Note the difference from earlier versions of the model in our choosing the centre-of-mass frame to determine how the new gas disc is built.
Not only will the two progenitor discs be mixed, but the surface density profile of the new disc is likely to look entirely different to both progenitors.
Similar to the case for minor mergers, if any retrograde gas is added to the new disc, we shrink it to ensure angular-momentum conservation is satisfied in the plane.
In earlier versions of the model, we triggered a specific starburst when a retrograde merger occurred.
We no longer do this, as this is effectively taken care of by the instability prescription.

All stars in the discs and bulges of both progenitors are moved to the merger-driven bulge of the descendant after a major merger, while the two IHS components of the progenitors are summed to form the IHS of the descendant.
We set the merger-driven bulge to have a specific angular momentum equal to the orbital specific angular momentum of the two-progenitor system in the centre-of-mass frame.
The velocity dispersion of the merger-driven bulge is updated consistently with how it is during minor mergers (where terms for the central that are only present for the satellite in equation \ref{eq:sigma_bulge} are added).
The size of the merger-driven bulge is calculated by solving the Jeans equation, in the same manner as for the instability-driven bulge, except this time accounting for the non-zero angular momentum:
\begin{equation}
\label{eq:a_mbulge}
a_{\mb} = 3 \left[ \frac{1}{\sigma^2_{\mb}} \left( \frac{{\rm d} \Phi}{{\rm d} R} - \frac{j_{\mb}^2}{R^3} \right) - \frac{1}{R} \right]^{-1} - R\,.
\end{equation}

The CGM and the ejected reservoir of the satellite are instead transferred to the fountain reservoir of the central, and the fountain time is updated when a major merger occurs.
The idea here is that the collision of the two comparable-mass CGMs should drive turbulence in the gas, causing a delay for it become available for cooling onto the remnant galaxy.


\section{Calibration of the model}
\label{sec:cali}

Our new version of \ds~stands out from most\,---\,possibly \emph{all}\,---\,other semi-analytic models of galaxy formation in its small number of free parameters.
We only allow three parameters to be calibrated in the model:
\begin{itemize}
\item $f_{\rm move}^{\rm gas}$, the fraction of Toomre-unstable gas that moves to adjacent annuli, as opposed to being consumed in a starburst with feedback (Section \ref{sec:instab}).
This strongly influences star formation rates, the structure and morphology of galaxies, and the rate at which black holes grow.
To our knowledge, no other semi-analytic model has a parameter that is directly analogous to $f_{\rm move}^{\rm gas}$.
\item $\mathcal{E}_{\rm SN}$, the average energy per supernova that is coupled to the ISM/CGM (Section \ref{sec:feedback}).
This sets the normalisation for the strength of stellar feedback, which other models usually rely on additional free parameters for.
\item $\epsilon_{\rm AGN}$, the radiative efficiency of black holes, which controls the strength of AGN feedback (Section \ref{sec:agn}).
Again, most other models have more than one parameter to control the overall strength of AGN feedback.
\end{itemize}
Our three free parameters are calibrated against three observables:
\begin{itemize}
\item the \zo~stellar mass function (SMF) from \citet{driver22};%
\footnote{The bins widths used here are half that published in \citet{driver22}.  
This better captures the shape of the knee of the SMF.  
These data were provided privately by S.~P.~Driver.
Note as well that these data are at an average redshift of $z\!\simeq\!0.072$, but we have calibrated to our \zo~output.}
\item the \zo~\HI~mass function (HIMF) from \citet{jones18};
\item and the cosmic star formation density history (CSFH) from \citet{dsilva23}.
\end{itemize}
The best-fitting values for our parameters, following the method we describe below, are provided in Table \ref{tab:params}.

We invoke a particle-swarm optimisation (PSO) method to calibrate \ds.
PSO was introduced by \citet{kennedy95} and first adopted to calibrate a semi-analytic model by \citet{ruiz15}.
Using PSO, we
find the parameter set that provides the minimum multiplicative reduced $\chi^2$ across the three observables.
We deliberately multiply these together instead of adding, as this circumvents issues of the quoted errors carrying different meaning across the constraints, and it ensures that an improvement in one reduced $\chi^2$ of a given \emph{factor} is treated the same as any other constraint.
That is, we minimize
\begin{equation}
\label{eq:chi2}
\chi^2_{\rm joint} \equiv \prod_{\rm var} \left(\frac{1}{N_{p,{\rm var}} - 3} \sum_{p = 1}^{N_{p,{\rm var}}} \left[ \frac{y_{p,{\rm var}}^{\rm obs} - y_{p,{\rm var}}^{\rm mod}}{\sigma_{p,{\rm var}}^{\rm obs}} \right]^2 \right) \,,
\end{equation}

\begin{table}[H]
\begin{tabular}{c c l}
\hline
Parameter & Search range & Best fit \\\hline
$f^{\rm gas}_{\rm move}$ & 0.010--0.999 & 0.82 \\
$\mathcal{E}_{\rm SN}/(10^{44}\,{\rm J})$ & 0.050--3.000 & 0.81 \\
$\epsilon_{\rm AGN}$ & 0.0378--0.4226 & 0.31\\\hline
\end{tabular}
\vspace{-0.2cm}\caption{
Search range of \ds's free parameters for the PSO calibration and the output best-fitting values to two significant figures.  
The latter are used in all results throughout this paper.
There is no prior expectation on the value of $f_{\rm move}^{\rm gas}$; as it is defined to have a value in $(0,1)$, its search range conservatively covers almost all possible values.
$\mathcal{E}_{\rm SN}$ is expected to have a value close to $10^{44}$\,J, based on literature canon, and the search range around this value is deliberately conservatively wide.
The search range of $\epsilon_{\rm AGN}$ matches the maximum possible range for a black-hole accretion model where mass moves quasi-statically through a centripetal disc until reaching the innermost stable circular orbit, given the range of possible black-hole spin values.
}
\label{tab:params}
\end{table}
%

\noindent where ${\rm var = SMF, HIMF, CSFH}$.
Each $y$ (evaluation) and $\sigma$ (uncertainty in $y$) term for point $p$ is calculated in log-space.
We match bin sizes in \ds~to the observational data where possible (SMF, HIMF), and interpolate \ds's inherent bins otherwise (CSFH), to ensure the fairest comparison of every point.
Note that we do not include an uncertainty for \ds~points in equation (\ref{eq:chi2}).
In principle, we could have applied a Poisson statistic here (as done by \citealt{ruiz15}), but this would lead to down-weighting points near and after the knee of the SMF and HIMF, which are arguably the most important\,---\,and, in practice, the hardest\,---\,parts of the mass functions to get right.
Counting uncertainties are also already incorporated into the observational uncertainties.%
\footnote{The errors from \citet{jones18} are exclusively Poisson uncertainties.  These are lower limits on the true errors of those points.}

The PSO code makes use of the {\sc pyswarm} package for {\sc python} and acts as a wrapper around \ds.
{\sc pyswarm} implements a PSO technique similar to that discussed by \citet{ruiz15}.
A core difference, however, is that the initial particle positions are randomly drawn from a uniform distribution corresponding to the search space boundary conditions (central column of Table \ref{tab:params}), rather than a Latin Hypercube configuration.
The PSO code we use was originally developed for the {\sc Shark} semi-analytic model, with a full description due to be published elsewhere.
Minimal edits were made from the PSO codebase in the {\sc Shark} repository%
\footnote{The code was adapted from here:~\url{https://github.com/ICRAR/shark/tree/1dc89e8ad8d76490f783c68a91798430605b6f70/optim}.  Note that the {\sc Shark} PSO codebase has since been updated.}
to allow the code to interface with \ds.

Once the initial conditions are set, the particle positions at the subsequent iteration are computed according to
\begin{subequations}
\begin{equation}
\vec{\mathcal{P}}_m^{[k+1]} = \vec{\mathcal{P}}_m^{[k]} + \vec{\mathcal{V}}_m^{[k]}\,,
\end{equation}
\begin{equation}
\vec{\mathcal{V}}_m^{[k]} = \frac{1}{2}\vec{\mathcal{V}}_m^{[k-1]} + \xi_1 \left( \vec{\mathcal{P}}_m^{[b]} - \vec{\mathcal{P}}_m^{[k]} \right) + \xi_2 \left(  \vec{\mathcal{P}}_q^{[l]} - \vec{\mathcal{P}}_m^{[k]} \right)\,.
\end{equation}
\end{subequations}
$\vec{\mathcal{P}}_m^{[k]}$ represents the $m$th particle's position in the $(f_{\rm move}^{\rm gas},\,\epsilon_{\rm AGN},\,\mathcal{E}_{\rm SN})$ parameter space at iteration $k$.
$\vec{\mathcal{V}}_m^{[k]}$ is the particle's corresponding `velocity.'
$\vec{\mathcal{P}}_m^{[b]}$ represents particle $m$'s position at the iteration of best fit (lowest $\chi^2_{\rm joint}$) across all previous iterations ($b \! \leq \! k$).
$\vec{\mathcal{P}}_q^{[l]}$ represents the position for the global best fit for \emph{all} particles across all previous iterations (that of particle $q$ at iteration $l$).
$\xi_1$ and $\xi_2$ are random numbers, each drawn from a uniform distribution in $[0,0.5]$.
The initial velocity of each particle, $\vec{\mathcal{V}}_m^{[0]}$, is set to a random fraction ($<\! 1$) of the length of the parameter space in each dimension, and also in a random direction.
Should the velocity of a particle move it to outside the search range, it is manually shifted to the edge of the search range.

\begin{figure}[H]
\centering
\includegraphics[width=0.9\textwidth]{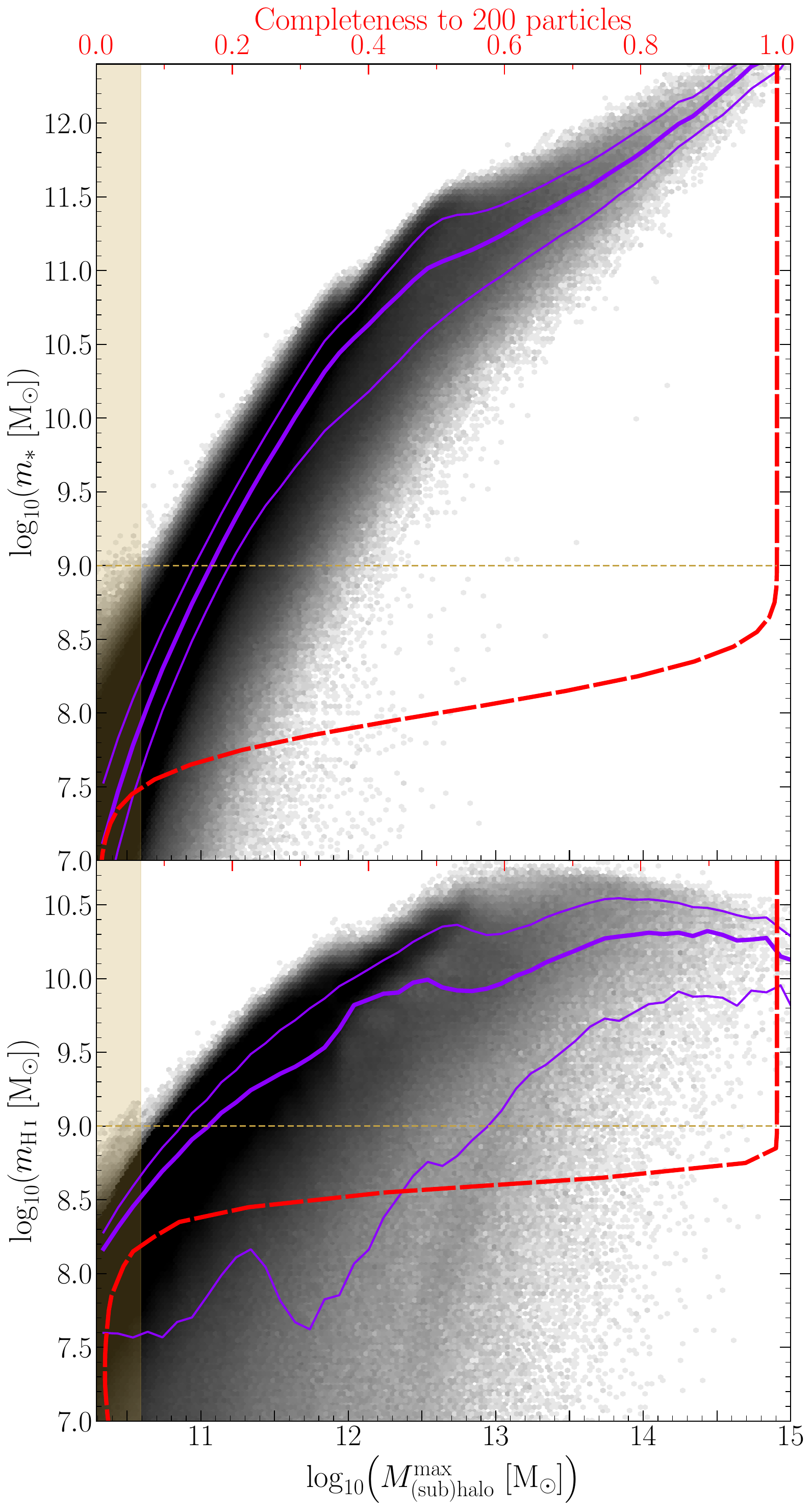}
\vspace{-0.2cm}\caption{
Historical maximum of the FoF or subhalo mass for \ds~galaxies versus their stellar mass (top panel) or \HI~mass (bottom panel) at \zo.
We define a galaxy as having 200 times the particle mass of MTNG on the $x$-axis as being resolved.
Calibration to the SMF and HIMF only occurs (approximately) above the respective masses where \ds~is complete to this mass resolution.
The long-dashed, red curves show the completeness as a function of stellar and \HI~mass.
The shaded region on the left is where galaxies are not reliably resolved by this definition.
The thin, dashed, horizontal lines are the masses above which we calibrate the SMF and HIMF.
As in other figures, hexbins and purple lines show the density of \ds~galaxies and the running 16th, 50th, and 84th percentiles.
}
\label{fig:completeness}
\end{figure}

We must also specify the range in the observational data to which the model is fitted.
We use the simulation's resolution to inform this decision.
Our loose criterion is that if galaxies of a given stellar or \HI~mass exclusively occupy (sub)haloes that have had at least 200 particles for one or more snapshots throughout their primary-progenitor branch in the merger trees, then we are `complete' at that mass, and can fit observations from there upwards.
200 particles is a semi-arbitrary number to require completeness to, but in the context of MTNG haloes having a minimum of 20 particles, our criterion strikes a balance between using as much information (as many haloes) as possible without giving too much trust towards MTNG's resolution limit.
The stellar or \HI~mass to which we are complete to 200 particles depends on the parameters of the model.
After exploring the parameter space, we found we could comfortably rely on being complete for $m_* \! \geq \! 10^9\,{\rm M}_\odot$ and $m_{\rm H\,{\LARGE{\textsc i}}} \! \geq \! 10^9\,{\rm M}_\odot$ (erring on the side of caution, especially in the case of the stellar mass).
These are the limits we adopt for the SMF and HIMF when calculating $\chi^2_{\rm joint}$.
In Fig.~\ref{fig:completeness}, we show stellar and \HI~mass distributions of \ds~galaxies as a function of their historical-maximum (sub)halo mass at \zo,%
\footnote{Although similar, Fig.~\ref{fig:completeness} is \emph{not} the same as a nominal stellar--halo or \HI--halo mass relation.  Those, instead, tend to have $M_{\rm 200c}$ on the $x$-axis and do not consider subhaloes as separate objects.}
and the 200-particle completeness as a function of each mass.

\begin{figure*}
\centering
\includegraphics[width=\textwidth]{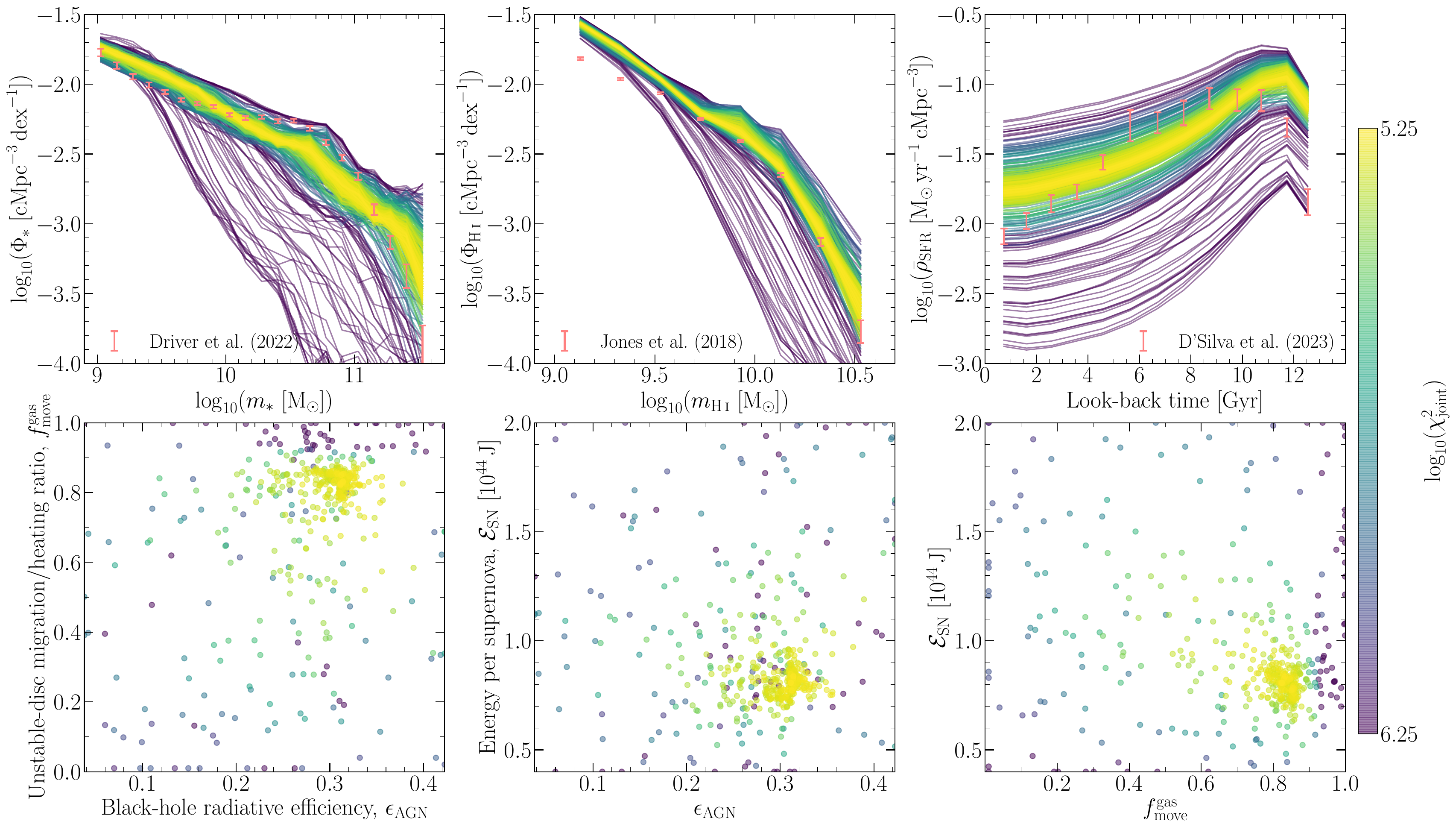}
\vspace{-0.2cm}\caption{
Three observed relations used to constrain the three free parameters in \ds.
\emph{Top left}:~stellar mass function at \zo~with observational data at  $z \! \simeq \! 0.072$ from \citet{driver22}.
\emph{Top middle}:~\HI~mass function at \zo~with observational data from \citet{jones18}.
\emph{Top right}:~cosmic star formation history with observational data from \citet{dsilva23}.
Each line in each of the panels corresponds to a run of \ds~with a different $(f_{\rm move}^{\rm gas},\,\epsilon_{\rm AGN},\,\mathcal{E}_{\rm SN})$ parameter set.
The colour indicates the combined reduced $\chi^2$ of the three constraints (equation \ref{eq:chi2}).
The \emph{bottom} panels show where each run lies in \ds's parameter space.
The best fit was found by applying a particle-swarm optimisation code.
See Section \ref{sec:cali} for further details.
}
\label{fig:cali}
\end{figure*}

In total, the PSO code went through two rounds of 16 average\footnote{One round went through 15 iterations, the other 17.  Both had converged.} iterations with 16 particles, meaning 512 parameter sets of \ds~were fully simulated with mini-MTNG.
Each instance of \ds~on mini-MTNG required $\sim$2.2\,CPUh of walltime.
It was important to do more than one run of the PSO code from start to finish to mitigate the possibility of finding a local minimum instead of the global minimum within the defined parameter space.
Fig.~\ref{fig:cali} shows the output of each run in the PSO chains against the calibration data, the relative goodness of fit of each run, and the particle positions in the model's parameter space.
We rounded the best-fitting parameters to two significant figures for the final run of the model on MTNG.

We note that \ds~can reproduce any of these three observables to higher accuracy if constrained exclusively by that dataset.
In principle, we could also have weighted the constraints in the joint $\chi^2$ calculation had we felt one was more important to reproduce.
But in the absence of an obvious reason why one constraint should have greater inherent value over another, we defaulted to having equal weights.
Evidently, there is some tension in the model with respect to simultaneously reproducing the SMF and HIMF. 
This could reflect (i) the simplicity of the molecular-fraction calculation in the model (Section \ref{ssec:fmol}); 
(ii) a deeper

\noindent limitation of the model's design;
(iii) uncertainties in the observational data, bearing in mind that no forward-modelling of any observational uncertainties has been done here;
or a combination thereof.
We leave a deeper assessment of this for future studies.

To highlight the significance of having so few free parameters, especially in the context of \ds~galaxies having multidimensional structure, we compare the number of free parameters and details of galaxy structure with other popular semi-analytic models in the literature in Table \ref{tab:pros}.
We count 15 free parameters for {\sc L-galaxies} based on the parameters listed that are not fixed in table 1 of \citet{henriques20}, i.e.~used in their Monte Carlo Markov Chain calibration procedure;
10 for {\sc Meraxes} based on table 1 of \citet{mutch16}, discounting the one parameter listed with a value of zero;
9 for SAG listed in table 1 of \citet{cora18} in their PSO calibration;
8 for SAGE based on inside knowledge and what is described as explicitly being calibrated/free in \citet{croton16};
and 18 for {\sc Shark} after cross-referencing the parameters listed in table 3 of \citet{shark} as explicitly described as free in the text and/or departing from their default value.
There are many more semi-analytic models in the literature that we could have added to Table \ref{tab:pros}, in principle.
However, we found some papers to be opaque in their description of how their model is parametrized, and we therefore felt that we could not make a fair comparison with them.
We expect those models fall in the same range as the others in Table \ref{tab:pros}.

\begin{table*}
    \begin{tabular}{l | c c c c c}
    & (i) & (ii) & (iii) & (iv) &(v) \\ \hline
    Model (year) & $N_{\rm param}^{\rm free}$ & $\Sigma_{\rm disc}(r)$ & $Z(r)$ & ${\rm d}\Sigma_*(r) / {\rm d}t_{\rm form}$ & Mass-loading \\\hline
    \ds~(this work) & 3 & Non-parametric & Non-parametric & Non-parametric & Local, non-parametric \\
    \ds~(2018) & 8 & Non-parametric & Non-parametric & None & Local, parametric \\
    {\sc L-galaxies} (2020)  & 15 & Non-parametric & Non-parametric & None & Local, parametric \\
    {\sc Meraxes} (2016) & 10 & Parametric & Flat & Flat & Global, parametric \\
    SAG (2018)  & 9 & Parametric & Flat & None & Global, parametric \\
    SAGE (2016) & 8 & Parametric & Flat & None & Global, constant \\
    {\sc Shark} (2018) & 18 & Parametric & Flat & Flat & Global, parametric \\ \hline
    \end{tabular}
    \vspace{-0.2cm}\caption{Stand-out features of \ds~compared to its previous version and other semi-analytic models in the literature.
Each row represents a different semi-analytic model.
Columns represent:~(i) the number of free parameters that model has; (ii) the model's treatment of radial disc structure; (iii) treatment of radial metallicity variation in galaxy discs, if any; (iv) treatment of stellar-age distributions as function of disc radius; (v) treatment of mass-loading from stellar feedback.
Only if an entry is `non-parametric' can that model make predictions for that property.
Note that the total number of nominal parameters a model may have is not what we equate as free parameters.
For a parameter to count as `free' by our definition, it has to have been described as being varied in a calibration procedure in the model paper.
The most relevant reference for each model is:~\ds~2018 \citep{stevens18}, {\sc L-galaxies} \citep{henriques20}, {\sc Meraxes} \citep{mutch16}, SAG \citep{cora18}, SAGE~\citep{croton16}, {\sc Shark}~\citep{shark}.
}
    \label{tab:pros}
\end{table*}

For a deeper dive into the complexity of calibrating (other) semi-analytic models, we refer the reader to \citet{carnage}.



\section{Closing remarks}

We have presented an overview of the salient features of the brand-new version of the \ds~semi-analytic model of galaxy formation.
\ds~bucks the trend of most semi-analytic models by numerically evolving the multidimensional structure of galaxy discs and by only having three free parameters that require calibration against observations.


The predictions from \ds~presented in this paper are but the tip of the iceberg as far as what the model can produce.
The code codebase is open to the community, which we hope will serve as a useful resource for research into galaxy formation and cosmology.


\section*{Acknowledgements}
All plots in this paper were built with the {\sc matplotlib} package for {\sc python} \citep{hunter07}.
Our analysis also made extended use of the {\sc numpy} package \citep{numpy} and {\sc pyswarm} (\url{https://github.com/tisimst/pyswarm}).

ARHS acknowledges receipt of the Jim Buckee Fellowship at ICRAR-UWA.
Parts of this research were supported by the Australian Research Council Centre of Excellence for All Sky Astrophysics in 3 Dimensions (ASTRO 3D), through project number CE170100013.

We used the OzSTAR supercomputer at Swinburne University of Technology to convert the MTNG merger trees and run \ds~for this project.
The OzSTAR supercomputing environment is managed through the Centre for Astrophysics \& Supercomputing with support from Swinburne IT. The supercomputing program receives continued financial support for operations from Astronomy Australia Limited and the Australian Commonwealth Government through the National Collaborative Research Infrastructure Strategy (NCRIS). The hardware for Ngarrgu Tindebeek was purchased through a grant from the Victorian Higher Education State Investment Fund (VHESIF).

We thank A.~Acharyya, S.~Bellstedt, J.~C.~J.~D'Silva, S.~P.~Driver, M.~G.~Jones, and A.~B.~Watts for supplying some of the observation-based data/results compared to in our figures.
ARHS further thanks C.~Power, the SU3 research group at ICRAR-UWA, and the {\sc Genesis} research group within ASTRO 3D for their comments in many meetings over the years;
R.~Tobar for help interfacing the PSO code with \ds;
V.~Springel and the wider MillenniumTNG collaboration for green-lighting the inclusion of the simulations in this project.

Finally, we thank the referee for finding several places in the paper where we had accidentally left out information.

\paragraph{Author contributions}
ARHS developed the model, led the code development, produced the data, and wrote the paper.
MS directly contributed to the code development, assisted in the model design, and interfaced \ds~with MTNG.
AR assisted in the development of the stellar-feedback model.
MWS developed and wrote the code for the initial PSO framework.
BH, CHA, and LH co-produced the MTNG simulations and merger trees.
All co-authors have read and supplied feedback on this text.

\paragraph{Data Availability Statement}
At the time of writing this paper, the MillenniumTNG data are private, but are due to be made publicly available in 2024.
We encourage anyone interested in obtaining access to \ds~data sooner to contact MS.

\end{multicols}

\end{document}